\documentclass[12pt]{article}

\advance\voffset by -2.0cm \advance\hoffset by -1.25cm
\usepackage{cite}
\usepackage{verbatim}
\usepackage{units}
\usepackage{mathtools}
\usepackage{amsmath}
\usepackage{amssymb}
\usepackage{amsthm}
\usepackage[shortlabels]{enumitem}
\usepackage{authblk}
\usepackage[bb=dsserif]{mathalpha}

\usepackage{wasysym}
\usepackage{esint}

\usepackage{color}
\usepackage[normalem]{ulem}

\usepackage{tikz}
\usetikzlibrary{patterns,decorations.pathmorphing,decorations.markings}
\usetikzlibrary{3d}
\usetikzlibrary {calc}
\usetikzlibrary{intersections}
\usepackage[colorlinks,linkcolor=blue]{hyperref}

\newcommand{\CC}{\mathbb{C}} 
\newcommand{\RR}{\mathbb{R}}

\newcommand{\ZZ}{\mathbb{Z}}

\newcommand{\A}{\mathcal{A}}
\newcommand{\B}{\mathcal{B}}
\newcommand{\C}{\mathcal{C}}

\newcommand{\Hh}{\mathcal{H}}

\newcommand{\Li}{\mathcal{L}}

\newcommand{\N}{\mathcal{N}}

\newcommand{\T}{\mathcal{T}}

\DeclareMathOperator{\Tr}{Tr}

\numberwithin{equation}{section}

\def\be{\begin{equation}}
\def\ee{\end{equation}}

\allowdisplaybreaks[3]
\title{A fresh view on string orbifolds}
\author{Stefano Giaccari\thanks{stefano.giaccari@pd.infn.it} }
\author{Roberto Volpato\thanks{volpato@pd.infn.it}}
\date{}
\affil{\small{Dipartimento di Fisica e Astronomia `Galileo Galilei' e INFN sez. di Padova\authorcr
Via Marzolo 8, 35131 Padova (Italy)}}
\begin{document}
\maketitle

\abstract{In quantum field theory, an orbifold is a way to obtain a new theory from an old one by gauging a finite global symmetry. This definition of orbifold does not make sense for quantum gravity theories, that admit (conjecturally) no global symmetries. In string theory, the orbifold procedure involves the gauging of a global symmetry on the world-sheet theory describing the fundamental string. Alternatively, it is a way to obtain a new string background from an old one by quotienting some isometry.\\
We propose a new formulation of string orbifolds in terms of the group of gauge symmetries of a given string model. In such a formulation, the `parent' and the `child' theories correspond to different ways of breaking or gauging all potential global symmetries of their common subsector. Through a couple of simple examples, we describe how the higher group structure of the gauge group in the parent theory plays a crucial role in determining the gauge group  and the twisted sector of the orbifold theory. We also discuss the dependence of this orbifold procedure on the duality frame.

}
\pagebreak
\tableofcontents

\section{Introduction}

In the last few years, there has been remarkable progress in refining the concept of global symmetry in quantum field theory (QFT) (see for example \cite{Gaiotto:2014kfa,Kapustin:2014gua,Cordova:2018cvg,Cordova:2022ruw,Gaiotto:2020iye,Freed:2022qnc} and references therein). The standard idea that symmetries form a group acting on the space of local operators in a QFT has been generalized in many different directions, so as to include higher form symmetries acting only on extended operators, higher group structures, where $p$-form symmetries of different degree $p$ are mixed in a non-trivial way, and non-invertible symmetries, where the group structure of symmetry operators is replaced by some suitable category of topological defects.

These developments have had consequences even for the study of string theory and, more generally, quantum gravity theory. There are various arguments, mostly based on black hole physics, supporting the conjecture that no global symmetry can exist in a consistent, UV complete theory of quantum gravity \cite{Polchinski:2003bq,Banks:2010zn}. This means that any symmetries arising in low energy effective descriptions of gravity must be either gauged, or broken at some higher energy scale.  This conjecture has been refined to include (or, rather, exclude) all higher form global symmetries, as well as topological defects \cite{Heidenreich:2020pkc,Heidenreich:2021xpr,McNamara:2021cuo,Rudelius:2020orz}. These refinements allow one to relate the absence of global symmetries to other conjectural properties of quantum gravity, such as the completeness of the spectrum \cite{Polchinski:2003bq,Banks:2010zn} or the cobordism conjecture \cite{McNamara:2019rup}.

Closely related to the idea of symmetry in QFT is the concept of orbifold. An orbifold is a way to obtain a new QFT from a `parent' one by gauging a discrete global symmetry. A considerable amount of works in the last years focused on generalizing this procedure to include all kind of new global symmetries \cite{Frohlich:2009gb,Carqueville:2012dk,Brunner:2013ota,Brunner:2013xna,Gaiotto:2014kfa,Kapustin:2014gua,Bhardwaj:2017xup,Tachikawa:2017gyf,Cordova:2018cvg}. 

The orbifold is also well defined in string theory, as a procedure to obtain a new string model from an old one. In fact, historically, many aspects of orbifolds in QFT were first described in the context of string theory \cite{Dixon:1985jw,Dixon:1986jc,Narain:1986qm,Hamidi:1986vh,Vafa:1986wx}.

This discussion led us to an apparent contradiction: a string model is supposed to provide a consistent UV complete theory of quantum gravity, and therefore there should be no global symmetries. Therefore, how can we understand the orbifold procedure in string theory as the `gauging of global symmetries'? This apparent paradox disappears when we look more closely at how an orbifold is defined in string theory.
One standard way to define a (perturbative) string theory model $A$ is by providing the two dimensional conformal field theory (CFT) $\C$ that describes the worldsheet of a fundamental string. Non-perturbative properties of the model can be deduced by consistency starting from the worldsheet theory. The worldsheet CFT is a quantum field theory \footnote{More precisely, the worldsheet theory is initially defined as a 2D (super-)gravity theory, and the CFT arises only after suitable gauge fixing. In this article, by `worldsheet theory' we will always refer to the latter CFT. Notice that the conjecture about the absence of global symmetries does not apply to gravity theories in two dimensions. } and can admit a group of global symmetries $\Gamma$. This group typically appears as a gauge group from the target space point of view. Then, it makes perfect sense to take the orbifold (in a QFT sense) of $\C$ by $\Gamma$, to obtain a new 2-D CFT $\C/\Gamma$. This new CFT can be taken as describing the worldsheet dynamics in a new string model $B$. We will call this procedure the \emph{worldsheet orbifold}.

Whereas this procedure is very well understood, nevertheless some of its features are not completely satisfactory. In string theory, the fundamental string is only one among various extended dynamical objects, including NS-branes, D-branes, etc. Non-perturbative string dualities exchange all such objects with each other. On the other hand, the worldsheet orbifold procedure that we have just described requires choosing one of these objects as a preferred one. There are of course other definitions of the orbifold in string theory, in particular in terms of the quotient of a string background by some isometry of the background. In general, though, such a description also depends on the choice of a particular duality frame.\footnote{There are some proposals (in particular exceptional field theory \cite{Hohm:2013vpa,Hohm:2013uia,Hohm:2014fxa,Blair:2018lbh,Berman:2020tqn}) to describe the low energy limit of string theory in a way that is manifestly duality invariant. It would be interesting to revisit the results of this article in this formalism; we will leave this to future work.}   It is natural to look for a procedure that brings us from string model $A$ to string model $B$ and that is `democratic' among all duality frames, and in particular that treats all dynamical objects on equal footing. 

There is a second puzzling aspect of string orbifolds. As mentioned above, in order to define the  orbifold of a quantum theory, one needs to choose a particular group of symmetries to `quotient' by. In fact, the reason why the orbifold procedure in QFT can be applied to a wide range of different  theories is that its definition only involves very few details about the theory itself -- essentially, just the structure of the symmetry group we want to divide by, and its action on the fields of theory.
In string theory, any group of symmetries is actually a gauge group, at least from the point of view of the target space. 
It is natural to ask whether one can simply regard string theory as a quantum theory  in spacetime with some gauge symmetry, and define a general `recipe' to obtain the orbifold theory, which depends only on the choice of a particular subgroup of the gauge group. One problematic aspect of this idea is that, unlike global symmetries, gauge symmetries are, in general, not intrinsic properties of a quantum theory, but might depend on the way the theory is described. This raises questions on whether the final outcome of any string orbifold procedure formulated in terms of  gauge symmetries might be description-dependent. On the other hand, in string theory there are various hints (the close relationship between gauge symmetries in target space and global symmetries on the worldsheet; the holographic correspondence between gauge symmetries in the bulk and global symmetries on the boundary; the completeness conjecture, which implies the existence of dynamical objects carrying all possible charges with respect to the gauge fields) that the presence of gauge symmetries is somehow a fundamental aspect of string models. As discussed further below, it is not unreasonable to expect that a description-independent notion of a string orbifold should exist.

\medskip

The goal of this article is to provide a first tentative proposal for such a procedure. Because it is formulated in terms of gauge symmetries in spacetime, rather than global symmetries on the worldsheet, we will call this procedure  a \emph{spacetime orbifold}. 

\medskip

The proposal is described in abstract terms in section \ref{s:stringorbifolds}, but it is best illustrated by the two simple examples that we consider in the subsequent sections: the orbifold  of type II superstring on a circle $S^1$ by a half-period shift (section \ref{s:halfperiod}), and   the orbifold of a toroidal compactification, again in type II, by the reflection along the torus coordinates (section \ref{s:reflection}).
In our approach, we focus on the structure of the gauge group $G$ (including the various higher form symmetries) of a given compactified string model. 
The dynamical objects are classified in terms of their couplings to  the corresponding gauge fields and in fact, the no-global symmetry conjecture allows to reconstruct the couplings in the worldvolume theories of extended objects relying only on the gauge group structure. The particular orbifold one wants to consider is specified by a choice of a finite subgroup $\Gamma$ in the $0$-form component of $G$. In passing from the parent to the orbifold theory, one needs to go through a sequence of modifications (restrictions and quotients) of the gauge group and of the spectrum of dynamical objects that are coupled to it. 
In particular, the procedure will require projecting out objects transforming in a non-trivial representation of $\Gamma$.  This projection is associated with the onset of non-endable line defects which in general link non-trivially with topological Gukov-Witten operators \cite{Rudelius:2020orz}.  Conjecturally, such topological operators should be absent in any consistent UV complete quantum gravity theory. This means that, in order to obtain a consistent string orbifold, in general one needs to gauge various (higher form) global symmetries that appear at intermediate stages of the procedure. One way quantum gravity avoids the appearance of global symmetries is through the presence of Chern-Simons couplings, that are closely related to the higher group structure of $G$. The projection must in general involve the gauge field configurations present in the Chern-Simons terms; as a consequence,
the way the gauge group can be consistently modified is strongly constrained by its higher group structure.

For example, in the case of type II on a circle $S^1$, considered in section $4$, the gauge group in the $9$-dimensional space-time contains a $U(1)^{A_1}\times U(1)^{B_1}$ $0$-form subgroup, associated with momentum and fundamental string winding along $S^1$. The group $\Gamma$ we want to orbifold by is the $\ZZ_2$ subgroup of $U(1)^{A_1}$, corresponding to half-period shifts along $S^1$. As a first guess, one might expect the gauge group in the orbifold theory to be simply given by a quotient of $U(1)^{A_1}\times U(1)^{B_1}$ by $\Gamma$. However, the corresponding $1$-form gauge fields $A_1$ and $B_1$ transform in a non-trivial $2$-group structure with the $2$-form Kalb-Ramond field $B_2$
\be A_1\to A_1+d\lambda\ ,\qquad B_1\to B_1+d\sigma\ ,\qquad B_2\to B_2+B_1\wedge d\lambda+d\Sigma_1\ ,
\ee where $\lambda$ and $\sigma$ are $0$-forms, and $\Sigma_1$ a $1$-form, with suitable quantization conditions. Simply quotienting out the subgroup $\Gamma\cong \ZZ_2\subset U(1)^{A_1}$ would make the quantization conditions on $\lambda$ incompatible with this transformation law. One is therefore forced to perform also a non-trivial $\ZZ_2$ extension of the group $U(1)^{B_1}$ -- this step can also be seen as taking a quotient of the higher form group that is the magnetic dual of $U(1)^{B_1}$. The charged objects with respect to this new $\ZZ_2$ gauge symmetry are exactly the twisted sector of the fundamental string. Therefore, one can argue for the presence of such twisted sector just from the requirement that the spectrum is `complete', i.e. that there exist dynamical objects carrying all possible charges compatible with Dirac quantization condition (in section \ref{s:halfperiod} we will provide an alternative argument as for why such twisted sector must be present). A similar mechanism occurs in the orbifold of type II superstrings on a torus $T^d$ by the reflection $\mathsf{C}$ of all coordinates of $T^d$, considered in section \ref{s:reflection}. In this case, the $0$-form component of the gauge group contains a non-abelian subgroup $G=\ZZ_2^C\ltimes \prod_{i=1}^d(U(1)^{A^i}\times U(1)^{B^i})$, where $\Gamma\cong \ZZ_2^C$ is the subgroup generated by $\mathsf{C}$, which acts on the gauge fields of the $U(1)$ factors by charge conjugation $A_1^i\to -A_1^i$, $B_1^i\to -B_1^i$. Naively, one might expect the gauge group of the orbifold theory to be obtained by first restricting $G$ to the abelian subgroup $G_{res}=\ZZ_2^C\times \ZZ_2^d\times \ZZ_2^d$ of symmetries that commute with $\mathsf{C}$, and then quotienting by $\Gamma\cong \ZZ_2^C$, to obtain the abelian group $\ZZ_2^d\times \ZZ_2^d$. However, this simple guess leads to the wrong answer, as can be seen, for example, from the worldsheet orbifold construction. From the space-time point of view, the problem is that we neglected the $2$-group structure relating the $1$-form gauge fields of $G$ to the $2$-form Kalb-Ramond field $B_2$. In fact, we will argue that, even when $A_1^i$ and $B_1^i$ are restricted to the $\ZZ_2$ subgroups of $U(1)^{A^i}$ and $U(1)^{B^i}$, there is a non-trivial action of $\mathsf{C}$ on the set of gauge fields by
$$(A_1^i,B_1^i,B_2)\stackrel{\mathsf{C}}{\longrightarrow} (A_1^i,B_1^i,B_2+2\sum_i A_1^i\wedge B_1^i)\ .
$$ A consequence of this transformation is that, once again, the $0$-form gauge group of the orbifold theory is a certain non-abelian $\ZZ_2$ central extension of $\ZZ_2^d\times \ZZ_2^d$, in agreement with the worldsheet analysis. The charged object with respect to this new central $\ZZ_2$ group is the twisted sector of the fundamental string, and the correct degeneracy of the twisted ground states can also be predicted. To summarise, in both examples we consider, the higher group structure of the original gauge group dictates how the naive restrictions and quotients of the $0$-form gauge group must be modified in order to get a consistent theory.  Combining these constraints with the requirement that the final theory has no global symmetries, one can deduce the main properties of the twisted sector of the orbifold theory.
In section \ref{s:conclusions}, we make use of the examples to discuss our construction in further detail. We stress that this article should be interpreted as a first sketch of our proposal: in this perspective, section \ref{s:conclusions} also contains some possible generalizations and open questions that we hope to analyze in the future.

\medskip

There are various reasons as for why such a procedure might be interesting.  First of all, in string theory the interplay between the worldsheet physics and the spacetime physics is often very subtle.\footnote{There exists an alternative approach to string orbifolds where the target space is described as a stack, rather than a geometric orbifold \cite{Pantev:2005rh,Pantev:2005wj,Pantev:2005zs}. This description allows one to recover all information about the twisted sector of the child theory from a purely space-time point of view, and can be generalized to include higher form symmetries. 
We thank E. Sharpe for bringing our attention to these constructions.} For example, worldsheet orbifolds might be inconsistent if global symmetries of the CFT have some 't Hooft anomalies. On the other hand, because the string theory is actually consistent, the corresponding gauge symmetries in spacetime cannot be anomalous. What is the spacetime interpretation of the obstructions to defining an orbifold, that are manifest in the worldsheet theory? In general, 't Hooft anomalies on the worldsheet are closely related to higher group structures for the spacetime gauge group. This means that the symmetries we would like to quotient by, sometimes do not really form a properly defined subgroup of the spacetime gauge group (see point $1$ in section \ref{s:conclusions} for a discussion about this point, and sections \ref{s:halfperiod} and \ref{s:reflection} for a couple of examples).

In fact, when it comes to consistency conditions for orbifolds, it is well known that a purely worldsheet perspective might not be sufficient. For example in order to make the string orbifold consistent, it is sometimes necessary to include non-perturbative degrees of freedom (e.g. spacetimes filling branes necessary for tadpole cancellation \cite{Sethi:1996es}) that cannot be easily interpreted in terms of an orbifold of a worldsheet QFT.

\medskip

A second motivation for introducing a `spacetime orbifold' is to get a well-defined procedure that does not single out the fundamental string as a `preferred' object. It is not a priori clear that such a democratic procedure does even exist. On the contrary, there are some arguments suggesting that it \emph{cannot} exist in general. These arguments are based on the fact that `orbifold does not always commute with duality' \cite{Vafa:1995gm,Sen:1995ff,Sen:1996na}. 

Consider, for example, type IIB superstring theory in $10$ flat dimensions. For generic string coupling constant, this theory has two $\ZZ_2$ symmetries, namely worldsheet parity $\Omega$ and $(-1)^{F_L}$, the symmetry acting by a minus sign on the R-R and R-NS sectors and acting trivially on the NS-NS and NS-R sectors. This symmetries are exchanged by non-perturbative S-duality. Now, if we take the orbifold by $\Omega$, we obtain type I string theory, while if we take the orbifold by $(-1)^{F_L}$ we obtain type IIA; these two theories are clearly not dual to each other. So, how are these examples compatible with the idea of a well-defined duality invariant orbifold procedure?

\medskip

There are two possible perspectives on the idea of a theory $A$ being dual to a theory $B$. One might consider $A$ and $B$ as two different theories that turn out to be (often non-trivially) equivalent to each other. From another point of view, we might consider $A$ and $B$ just as two different descriptions of the same theory. 

From the first point of view, the fact that the orbifolds $A/\Gamma$ and $B/\Gamma$ of two dual theories are still equivalent is a highly non trivial statement, and one should provide a reasonable argument (for example, the adiabatic argument of \cite{Vafa:1995gm}) to explain why this equivalence is true in most cases. On the other hand, from the second perspective, because the underlying quantum gravity theory is unique, the fact that one might possibly get different theories after applying the same orbifold procedure looks quite puzzling.
Indeed, the fact that we can obtain different orbifolds starting from different duality frames implies that the orbifold procedure is inherently description-dependent, and that it admits no unambiguous definition from the point of view of the `intrinsic' quantum theory of gravity. 
This conclusion about the description-dependence of string-theoretical orbifolds would be in stark contrast with orbifolds in QFT, that are completely based on intrinsic properties of the theory, such as its group of global symmetries.

\medskip

The (somehow tautological) point of view that we are going to adopt in this article is to declare that, by definition, the string theoretical orbifold must be independent of the duality frame. How is this compatible with the observation that `orbifolds do not always commute with duality'?
A way out of this conundrum is to allow the orbifold by a given group of symmetries not to be unique, but to depend on some subtle choices to be made. Some of these choices might be simpler or more natural in one description or another, which is the reason why one might obtain
two different quantum gravity theories $A/\Gamma$ and $B/\Gamma$. 
Nevertheless, if the duality frames are really equivalent descriptions of the same quantum gravity theory, there should be the possibility to describe any possible outcome of the orbifold procedure in each such duality frame -- it might however happen that some outcomes are more `exotic' than others in some descriptions. 
This hypothesis is much closer in spirit to the QFT orbifold, where for a given group of global symmetries, the resulting orbifold QFT is not necessarily unique. This phenomenon is known as discrete torsion, and the different possibilities are in finite number and can be fully classified.

In fact, in the above type IIB example, an alternative definition of the orbifold by $(-1)^{F_L}$ has been proposed in \cite{Hull:1998he,Bergshoeff:1998re}. In this case, one is instructed to include in the theory $32$ spacetime filling NS$9$-branes (the objects S-dual to D$9$-branes), and it is argued that the resulting theory is $SO(32)$ heterotic string. This result is the exact S-dual analogue of constructing type I theory by orbifolding worldsheet parity symmetry.

A natural question is how to describe the orbifold in a way that is manifestly description independent and duality invariant.
The ideal framework for this definition would be an intrinsic non-perturbative description of the quantum gravity theory, for example in terms of algebras of observables. Unfortunately, such a description of quantum gravity is not available.\footnote{One might argue that, at least for asymptotically AdS spacetime, holography does provide such a definition, in terms of the QFT on the boundary. On the other hand, when no semiclassical description of the bulk theory is available, it is in general very difficult to interpret the properties of the boundary QFT in terms of a quantum gravity theory. We comment more on this in section \ref{s:conclusions}.} 
Our goal is to define the orbifold through a sequence of abstract rules that make sense and can be applied whenever an explicit description of a string theory model is available, but that does not depend on the particular duality frame where such a description is formulated. This sequence of rules is what we call the `spacetime orbifold' procedure, in contrast with the `worldsheet orbifold' which is based on the choice of one dynamical object (the fundamental string) as a preferred one.

Based on our discussion above, it would be natural to test our orbifold procedure in the cases where `duality does not commute with orbifolds'. Unfortunately, a detailed discussion of such examples needs some technical refinements of our approach that we are not going to discuss in this paper (see paragraph 9 in section \ref{s:conclusions}); we hope to address these issues in future work. Here, let us just mention that our hypothesis that the potential outcomes of an orbifold procedure does not depend on the duality frame leads to some highly non-trivial consequences. For example, in the case orbifolds of $10$-dimensional type IIB theory, one should be able to obtain M-theory compactified on a large circle from a `spacetime' orbifold of perturbative type IIB by worldsheet $\Omega$. This example would show that the `spacetime orbifold' that we are proposing must be a more general procedure than the standard `worldsheet orbifold'. This opens up the possibility of considering brand new ways of constructing a new consistent string model from a `parent' one.

\section{Orbifolds in quantum field theory}\label{s:QFTorbifolds}

In quantum field theory (not coupled to gravity), the orbifold procedure is a way to obtain a new theory from a `parent' theory by `gauging a global symmetry'. In this section, we review the features of (generalized) global symmetries and orbifolds in QFT, that will be useful in the following sections. 

\subsection{Examples of symmetries and orbifolds in QFT}

\subsubsection{Orbifolds in 2D CFT}\label{s:2DCFT}

The best studied examples of orbifolds are in two-dimensional conformal field theory (2D CFT) \cite{Dixon:1985jw,Dixon:1986jc,Narain:1986qm,Hamidi:1986vh,Vafa:1986wx}. In this section, we review the main properties of such orbifolds.

Let us consider a bosonic unitary Euclidean 2D CFT $\C$, with a single vacuum and with a finite group of (global) symmetries $\Gamma$. This means that the space of states/operators $\Hh$ admits a unitary representation $\rho$ of $\Gamma$, that fixes the vacuum, commutes with the Virasoro algebra, and preserves the operator product expansion (OPE).

The orbifold procedure is a way to define a new CFT $\C/\Gamma$ (the orbifold of $\C$ by $\Gamma$) from the parent one. For simplicity, let us focus on the case where $\Gamma$ is abelian. The orbifold can be described as a two-steps procedure.
\begin{itemize}
    \item \emph{Projection.} The first step is to project onto the subspace $\Hh^\Gamma$ of states/operators  that are invariant under $\Gamma$. This subspace contains the vacuum state and the Virasoro algebra, and is closed under OPE.
    \item \emph{Twisted sector.} Next, for each $g\in \Gamma$, we consider the twisted sector space $\Hh_g$. This is a space of operators that are not mutually local with respect to the operators in $\Hh$ (and also with respect to other twisted operators), but rather create some branch cuts in correlation functions. More precisely, consider a correlation function $\langle O_g(0) O(w)\cdots \rangle $, where $O_g\in \Hh_g$, $O\in \Hh$, and $\cdots$ denote other insertions farther from the origin than $O$. Then, if we analytically continue $w\to e^{2\pi i}w$ along a small circle around $O_g(0)$, the correlation function is not single-valued, as the operator $O$ transforms as $O(e^{2\pi i}w)=(\rho(g)\cdot O)(w)$. Each $\Hh_h$, $h\in \Gamma$, affords a (possibly projective) representation $\rho_h$ of $\Gamma$, and we have an analogous transformation $O_h(e^{2\pi i}w)=(\rho_h(g)\cdot O_h)(w)$ in the (analytically continued) correlation function as $O_h$ moves along a small circle around $O_g\in \Hh_g$.\\
    The space of states $\tilde \Hh$ of the orbifold theory $\C/\Gamma$ is obtained as the direct sum of 
    $\Hh^\Gamma\equiv \Hh_{g=1}^\Gamma$ with the $\Gamma$-invariant subspaces of each twisted sector
    \be \tilde \Hh=\oplus_{g\in \Gamma} \Hh_g^\Gamma\ .
    \ee The projection onto the $\Gamma$-invariant operators is a necessary condition for the OPE to be local.
\end{itemize}
This construction generalizes easily to the case where $\Gamma$ is non-abelian. In this situation, for each conjugacy class $[g]$ of $\Gamma$, there is a (projective) action of $\Gamma$ on $\Hh_{[g]}:=\oplus_{h\in [g]} \Hh_h$, the direct sum of the $h$-twisted sectors for all the elements $h$ in the conjugacy class. The orbifold Hilbert space is then given by the sum $\oplus_{[g]}\Hh_{[g]}^\Gamma$ over all conjugacy classes $[g]$ of the $\Gamma$-invariant subspaces $\Hh_{[g]}^\Gamma\equiv(\Hh_{[g]})^\Gamma$; in fact, one can show that $\Hh_{[g]}^\Gamma$ is isomorphic to $\Hh_g^{C_\Gamma(g)}$ for any $g$ in the conjugacy class, where $C_\Gamma(g)$ is the centralizer of $g$ within $\Gamma$.

The inclusion of the twisted sector is necessary in order to obtain a theory with modular invariant torus partition function. The procedure outlined above does not always yield a consistent CFT. In particular, it might be impossible to obtain a modular invariant partition function, due to the failure of the level matching condition, or it might be impossible to define a local OPE among the twisted operators \cite{Vafa:1986wx,Hamidi:1986vh,Dijkgraaf:1989pz,Roche:1990hs,Bantay:1990yr}. On the other hand, if a consistent orbifold exists for a given group $\Gamma$, it might not be unique \cite{Vafa:1986wx}.

In order to understand these points, let us focus on the OPE between twisted operators. In general one expects a $g$-twisted sector $\Hh_g$ to be an irreducible ($g$-twisted) module with respect to the OPE with operators in $\Hh$, and to be the unique (up to isomorphisms) such irreducible $g$-twisted module.\footnote{An analogous statement can be rigorously proved in the context of vertex operator algebras (VOAs) under standard conditions for the VOA \cite{DLM1997,DLM1998,DLM1998b,Carnahan:2016guf}
.} The OPE between operators $O_g\in \Hh_g$ and $O_h\in \Hh_h$ will produce $gh$-twisted operators, so that the space $\Hh_g\otimes \Hh_h$ can be identified with $\Hh_{gh}$. This identification requires choosing an isomorphism
\be\label{isom2} \varphi_{g,h}:\Hh_g\otimes\Hh_h\to \Hh_{gh}\ ,
\ee of $gh$-twisted modules, which is unique only up to a phase. Therefore, in order to define the OPE in the orbifold theory, one needs to choose the isomorphisms $\varphi_{g,h}$ for each pair $g,h\in \Gamma$. 

For a given choice of $\{\varphi_{g,h}\}_{g,h\in \Gamma}$, the two isomorphisms $\varphi_{gh,k}\circ(\varphi_{g,h}\otimes 1):\Hh_g\otimes\Hh_h\otimes \Hh_k\to \Hh_{ghk}$ and $\varphi_{g,hk}\circ(1\otimes \varphi_{h,k}):\Hh_g\otimes\Hh_h\otimes \Hh_k\to \Hh_{ghk}$ might differ by a phase, i.e. there might be a non-trivial associator $\alpha_{g,h,k}:\Hh_{ghk}\to\Hh_{ghk}$, such that
	\be\label{noass} \varphi_{g,hk}\circ(1\otimes \varphi_{h,k})=\alpha_{g,h,k}\circ  \varphi_{gh,k}\circ(\varphi_{g,h}\otimes 1)\ .
	\ee The associator must be proportional to the identity $\alpha_{g,h,k}=\alpha(g,h,k)\mathrm{id}_{\Hh_{ghk}}$, where $\alpha:\Gamma\times\Gamma\times\Gamma\to U(1)$ satisfies a certain cocycle condition
	\be\label{3cocyle} \alpha(g,h,k)\alpha(g,hk,l)\alpha(h,k,l)=\alpha(gh,k,l)\alpha(g,h,kl)\ ,
	\ee for all $g,h,k,l\in \Gamma$.
	If $\alpha$ is non-trivial, the OPE among twisted sectors is not associative and the orbifold theory is not a consistent CFT. One can try to modify the isomorphisms $\varphi_{g,h}$ by
	\be\label{betadiff} \varphi_{g,h}\to \tilde \varphi_{g,h}=\varphi_{g,h}\beta(g,h)\ ,
	\ee for some $\beta:\Gamma\times \Gamma\to U(1)$, and this modifies the associator by
	\be \alpha\to \tilde \alpha=\alpha\cdot \partial\beta\ ,
	\ee
	where $\partial \beta$ is a $3$-coboundary
	\be (\partial \beta)(g,h,k) =\frac{\beta(g,h)\beta(gh,k)}{\beta(g,hk)\beta(h,k)}\ .
	\ee Therefore, the obstruction to defining an associative OPE is represented by a $3$-cocycle $\alpha$ modulo $3$-coboundaries $\partial \beta$, which defines a class $[\alpha]$ in $H^3(\Gamma,U(1))$. Different choices of the collection $\{\varphi_{g,h}\}_{g,h\in \Gamma}$ in a given CFT lead to different cocycles $\alpha$ in the same class. 
	Only when the class $[\alpha]$ is trivial, one can choose the isomorphisms $\varphi_{g,h}$ in such a way that $\alpha_{g,h,k}$ is the identity for all $g,h,k\in \Gamma$, so that the OPE of twisted operators is associative. Triviality of $[\alpha]$ also implies that the torus partition function of the orbifold theory is modular invariant, and in particular that the level matching condition is satisfied for all $g$-twisted sectors. In fact, when the group $\Gamma$ is cyclic, the triviality of $[\alpha]$ is equivalent to the level matching condition for all $g$-twisted sectors. On the other hand, for more general (in particular, non-abelian) groups $\Gamma$, it might happen that the level matching condition is satisfied for all $g\in \Gamma$, but the class $[\alpha]$ is non-trivial, and the orbifold, therefore, is inconsistent.
	
	If $[\alpha]$ is trivial, there might 
	still
	be different choices of $\{\varphi_{g,h}\}_{g,h\in \Gamma}$ for which $\alpha=1$. Such choices are related to each other by transformations \eqref{betadiff} for some $\beta$ satisfying a $2$-cocycle condition $\partial \beta=1$. Different $\{\varphi_{g,h}\}$ will lead to equivalent CFTs only if they are related by redefinitions of twisted operators
	\be \gamma(g):\Hh_g\to \Hh_g\ ,
	\ee for some $\gamma:\Gamma\to U(1)$. Thus, nonequivalent orbifold theories are in one-to-one with classes $[\beta]\in H^2(\Gamma,U(1))$, that are represented by $2$-cocyles $\beta$ modulo
	\be \beta\to \beta\cdot \partial \gamma\ ,
	\ee where $\partial \gamma$ is a $1$-coboundary
	\be (\partial \gamma)(g,h)=\frac{\gamma(gh)}{\gamma(g)\gamma(h)}\ .
	\ee 
	The possibility of having multiple consistent orbifold theories associated with the same `parent' theory and symmetry group $\Gamma$ is known as discrete torsion.
	
	\medskip
	
	A useful way to describe the orbifold procedure in CFT is in terms of defects \cite{Frohlich:2006ch,Frohlich:2009gb,Carqueville:2012dk,Brunner:2013ota,Brunner:2013xna,Bhardwaj:2017xup,Tachikawa:2017gyf,Burbano:2021loy}. With each symmetry generator $g\in \Gamma$, one can associate a topological defect $\Li_g(\gamma)$ supported on an oriented $1$-dimensional curve $\gamma$. Inserting $\Li_g(\gamma)$ in a correlation function creates a discontinuity of the fields at $\gamma$, with the prescription that the fields on the left of $\gamma$ are related to the field on the right of $\gamma$ by the action of $g$. Here, the `left' and `right' side of the defect are defined with respect to the orientation of $\gamma$, so that reversing the orientation of $\gamma$ is equivalent to exchanging $\Li_g$ with $\Li_{g^{-1}}$. The defects $\Li_g(\gamma)$ are topological, in the sense that correlation functions are invariant under small deformations of the support $\gamma$, so long as $\gamma$ does not cross the support of some other operator.  The twisted sector $\Hh_g$ is naturally interpreted as the space of point-like operators where an (outgoing oriented) $\Li_g$ defect can start. Similarly, $(\Hh_g)^*\cong \Hh_{g^{-1}}$ is the space of point operators where an incoming $\Li_g$ defect can end. When two parallel defects are brought very close to each other, they can `fuse' into a single defect. The fusion of defects respects the group-like structure of $\Gamma$, so that two defects $\Li_g$ and $\Li_h$ fuse into $\Li_{gh}$. One can also consider $k$-fold junctions among defects, i.e. point-like operators with $k$ outgoing defects attached. A junction operator is called topological if it can be moved without changing the correlation function, as long as no other insertion is crossed. In particular, a $g$-twisted operator can be interpreted as a $1$-fold junction, which is usually not topological. There is always a $1$-dimensional space of topological $3$-fold junctions $\varphi_{g,h}$ with two incoming defects $\Li_g$ and $\Li_h$ and one outgoing defect $\Li_{gh}$. The choice of a topological junction $\varphi_{g,h}$ is equivalent to the choice of an isomorphism \eqref{isom2}. More precisely, the isomorphism, seen as a map $\Hh_g\otimes \Hh_h\otimes (\Hh_{gh})^*\to \CC$, is determined by a $3$-point correlation function on the sphere, with the insertion of a $g$-twisted, a $h$-twisted and a $(gh)^{-1}$-twisted operators, with the corresponding defects joining at a topological junction $\varphi_{g,h}$. 
	
	Suppose we choose a collection $\{\varphi_{g,h}\}_{g,h}$ of topological junctions. The phase $\alpha$ in \eqref{noass} appears when one tries to deform a network of defects. 	
\begin{figure}
\begin{tikzpicture}[line width=0.8 pt,>=latex]
\filldraw[pattern color=green,pattern=dots,draw opacity=0.2] (2,0) -- (3,1) -- (2,2) -- (1,1);
\draw (0,0) node[above] {g} -> (0.5,0.5) ;
\draw[>-]  (0.5,0.5) -- (1,1);
\draw (2,0) node[left] {h}  -> (1.5,0.5) ;
\draw[>-]  (1.5,0.5) -- (1,1);
\draw (4,0) node[above] {k} -> (3,1) ;
\draw[>-]  (3,1) -- (2,2);
\draw (1,1)  -> (1.5,1.5) node[above left] {gh};
\draw[>-]  (1.5,1.5) -- (2,2);
\draw (2,2)  -> (2,2.5) node[right] {ghk};
\draw[>-]  (2,2.5) -- (2,3);
\draw[dashed,thin] (2,0) -- (3,1);
\useasboundingbox (0,-1) rectangle (5,4);
\end{tikzpicture}
\begin{minipage}[b][2cm][t]{.15\textwidth}
$=\alpha(g,h,k)$
\end{minipage}
\begin{tikzpicture}[line width=0.8 pt,>=latex]
\filldraw[pattern color=green,pattern=dots] (2,0) -- (3,1) -- (2,2) -- (1,1);
\draw (4,0) node[above] {k} -> (3.5,0.5) ;
\draw[>-]  (3.5,0.5) -- (3,1);
\draw (2,0) node[right] {h}  -> (2.5,0.5) ;
\draw[>-]  (2.5,0.5) -- (3,1);
\draw (0,0) node[above] {g} -> (1,1) ;
\draw[>-]  (1,1) -- (2,2);
\draw (3,1)  -> (2.5,1.5) node[above right] {hk};
\draw[>-]  (2.5,1.5) -- (2,2);
\draw (2,2)  -> (2,2.5) node[right] {ghk};
\draw[>-]  (2,2.5) -- (2,3);
\draw[dashed,thin] (2,0) -- (1,1);
\useasboundingbox (-1,-1) rectangle (3,4);
\end{tikzpicture}
	\caption{\small Two different ways of fusing three defects $\Li_g$, $\Li_h$, and $\Li_k$ into a single defect $\Li_{ghk}$ through a sequence of $3$-pronged junctions. The dashed lines represent the identity defect, that we added for later convenience. One can continuously deform the left configuration into the right one; along this transformation, there is a point where the defects $\Li_g$, $\Li_h$, $\Li_k$, and $\Li_{ghk}$ are connected to a single $4$-pronged junction.  Passing across that point, a correlation function might get a non-trivial phase $\alpha(g,h,k)$. Equivalently, the left and the right-hand side are related by a local gauge transformation acting by a group element $h$  in the green region; this picture makes it clear that $\alpha(g,h,k)$ is a  't Hooft anomaly -- correlation functions pick up a non-trivial phase under gauge transformations  of the background gauge field.}\label{f:defjunct}
\end{figure}
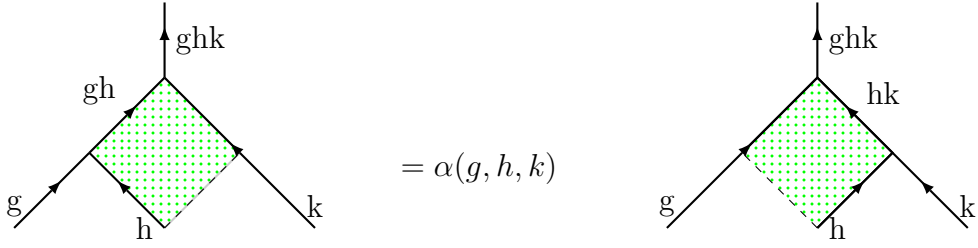
In particular, consider a correlation function with the insertion of a network of defects as in the left of figure \ref{f:defjunct}. Upon moving a topological junction $\varphi_{g,h}$ across a topological junction $\varphi_{gh,k}$, we obtain a new correlation function with the insertion of the network on the right, containing the junctions $\varphi_{h,k}$ and $\varphi_{g,hk}$. The two correlators are equal only up to the overall phase $\alpha(g,h,k)$.

	The correlation functions in the orbifold theory can be obtained in terms of correlation functions in the original theory with the insertion of a network of defects $\Li_g$. More precisely, suppose we want to compute a $n$-point correlation function $\langle \prod_i O_i(z_i)\rangle$ in the orbifold theory, where $O_i\in \bigoplus_g \Hh_g$. We can choose a triangulation of the worldsheet, such that the insertion points of the operators coincide with some of the vertices of the triangulation. We can also assume that the triangulation has only trivalent vertices. Then, the orbifold correlation function corresponds to a correlation function in the original theory, where we insert in each edge a topological defect $\Li_g$, $g\in \Gamma$, and in each vertex a topological junction $\varphi_{g,h}$ from an appropriately chosen collection $\{\varphi_{g,h}\}_{g,h}$, and we sum over all (consistent) possibilities.  The sum over all defect networks implements a projection, so that the correlation function is non-zero only if all $O_i$ are in the $\Gamma$-invariant subspace $(\bigoplus_g \Hh_g)^\Gamma$. Equivalently, one can describe the orbifold as the insertion of a superposition $\Li_\Gamma:=\sum_{g\in \Gamma} \Li_g$ of all possible symmetry defects in each edge of the triangulation.
	This prescription is well-defined, and in particular is independent of the choice of the triangulation, if and only if all the phases $\alpha(g,h,k)$ are trivial. 
	
	The insertion of a network of defects $\Li_g$, $g\in \Gamma$, in a correlation function of the original theory can also be interpreted as a coupling to a background gauge field for the group $\Gamma$. Since $\Gamma$ is finite, and therefore discrete, the gauge connection can only be flat -- in particular, there is no curvature and no propagating degrees of freedom. Nonequivalent (flat) gauge configurations on a worldsheet $X$ are represented by classes in $H^1(X,\Gamma)$. For a given triangulation of $X$, we can regard the faces as corresponding to open patches $U_i$, the edges as double intersections $U_i\cap U_j$, and the vertices as triple intersections $U_i\cap U_j\cap U_k$. The insertions of defects supported at the edges correspond to the choice of (constant) transition functions $\phi_{ij}\in \Gamma$ in each double intersection. Consistency conditions $\phi_{ij}\phi_{jk}\phi_{ki}=1$ on triple intersections are assured by the requirement that there must exist some non-zero topological junction operator at each vertex. A local gauge transformation can be interpreted as a change of the triangulation. Therefore, the sum over network of defects that defines the correlation functions in the orbifold theory is just a sum over all possible configurations for the background $\Gamma$-gauge field. Performing such a sum amounts to making the background gauge field dynamical. In this sense, the orbifold procedure can be seen as the gauging of a discrete group.
	
	This construction suggests that the notion of global symmetries in 2D CFT (and in more general QFTs, see \cite{Choi:2021kmx,Kaidi:2021xfk,Choi:2022zal}) can be extended by including all kind of topological defects in the theory. In this generalization, the group multiplication gets replaced by the fusion product of defects, and the ordinary symmetries are characterized as the subset of defects that are invertible with respect to fusion.  The notion of orbifold can be generalized accordingly, by replacing the superposition defect $\Li_\Gamma=\sum_{g\in \Gamma} \Li_g$ by some other suitable topological defects, not necessarily defined in terms of (ordinary) symmetries \cite{Frohlich:2006ch,Frohlich:2009gb,Carqueville:2012dk,Brunner:2013ota,Brunner:2013xna,Bhardwaj:2017xup,Tachikawa:2017gyf,Burbano:2021loy}.

\subsubsection{Orbifolds of higher form symmetries}\label{s:highform}

Considering orbifolds of QFTs in $d$ spacetime dimensions naturally
leads to the notion of \emph{generalized global symmetries} which
has been introduced in \cite{Gaiotto:2014kfa}. 
As for the discussion in this subsection, we assume $p$-form global symmetry transformations form a group $G^{(p)}$,
and in particular we will ignore the possible mixing of several $q$-form symmetry groups with possibly different $q$'s. These transformations can be expressed in terms of \emph{charge  operators} (or \emph{defects}\footnote{We follow a common sloppiness and use the words `defect' and `operator' somehow interchangeably, although strictly speaking in Lorentzian signature the latter should only refer to the case where the support is space-like.} using the nomenclature as in section \ref{s:2DCFT}) $U_{g}(M^{(d-p-1)})$ supported
on closed manifolds $M^{(d-p-1)}$ of dimension $d-p-1$ and labeled
by elements $g$ of the group $G^{(p)}$, with the group multiplication
law 
\begin{equation}
U_{g}(M^{(d-p-1)})\times U_{g'}(M^{(d-p-1)})=U_{gg'}(M^{(d-p-1)})\,.
\end{equation}
In particular, every charge operator
\(U_{g}\) has an inverse \(U_g^{-1}=U_{g^{-1}}\), such that \(U_{g}\times U_{g^{-1}}=1 \), the identity operator. Charge operators 
are therefore \textit{invertible} and \textit{topological} in the sense that small continuous deformations of \(M^{(d-p-1)}\) do not affect any physical observables unless \(M^{(d-p-1)}\) crosses any charged operators. Given an operator $V(C^{p})$ supported on a manifold $C^{p}$ of
dimension $p$ and transforming in a representation $g(V)$ of $g$,
if $S^{d-p-1}$ is a small sphere linking once with $C^{p}$ , we have the Ward identity
\begin{equation}
U_{g}(S^{d-p-1})\,V(C^{p})=g(V)\,V(C^{p})\,.
\end{equation}
\(g(V)\) is simply a phase when the global symmetry group is abelian, which is always the case for \(p\geq 1\). The equation above should be understood as an operator equation, valid within general correlation functions provided that there are no other operator insertions that link nontrivially with \(S^{d-p-1}\) or \(C^{p}\). If the support of the charged operator  \(V\) is allowed to be a manifold \(C^{(p)}\) with a boundary, we say that \(V\) is \textit{endable}. It is easily proven that any endable operator must link trivially with any topological operator so that any endable operator of dimension $p$ cannot carry charge under a $p$-form global symmetry.

An example that is particularly useful to illustrate these ideas is
a $4d$ non-abelian gauge theory based on a Lie algebra $\mathfrak{g}$,
which admits 't Hooft-Wilson line defects. These lines can be thought of as the worldvolume of very massive probe (i.e., non-dynamical) particles charged under  $\mathfrak{g}$. In \cite{Kapustin:2005py,Goddard:1976qe}
it was shown that 
't Hooft-Wilson line defects are classified by equivalence classes $\left(e,m\right)\in\varLambda_{e}^{w}\times\varLambda_{m}^{w}$, defined modulo the action of the Weyl group $\mathcal{W}$. Here  $\Lambda_{e}^{w}=\mathbf{t}^{\ast}$ is the weight lattice of  $\mathfrak{g}$  and  $\varLambda_{m}^{w}$ is the so-called magnetic weight lattice, namely the Cartan subalgebra $\mathbf{t}\subset\mathfrak{g}$, which is the weight lattice of the Langlands-dual Lie algebra $\mathfrak{\hat{g}}$. We should also notice this labeling contains more information than simply the representations of $\mathfrak{g}$ and  $\mathfrak{\hat{g}}$ since, chosen a  $\mathcal{W}$-orbit of a magnetic weight $B\in\mathbf{t}$ , an allowed electric representation is determined by a $\mathcal{W}_{B}$-orbit of a weight $\mu\in\varLambda_{e}^{w}=\mathbf{t}^{\ast}$, with $\mathcal{W}_{B}$ the stabilizer subgroup of $B$ in $\mathcal{W}$. Following \cite{Heidenreich:2021xpr}, we say that a  't Hooft-Wilson line is endable if it may be defined on an open curve with endpoints supporting local charged operators in the same representation. On top of this, we should take into account that in general two 't Hooft-Wilson line defects labelled by $\left(e,m\right)$ and $\left(e',m'\right)$ are not pairwise local. In fact, the Dirac quantization condition relates mutual locality to the condition of integrality of the Dirac pairing $m\cdot e'-m'\cdot e$. 

The group algebra $\mathfrak{g}$ does not contain all the information about the lines. In fact,  it is shared  by all gauge groups of the form $G^{(0)}=\mathcal{G}^{(0)}/\Gamma^{(0)}$, with $\mathcal{G}^{(0)}$ the unique simply-connected gauge group with the Lie algebra $\mathfrak{g}$ and $\Gamma^{(0)}\subset Z\left(\mathcal{G}\right)$ a subgroup of the center $Z\left(\mathcal{G}\right)$. Indeed, we should allow for all Wilson lines $\left(e,0\right)$ with $e$ belonging to the electric weight sublattice $\varLambda_{e}^{w\,G}$ labelling the representations of $G$. This sublattice needs contain the root lattice $\varLambda_{e}^{r}$. On the other hand, if the gauge group is  $\mathcal{G}$,
$e$ is any weight in $\Lambda_{e}^{w} $ and hence because of Dirac quantization $m$ must be in the magnetic root lattice $\varLambda_{m}^{r}$, i.e. the coroot lattice of $\mathfrak{g}$. We can therefore consider the charges $\left(e,m\right)$ modulo $\varLambda_{e}^{r}\times \varLambda_{m}^{r}$, and realize they encode the information about the topological linking of  't Hooft-Wilson lines to
the \textit{Gukov-Witten operators}  \cite{Gukov:2006jk,Gukov:2008sn},
which are codimension-2 defect operators labeled by conjugacy classes $[g]$ in  ${G}$. Analogously to 't Hooft-Wilson lines, Gukov-Witten operators can be interpreted as insertions of probe (non-dynamical) \textit{vortices}, codimension 2-objects defined by the nontrivial gauge holonomy around their worldvolume. Gukov-Witten operators that are topological cannot link nontrivially with endable 't Hooft-Wilson lines. 
In a pure ${G}$ gauge theory the endable Wilson lines are precisely those corresponding to representations built from the adjoint under taking tensor products and sub-representations, and  the topological Gukov-Witten operators correspond precisely to conjugacy classes 
which act trivially on the adjoint representations, i.e. contained in the centralizer $Z\left(G_0\right)$ of the identity component $G_0$ of the group $G$ \cite{Heidenreich:2021xpr}. So
a pure ${G}$-gauge theory has an electric one-form symmetry valued in  $Z\left(G_0\right)$, which shifts the gauge fields by a flat $Z\left(G_0\right)$-valued gauge field. Therefore, for the pure $\mathcal{G}$-gauge theory 't Hooft-Wilson line defects have one-form charges in the character group
$\widehat{Z}\left(\mathcal{G}\right)={\varLambda_{e}^{w}}/{\varLambda_{e}^{r}}$.
They are also charged under a magnetic one-form symmetry with charges in $Z\left(\mathcal{G}\right)={\varLambda_{m}^{w}}/{\varLambda_{m}^{r}}$,
so that the charges can be labeled by
\begin{equation}
Z^{\sharp}=\widehat{Z}\left(\mathcal{G}\right)\times Z\left(\mathcal{G}\right).
\end{equation}
However two general 't Hooft-Wilson line defects  labeled by $\left(z^{e},z^{m}\right)$ and $\left(z'^{e},z'^{m}\right)$ are pairwise local only under the assumption of integrality of the Dirac pairing \(z^{m}\cdot z'^{e}-z'^{m}\cdot z^{e}\). Since purely electric line defects  $\left(z^{e},0\right)$ with \(z^{e}\in \widehat{Z}\left(\mathcal{G}\right)\) unconstrained are allowed, this condition implies that no non-trivial magnetic charge is permitted and the  $\mathcal{G}$-gauge theory has no  magnetic one-form symmetry. More generally, an important observation is that the set of lines that are charged under the one-form symmetry is a way to discriminate between gauge theories based on the group $\mathcal{G}$ or on the quotient $G=\mathcal{G}/\Gamma$ by some subgroup of the center, even in the absence of matter  in a nontrivial representation of $\Gamma$. For concreteness we may consider the case $\mathcal{G}=SU\left(N\right)$, $Z\left(\mathcal{G}\right)=\mathbb{\mathbb{Z}}_{N}$, for which the representations of lines are specified by 
\begin{equation}
\left(z^{e},z^{m}\right)=\left(n,m\right) \mod N\,,
\end{equation}
with $n,m\in \ZZ$. Mutual locality of two line operators labeled by $\left(z^{e},z^{m}\right)$
and $\left(z'^{e},z'^{m}\right)$ implies the Dirac quantization condition
\begin{equation}
z^{m}\cdot z'^{e}-z'^{m}\cdot z^{e}=0\mod N.\label{eq:DiracQuant}
\end{equation}
The allowed purely electric line operators are therefore labeled by $\left(z^{e},0\right)$, with $z^{e}=0,\ldots,N-1$. Clearly, the locality condition \eqref{eq:DiracQuant} shows that all the magnetically charged lines must be associated with the root lattice and so have $z^{m}=0\mod N$. On the other hand, if $G={SU\left(N\right)}/{\mathbb{\mathbb{Z}}_{N}}$, the purely electric line operators must be in a trivial representation of $\ZZ_N$, that is  $\left(z^e,z^m\right)=\left(0,0\right)$. Under a completeness requirement, it can be easily shown \cite{Aharony:2013hda} by means of \eqref{eq:DiracQuant} that the allowed lines belong to classes
\begin{equation}
L_n=\left\{\left(z^e,z^m\right)=\left(nm,m\right) \mod N\right\}\,,
\end{equation}
with $m,n=0,1,\ldots,N-1$, which means that for every $n$ we have a distinct theory $\left( SU\left(N\right)/\mathbb{Z}_N\right)_n$, whose line operators have charges in $L_n$.  An interesting observation is that, because of the Witten effect, a shift of $\theta\rightarrow \theta + 2\pi$ changes the electric charge carried by a line operator,
\begin{equation}
\left(z^e,z^m\right) \rightarrow \left(z^e + z^m ,z^m\right)\,,
\end{equation}
so that the sets $L_n$ are transformed as $L_n\rightarrow L_{n+1}$.  This can be understood as  the fact that whereas for $G=SU\left(N\right)$, $\theta\in \left[0,2\pi\right)$, for $G=SU\left(N\right)/\mathbb{Z}_N$  $\theta\in \left[0,2 \pi N\right)$ so that $\theta\rightarrow \theta + 2\pi$ is no more a symmetry of the theory but only $\theta\rightarrow \theta + 2\pi N$. This in turn is related to the fact that the $SU\left(N\right)$ theory has several vacua mapped into each other by a (spontaneously broken) global symmetry which is absent in the $SU\left(N\right)/\mathbb{Z}_N$ theory. More in general, one can observe that, when the gauge group is $SU\left(N\right)/\mathbb{Z}_k$, with $k$ a divisor of $N$, $k k' =N$ , the allowed sets of charges of line operators are
\begin{equation}
L_{k,n}=\left\{\left(z^e,z^m\right)=e\left(k,0\right)+m\left(n,k'\right) \mod N\right\}\,,
\end{equation}
with $e$ and $m$ integers and $n=0,1,\ldots,k-1$. The shift  $\theta\rightarrow \theta + 2\pi$  sends  $L_{k,n}\rightarrow L_{k,n+k'}$, which is interpreted as a mapping between different theories $\left(SU\left(N\right)/\mathbb{Z}_k\right)_n$. Only theories with the same $n \mod l$, where $l=\gcd\left(k,k'\right) $, are related by the shift of the $\theta$-angle, and there are $l$ sets of theories that are distinguished by a discrete analogue of  the original $\theta$-angle.

Gauging a subgroup $\Gamma\subset Z\left(\mathcal{G}\right)$ of the global one-form symmetry reduces the electric one-form symmetry from  $Z\left(\mathcal{G}\right)$ to \(Z\left(\mathcal{G}\right)/\Gamma\) and consequently restricts the allowed Wilson lines to be in $\widehat{Z\left(\mathcal{G}\right)/\Gamma}\subset \widehat{Z}\left(\mathcal{G}\right)$ , which in its turn implies that additional lines can now be introduced in accordance with the integrality of the Dirac pairing. Under the assumption that the set of lines be maximal and complete, the gauging procedure therefore involves the choice of a maximal Lagrangian subgroup \(L\subset Z^{\sharp}\). As shown in \cite{Aharony:2013hda},  such a choice is not unique and is associated with a discrete theta parameter. For trivial discrete theta parameter, the additional lines have magnetic charges \(m\in\Gamma\subset Z\left(\mathcal{G}\right)\) , while their electric charges can be screened by the allowed Wilson lines. This means a new one-form magnetic symmetry valued in \(\widehat{\Gamma}\), the character group of $\Gamma$, will appear and that the full one-form global symmetry is \(Z\left(\mathcal{G}\right)/\Gamma\times\widehat{\Gamma}\).

The gauging of a global $\ZZ_k$ $1$-form symmetry can be described explicitly in terms of a coupling to a $\ZZ_k$ gauge theory. If $\mathcal{G}=SU\left(N\right)$, $Z\left(\mathcal{G}\right)=\mathbb{Z}_{n}$, we can follow \cite{Kapustin:2013qsa,Kapustin:2014gua,Gaiotto:2014kfa}, and introduce the gauge field for a $\left(SU(N)\times U(1)\right)/\mathbb{Z}_{k}$ gauge theory,
\begin{equation}
a_1+\frac{1}{k}\widetilde{A}_1\mathbb{1}\,,
\end{equation}
where $a_1$ is the $SU\left(N\right)$ traceless gauge field, $\widetilde{A}_1$  a $U(1)$ gauge field and $\mathbb{1}$ the unit matrix. We can then remove these $U(1)$ degrees of freedom by imposing the $1$-form gauge symmetry
\begin{equation}
  \widetilde{A}_1 \rightarrow \widetilde{A}_1 - k\, \Sigma_1 \,, 
  \label{cont1formsymmetry}
\end{equation}
where $\Sigma_1$ is a $U(1)$ gauge field with standard normalization. We are left with a $\ZZ_k$ $1$-form global symmetry, which we gauge by adding to the action \cite{Gukov:2013zka}
\begin{equation}
 2\pi i F_2 \wedge \left(d\widetilde{A}_1 +k B_2\right)+\pi ipk B_2\wedge B_2   \,,
\end{equation}
with $p\sim p+k$ an integer corresponding to a discrete $\theta$ angle. $B_2$ is the gauge field for \eqref{cont1formsymmetry} transforming as $B_2\rightarrow B_2 + d \Sigma_1$, and $F_2$ is a $2$-form Lagrange multiplier transforming as $F_2\rightarrow F_2-p\, d\Sigma_1$. Upon integrating out $\widetilde{A}_1$ and expressing the result in terms of the dual gauge field $A_1$ such that $F_2=dA_1$ with $A_1\rightarrow A_1 -p\, \Sigma_1$, we get
\begin{equation}
 2\pi ik B_2 \wedge d A_1 +\pi ipk B_2\wedge B_2\,,     
\end{equation}
which is a standard action for a $\ZZ_k$ gauge theory.

\subsubsection{Higher group symmetries}
\label{HGS}

So far we have considered each higher form symmetry independently, which is justified in situations where the full global symmetry is given by a standard product between $p$-form symmetry groups for different degrees $p$. However, in general, when higher form symmetries groups of different degrees are present in the same theory, they can combine in a very non-trivial way.

Let us consider the case of a QFT with a $0$-form group $G^{(0)}$ and a $1$-form group $G^{(1)}$. 
First of all, the codimension $1$ defects realizing the $G^{(0)}$ symmetry group can act on the codimension $2$ defects realizing $G^{(1)}$ by a group homomorphism \begin{equation}
\rho : G^{(0)} \rightarrow \mathrm{Aut}\left(G^{(1)}\right)\,.
\end{equation}
This means that when a codimension $2$ defect $U^{(d-2)}_h$, $h\in G^{(1)}$, crosses a codimension one defect $U^{(d-1)}_g$, $g\in G^{(0)}$, it emerges on the other side as $U^{(d-2)}_{\rho(g)\cdot h}$. 

In general, the $0$-form group $G^{(0)}$ also acts on line operators by a (possibly projective) representation. In the presence of a $1$-form symmetry $G^{(1)}$, the class of such projective representation becomes ambiguous. Indeed, as the $G^{(0)}$ symmetry group is realized by topological codimension-one defects, their junctions can be decorated by codimension-two topological defects for the  $1$-form symmetry group $G^{(1)}$. These decorations are invisible for local (point-like) operators, but they can change by a phase the action of $G^{(0)}$ on lines that carry charge under $G^{(1)}$. This can shift the cohomology class of the projective representations of $G^{(0)}$ on the lines.
As discussed in \cite{Delmastro:2022pfo,Brennan:2022tyl}, consistency of fusion of the junctions and $1$-form implies that distinct choices are related to each other by a class in a twisted cohomology group
\begin{equation}
H^2_\rho\left(G^{(0)},G^{(1)}\right)\,,
\end{equation} where $H^2_\rho$ denotes the cohomology with respect to a twisted differential $d_\rho$ that depends on the homomorphism $\rho$ (see for example \cite{Benini:2018reh,Bhardwaj:2022scy} for details).
More precisely, given a flat background $A_1$ for $G^{(0)}$, i.e. a $G^{(0)}$-valued $1$-cocycle on the $d$-dimensional spacetime manifold $M_{d}$
\begin{equation}
A_1\in Z^1\left(M_{d}, G^{(0)}\right)\,,
\end{equation}
a background for  the $1$-form symmetry $G^{(1)}$ is specified by 
\begin{equation}
B_2\in Z^2_{A_1}\left(M_{d}, G^{(1)}\right)\,,
\end{equation}
which is a $G^{(1)}$-valued $A_1$-twisted $2$-cocycle. The class $[\eta]\in H^2_\rho\left(G^{(0)},G^{(1)}\right)\equiv H^2_\rho\left(BG^{(0)},G^{(1)}\right)$ constrains the background fields by 
\begin{equation}
\label{fracclass}
    [B_2]=[A_1^*\eta]\ ,
\end{equation} where $[B_2]$ is the cohomology class of the cocyle $ B_2$, the flat background $A_1$ is interpreted as a map $A_1:M_{d}\to BG^{(0)}$ from spacetime to the universal classifying space $BG^{(0)}$, and $A_1^*\eta$ is the pull-back of a representative of the class $[\eta]$ via this map.

 The phenomena that we just described only affect the action of the group $G^{(0)}\times G^{(1)}$ on various operators of the theory. 
 The interplay between $p$-form symmetries of different degrees can lead to more radical modifications of the structure of symmetries, that might be combined into higher-categorical structures known as $n$-groups \cite{Sharpe:2015mja,Tachikawa:2017gyf}. In particular, let us focus a QFT with a $0$-form group $G^{(0)}$ and a $1$-form group $G^{(1)}$ giving rise to a $2$-group \cite{Cordova:2018cvg,Benini:2018reh,Bhardwaj:2022dyt}.
The $2$-group structure  can be understood as a failure of associativity when we consider a junction of three $0$-form symmetry defects $U_g$, $U_h$, $U_k$ into $U_{ghk}$. This is similar to the situation illustrated by figure \ref{f:defjunct} in $2$ dimensions. In this case, the invariant data encoding such a failure is a class $[\alpha]$ (Postnikov class), which is an element of the third group-cohomology group of $G^{(0)}$ with values in $G^{(1)}$:
\begin{equation}
\label{Postclass}
[\alpha]\in H^3_\rho(BG^{(0)},G^{(1)})\,.
\end{equation} The $2$-group structure implies that, given a standard $G^{(0)}$-connection $A_1$, the appropriate background field $B_2$ for $G^{(1)}$ has its coboundary fixed by $\alpha$ as
\begin{equation}
d_{\rho(A_1)} B_2 = A_1^\ast \alpha\,,
\label{coboundary2group}
\end{equation}
where $A_1$ is viewed as a map from spacetime to the classifying space $BG^{(0)}$. When $\alpha=0$, \eqref{coboundary2group} is in fact equivalent to the cocycle condition $d_{\rho(A_1)} B_2 = 0$, whose solutions are given by \eqref{fracclass}. When the pull-back $A_1^*[\alpha]$ of the Postnikov class  is non-trivial, this means that the background $B_2$ satisfying \eqref{coboundary2group} is defined only locally,
and that under a $0$-form symmetry transformation with parameter $\lambda$, $B_2$ acquires a non-trivial transformation
\begin{equation}
    A_1 \rightarrow A_1^\lambda\,,\quad B_2 \rightarrow B_2 + \zeta \left(\lambda,A_1\right)\,,
\end{equation}
with nonequivalent choices of $\zeta$ classified by \eqref{Postclass}. Given a solution $B_2$ to \eqref{coboundary2group}, there is still an ambiguity corresponding to shifting $B_2$ by a class in $H^2_\rho(G^{(0)},G^{(1)})$.

There is an important difference between the case of a $2$-group and the 't Hooft anomalies discussed in the previous section. In fact, $\alpha$ in this case is not just a phase, but a nontrivial operator of the theory -- it corresponds to a modification of the $B_2$ background, i.e. to the insertion of a codimension $2$ defect for the $1$-form symmetry $G^{(1)}$. Taking all these data into account we can describe a $2$-group global symmetry as
\begin{equation}
{G}=\left(G^{(0)},G^{(1)},\rho ,[\alpha]\right)\,.
\end{equation}

As we already saw for higher form global symmetry, we can also probe the $2$-group symmetry coupling the theory to background gauge field, namely to the connections of $2$-group gauge theory \cite{Baez:2004in,Baez:2005qu,Brennan:2020ehu,Hidaka:2020iaz, Hidaka:2020izy}.  In particular, codimension-$1$ $0$-form symmetry defects can be viewed as transition functions connecting couples of locally trivial patches in a principal $G^{(0)}$-bundle, and codimension-$2$ $1$-form symmetry defects are associated with transition functions for a $G^{(1)}$-gerbe. So in this language the appropriate background fields are a $1$-form gauge field for $G^{(0)}$ and a $2$-form gauge field for $G^{(1)}$, whose gauge transformations are controlled by $\rho$ and $[\alpha]$. 
In fact, describing the gauge transformations of the background fields is a very convenient way to encode the invariant information $\rho$ and $[\alpha]$ defining the $2$-group structure. As a simple example of this approach, we consider a continuous $2$-group where $\rho$ is trivial, $G^{(0)}=\prod_I U(1)^{(0)I}$ with background gauge fields $A_1^{I}$ that transform as $A_1^{I}\rightarrow A_1^{I} +d \lambda^I$, and $G^{(1)}=U(1)^{(1)}$ with the background $U(1)^{(1)}$-gerbe $ B_{2}$ transforming as
\begin{equation}
 B_{2} \rightarrow  B_{2} + d \Sigma_{1} +\frac{1}{2\pi}\sum_{I,J}\hat{\kappa}_{IJ}\lambda^I F_{2}^J\,,\quad F_{2}^I = d A_1^{I},\quad \hat{\kappa}_{IJ}\in \ZZ\,.
\end{equation}
The integer matrix $\hat{\kappa}_{IJ}$ can be identified with the Postnikov class $[\alpha]$. The gauge invariant field strength $H_{3}$ is consequently defined as
\begin{equation}
    d B_{2}=H_{3} + \frac{1}{2\pi}\sum_{I,J}\hat{\kappa}_{IJ} A_1^I F_{2}^J\,.
\end{equation}

Gauge fields with gauge transformations of this form are ubiquitous in string theory. From the target space viewpoint, they are actually \emph{dynamical} gauge fields, but they can also appear as background fields for the worldsheet theory describing the fundamental string.

\subsection{General features of orbifolds in QFT}\label{s:generalorb}

Let us now summarize some of the general features of orbifolds in quantum field theory.
\begin{itemize}
	\item {\it Anomalies.} Before even trying to construct the orbifold, one needs to check that it is consistent to gauge $\Gamma$. The obstruction to doing this are given by 't Hooft anomalies. In order to detect whether a given group $\Gamma$ has such an anomaly, one needs to consider the theory in the background of an external (non-dynamical) gauge field $\A$ for the group $\Gamma$. Formally, this means that one chooses a principal $\Gamma$-bundle with a connection $\A$ in the spacetime where the QFT is defined, and considers the partition function $Z[\A]$ obtained by requiring the fields of the QFT to be sections of vector bundles (for the appropriate $\Gamma$-representation) associated with the $\Gamma$-bundle of $\A$. Notice that when $\Gamma$ is discrete, and in particular when it is  finite, then the connection $\A$ is necessarily flat. A 't Hooft anomaly is a failure of the partition function $Z[\A]$, seen as a functional on the space of $\Gamma$-connections, to be invariant under $\Gamma$ gauge transformations of $\A$. More precisely, it is an obstruction to choosing suitable local counterterms in the action to make the partition function invariant. When such an anomaly is present, one cannot gauge $\Gamma$, i.e. one cannot consistently make the gauge connection $\A$ a dynamical field in theory.\\
	In the example of a bosonic 2D CFTs, 't Hooft anomalies for a group $\Gamma$ are classified the class $[\alpha]$ in the cohomology group $H^3(\Gamma,U(1))\cong H^4(\Gamma,\ZZ)$. 
	't Hooft anomalies can be defined even for higher form symmetries and for $n$-groups, see for example \cite{Kapustin:2014gua,Cordova:2018cvg,Benini:2018reh}.

	\item {\it Gauging.} If there are no 't Hooft anomalies for the group $\Gamma$, then one can promote the connection $\A$ to a dynamical field on the theory. This amounts to summing over all possible gauge bundles and connections $\A$, possibly with suitable weights. This sum is the path integral description of the partition function of the new theory. As for the operatorial description of the new orbifold theory, suppose that the spacetime is of the form $\Sigma\times \RR_t$, where $\RR_t$ is a time-like direction and $\Sigma$ is space-like, and let $\Hh$ be the Hilbert space of states on $\Sigma$. Then, the orbifold procedure can be thought of as consisting of two steps. First, one projects onto a `subtheory' whose operators are invariant under (local) $\Gamma$-transformations. Such an intermediate step might not necessarily satisfy all properties that we require for a QFT -- for example, in the case of 2D CFT, this first step alone does not lead to a modular invariant torus partition function. Next, one `completes' the subtheory by introducing new sectors in the space of states of the theory (twisted sectors), as well as the corresponding twist operators. Roughly speaking, the insertion of a twist operator at a certain space slice $\Sigma$ at fixed time creates a non-trivial $\Gamma$-bundle from the vacuum.
	\item {\it Discrete torsion.} For a given symmetry group $\Gamma$, there might be many consistent ways of introducing the twisted sectors, leading to different orbifolds. This freedom is known as discrete torsion \cite{Vafa:1986wx,Kapustin:2014gua}. In the path integral definition of the orbifold partition function, these different possibilities correspond to nonequivalent ways of weighing the contributions $Z[\A]$ for various choices of the gauge bundle and connection. \\
	In the 2D CFT example, discrete torsion corresponds to nonequivalent choices of the collection $\{\varphi_{g,h}\}$, leading to a trivial associator $\alpha_{g,h,k}$; such choices are in one-to-one correspondence with classes in $H^2(\Gamma,U(1))$.
	\item {\it Quantum symmetry and invertibility.} Suppose that the `child' theory $B$ is obtained from a `parent' theory $A$ via the gauging (orbifold) of some global $p$-form symmetry $\Gamma$. Then, the child theory $B$ always has a global $(D-p-2)$-form symmetry, sometimes known as the `quantum symmetry', that acts non-trivially only on the twist operators \cite{Kapustin:2014gua}. By gauging the quantum symmetry in theory $B$, one simply gets back the original theory $A$. In this sense, the orbifold procedure is invertible. We stress that, even if one starts with an ordinary symmetry group $\Gamma$ in the theory $A$, the quantum symmetry might in general be a `generalized' kind of symmetry, such as, for example, a (possibly non-invertible) object in a category of topological defects (see for example \cite{Bhardwaj:2017xup} for a recent discussion in 2D CFT).
	\item {\it Orbifolds of families of QFTs.} Rather than considering the orbifold of a single quantum field theory, one can consider the orbifold of a whole family of QFTs with some given symmetry group $\Gamma$. It is actually very difficult to provide a rigorous definition of what a `family of QFTs' should be, so here we will just sketch some basic ideas. Roughly speaking,  the correct definition should involve a fibration over a (topological) space of parameters $B$ (the base), where the fiber is a pair $(\T,\Gamma)$ of a quantum field theory $\T$ and a group $\Gamma$ of global symmetries of $\T$ isomorphic to a fixed abstract group (equivalently, an action of the abstract group $\Gamma$ as a group of global symmetries of $\T$). One might require proper notions of continuity and some choice of equivariant connection on such a family. 
	Let us assume that the 't Hooft anomaly for $\Gamma$ vanishes over the whole family. Then, we expect the orbifold procedure to define a second family over the same base $B$ and with fiber $(\T/\Gamma,Q)$, where $\T/\Gamma$ is the orbifold of $\T$ by $\Gamma$ and $Q$ is the quantum symmetry group. The procedure should be invertible, in the same sense as in the previous point. In case of non-trivial discrete torsion, the base space of the orbifold family might be a covering of the original base space $B$.
\end{itemize}

\section{Orbifolds in string theory}\label{s:stringorbifolds}

Our goal is to define an orbifold procedure in a theory of quantum gravity (and, in particular, in string theory), that resembles as much as possible the properties of orbifolds in QFT summarized in the previous subsection. 

As discussed in the introduction, there is an immediate problem with this program: in all known string theory models (and, conjecturally, in all theories of quantum gravity), there are no global symmetries. Therefore, the very definition of `orbifold' as a gauging of a (discrete) global symmetry simply does not make sense in quantum gravity. Nevertheless, there is a well known procedure in string theory that allows one to obtain a new string theory model starting from an old one. This orbifold procedure in string theory can be described from two different (but, ultimately, equivalent) perspectives:
\begin{itemize}
	\item \emph{Worldsheet orbifold.} One can focus on the 2D CFT describing the worldsheet of a fundamental string. Quite generally, what from the spacetime point of view appears as a gauge symmetry corresponds to a global symmetry for the worldsheet CFT. Therefore, it makes sense to consider the gauging of the worldsheet CFT by this global symmetry, and obtain a new 2D CFT. This is really an orbifold in the QFT sense, as described in the previous section, but the theory to which the orbifold procedure is applied is not the quantum gravity theory in spacetime, but rather the worldsheet QFT.
	\item \emph{Background orbifold.} Alternatively, one can consider the orbifold as the quotient of some geometric string background by isometries. The starting point is a string theory model, defined by choosing a topology for the $10$-dimensional spacetime (usually of the form $K_{10-D}\times M_D$, where $K_{10-D}$ is a compact manifold), and a consistent background for the metric, the dilaton and all the other fields appearing in the low energy effective field theory. If the chosen background is invariant under some finite group $\Gamma$ of isometries, one can define a new background by identifying the points in spacetime that are related by the action of $\Gamma$. The spacetime obtained in this way is a \emph{geometric} orbifold, possibly with some (mild) singularities at the loci of points that are stabilized by some non-trivial subgroup of $\Gamma$. This definition of the orbifold procedure makes sense in spacetime, but in general it depends on the geometry and topology of the compact submanifold $K_{10-D}$. In turn, these data might depend on the particular duality frame one is choosing to describe the theory. 
\end{itemize}

The group $\Gamma$ of isometries of a consistent string background induces a group $\Gamma$ of global symmetries on the worldsheet CFT $\C$ of the fundamental string. In this case, if one considers the string model obtained as the background orbifold by isometries, the corresponding worldsheet CFT is just the worldsheet orbifold $\C/\Gamma$, i.e. the one obtained from $\C$ by gauging the group $\Gamma$ of global symmetries. In this sense, the two points of view described above are equivalent, in the sense that they just lead to the same final string model.\footnote{The distinction between worldsheet orbifold and background orbifold procedure is very schematic. In practice, one often uses a mixture of geometric and worldsheet intuition in order to describe the resulting orbifold theory.}

The worldsheet point of view has the drawback that it singles out one of many dynamical objects in string theory. The properties of the other objects are then deduced by consistency of their interaction with the fundamental string -- for example one can derive the properties of D-branes by analyzing the boundary states in the worldsheet orbifold CFT.

The `background orbifold' viewpoint is apparently more `democratic' among the different dynamical objects. However, the description of the $10$-dimensional spacetime geometry is usually only valid in one specific duality frame, so that the fundamental string in this particular frame is again singled out among all dynamical objects. In a sense, the background just represents the geometry as `seen' by the fundamental string.

To better illustrate this point, consider type IIA compactified on K3$\times \RR^{5,1}$, giving rise to a quantum gravity theory in $6$ extended dimensions with $16$ spacetime supersymmetries and a gauge group of rank $24$ (generically, $U(1)^{24}$). This theory is dual to the heterotic string on $T^4\times \RR^{5,1}$, with the duality exchanging the type IIA fundamental string with the heterotic $5$-brane wrapping $T^4$, and the IIA NS5-brane wrapping K3 with the fundamental heterotic string. Describing the corresponding background as type IIA on K3, rather than heterotic on $T^4$,
singles out the type IIA fundamental string among all dynamical objects. In fact the internal K3 manifold is the geometry `seen' by this particular object, whereas a type IIA NS5-brane wrapping K3 would be better described as a heterotic string, and it would `see' an internal $T^4$ geometry.

Understanding the dependence (or independence) of the orbifold procedure on the duality frame is one of our main motivations for this work. It has long been known that `orbifolds do not always commute with dualities' \cite{Vafa:1995gm,Sen:1995ff,Sen:1996na}. Consider two string theory models $A$ and $B$ related by a duality. Suppose that $A$ has a symmetry $\Gamma$; by duality, the model $B$ must admit an isomorphic group $\Gamma'\cong \Gamma$ of symmetries. There are known examples where the orbifold of $A$ by $\Gamma$ is not dual to the orbifold of $B$ by $\Gamma'$. One example, already mentioned in the introduction, is the orbifold of 10d type IIB theory by worldsheet parity $\Omega$, which gives type I string theory. The S-dual of $\Omega$ is $(-1)^{F_L}$, the symmetry acting by $-1$ on the R-NS and R-R sectors and acting trivially on the NS-NS and NS-R sectors. The orbifold of type IIB  by $(-1)^{F_L}$ is type IIA, that clearly is not dual to type I. Another example arises within the pair of dual theories that we mentioned above, namely type IIA on K3 and heterotic on $T^4$. The orbifold of type IIA on K3 by the symmetry $(-1)^{F_L}$ gives rise to type IIB on K3. On the other hand, the corresponding symmetry on the worldsheet heterotic string has a 't Hooft anomaly, so that the orbifold is not even well defined!    

This phenomenon is perfectly compatible with the worldsheet description of the orbifold, if the duality is non-perturbative. Indeed, this description is based on gauging a global symmetry on the worldsheet of a fundamental string, a procedure that only makes sense in a duality frame where string perturbation theory is reliable. Non-perturbative dualities map the fundamental string to some different object and exchange the weak and the strong coupling limit of string theory. Therefore, we are not guaranteed that the orbifold procedure in the dual theory gives the same result.

It is natural to ask whether one can define a \emph{spacetime orbifold} procedure in string theory that treats all dynamical objects in a `democratic' way, and that does not depend on any specific effective description or duality frame. In the example of the IIA on K3 and heterotic on $T^4$ dual pairs, we can just describe the resulting theory as a six dimensional quantum gravity theory, with asymptotically flat geometry, with a given gauge group $H$ and a collection of dynamical (point-like or extended) objects that are charged with respect to $H$ (see \cite{Fraiman:2022aik} for a recent description of the landscape of such models and their orbifolds). The finite group $\Gamma$ we want to quotient by appears at some special loci in the moduli space, where it combines with the generic group $H$ to form an enhanced gauge group.  We stress that there is no need to choose a specific duality frame in order to describe the gauge group. We can also classify all stable dynamical objects in the theory just in terms of their couplings to the various gauge fields, again without making a specific choice between the type IIA and the heterotic frame.  In fact, only if we take this `democratic' perspective we are able to match the action of the group $\Gamma$ on the two sides of the duality. The orbifold procedure consists in projecting out some of the dynamical objects of the theory, modifying the gauge group of the theory via a sequence of restrictions, quotients and extensions, and introducing new objects and fields (the twisted sector) which interact in a consistent way with the ones of the parent theory. 

These arguments suggest the existence of a `spacetime orbifold' procedure that satisfies the following properties:
\begin{itemize}
	\item All dynamical objects and all effective descriptions of the theory are taken on the same footing.
	\item It is defined by specifying some finite subgroup $\Gamma$ of the gauge group, that must get quotiented out in the final theory. As suggested by the examples in section \ref{s:highform}, quotienting of gauge groups can often be described in terms of the gauging of some higher form finite global symmetry in spacetime.
	\item The sequence of steps describing the orbifold depends very little on the dynamics of the theory, and is mainly based on the symmetries of the theory itself and on how the dynamical objects of the theory are coupled to them. This property should allow us to extend the procedure to families of models sharing the same group of symmetries.
	\item It is equivalent (at least, in most cases) to the standard definitions of orbifolds in string theory that are given above, in the sense that in most cases it just provides the same orbifold theory. In fact, we expect this procedure to be a generalization of the usual string orbifolds.
\end{itemize}

If such a `spacetime orbifold' procedure exists, then one can reasonably expect it to always commute with dualities. On the other hand, this seems to be in contradiction with the very examples we have described above: we know already that there are cases where duality and orbifolds do not commute! As discussed in the introduction, there are various possible resolutions to this paradox:
\begin{itemize}
	\item The most obvious possibility is that any orbifold procedure is intrinsically related to a given choice of duality frame or effective description. Therefore, there is no meaningful way to define the orbifold that does not make any reference to a choice of duality frame. 
	\item A second possibility is that, in some cases, there might be more than one possible way to take an orbifold by a given group $\Gamma$. In QFT, we already know that this is the case, due to discrete torsion. It might very well be that, if one focuses on a specific dynamical object or duality frame, not all these possibilities can be described in a natural way in terms of an orbifold on a worldsheet theory. That would mean that only one of the possible ways of taking an orbifold (in spacetime sense) is `visible' in a given duality frame, and a different way of taking the orbifold is visible in another duality frame. When this happens, then it would appear as if `the orbifold does not commute with duality'. What is really happening is that the worldsheet orbifolds on the two sides of the duality are not the same operation from a spacetime point of view. This is the possibility that we will consider in this article.
\end{itemize}

In the following, we will describe a rough proposal for this `spacetime orbifold' procedure. The proposal will be then illustrated in a couple of examples in the following sections. In particular, we want to verify that, at least for these simple examples, the spacetime orbifold proposal reproduces the orbifold as obtained from either the worldsheet or the geometric background perspective.

\subsection{Initial data: the parent theory}

Let us start by describing the parent theory, i.e. the string theory model for which we want to take the orbifold. Our goal is to express this `initial data' in a way that does not depend on the choice of a particular duality frame, and that is mainly based on the symmetries of a given theory.

Let us consider the compactification of some superstring theory on a consistent background. Quite generally, we can describe this model as a quantum gravity theory in $D$ extended dimensions with a given asymptotic geometry -- for simplicity, we will only consider asymptotically flat spacetime. 
We do not fix the topology of the spacetime away from the asymptotic boundary, since we want to allow for dynamical topology changes. The theory has a gauge group $G$, where `group' should be meant in a  very general sense: it might comprise subgroups $G^{(p)}$ of $p$-form gauge symmetries for different degrees $p$, with non-trivial mixings among various values of $p$, giving rise to structures such as $n$-groups or generalizations. Furthermore, there might be some matter fields or some (possibly extended) dynamical objects (fundamental strings, branes, monopoles,...) that can be charged with respect to the various gauge fields.\footnote{Of course, in string theory the distinction between smooth extended solitons and fundamental branes is just an artefact of an effective description -- they are just different limits of the same object, and we will refer to them generically as branes.} On the other hand, we do not include $D$-dimensional spacetime diffeomorphisms in the gauge group $G$; here, we are implicitly assuming that there is an (approximate) description of our theory where this separation between $D$-dimensional spacetime diffeomorphisms and an `internal' gauge group $G$ makes sense. In particular, in the following, when we talk about gauge invariant operators, we only mean invariant under local $G$-transformations.

In general, in QFT, the gauge group is not an intrinsic property of a given theory: there might exist different descriptions of the same QFT with different gauge groups. For example, pure $U(1)$ Maxwell theory in three dimensions can be described in terms of a dual scalar field, with trivial gauge group. This can happen because physical states and observables, that characterize intrinsically a QFT, are all invariant under local gauge transformations.

On the other hand, physical states might not be invariant under asymptotic gauge transformations, i.e. such that the gauge parameters do not vanish at infinity (see for example \cite{Harlow:2018jwu,Harlow:2018tng}
 for a similar discussion). 
Suppose, for example, that our string model admits, in some semiclassical description, a (ordinary $0$-form) $U(1)$ gauge group. We want to allow for our theory to contain states whose total electric and/or magnetic $U(1)$ charge is different from zero, and which therefore belong to non-trivial representations with respect to constant $U(1)$ transformations. Similarly, when the gauge group contains a discrete subgroup $H$, we expect the quantum Hilbert space to contain states transforming in non-trivial representations of $H$ under the action of the asymptotic gauge group.

If there is some charged state in a faithful representation of a gauge group $G$, then all possible descriptions of the theory must contain $G$ as a group of symmetries\footnote{In principle, we cannot exclude that, in some particular description, $G$ is extended to some larger gauge group. If such situations occur, then $G$ should be regarded as the quotient of the gauge group by the subgroup acting trivially on all states.}. In all known string theory models, and conjecturally in all consistent quantum gravity theories, the spectrum of charges is complete, i.e. contains all possible representations of a gauge group. This means that there are physical states transforming faithfully under asymptotic gauge transformations, so that the gauge group is not a property that depends on the particular description (for example, the duality frame) of the theory. 

We emphasize that, for this description to be valid, the group $G$ should be in a Coulomb (or free charge, in the terminology of \cite{Harlow:2018jwu,Harlow:2018tng}) phase. In particular, $G$ should not include any spontaneously broken symmetry. This means that if there is a moduli space of string models, the group $G$ will, in general, depend on the moduli. 
On the other hand, there might exist continuous families of different string models, parametrized by some moduli, with the same unbroken gauge group $G$. This can happen, in particular, when the marginal operators corresponding to small deformations of (some of) the moduli are invariant under $G$. In this case, we expect our orbifold procedure to map the whole family of parent theories to a family of orbifold theories with the same base -- see the discussion at the end of section \ref{s:generalorb}.

Finally, in order for the definition to be duality invariant, one should include `electric' and `magnetic' gauge fields in a democratic way. For example, in ten dimensional type IIA theory, $G$ will contain the Ramond-Ramond $p$-form $U(1)$ gauge group for all possible odd values of $p$, even though the $p$-form and $(8-p)$-form gauge fields are actually dual to each other.

In any semiclassical description of the theory, there will be extended dynamical objects (which, depending on the duality frame, will be called D-branes, NS5-branes, etc.) that are coupled in various ways to the gauge fields. There are usually degrees of freedom localized on the worldvolume of such extended objects, giving rise to worldvolume currents that are sources for $p$-form gauge fields in the bulk. The gauge couplings also dictate how these extended objects are allowed to end onto each other. 
As discussed in \cite{Heidenreich:2020pkc}, most of these properties are strongly constrained once we require the absence of all kind of global symmetries (in particular, what are called Chern-Weil symmetries in \cite{Heidenreich:2020pkc}) in the complete quantum gravity theory. In fact, in \cite{Heidenreich:2020pkc} it is shown how one can reconstruct the various couplings in the worldvolume theory of extended dynamical objects, starting from the low energy effective action, and in particular from the Chern-Simons terms, for the massless $p$-form gauge fields in spacetime. In turn, these Chern-Simons terms are fixed by the higher group structure of the gauge group, which is part of the initial data we are considering. 
This argument suggests that the full information about the structure of the gauge group $G$, together with the requirement of the absence of global symmetries, is sufficient to provide very detailed information about a given string theory model.

\subsection{The orbifold projection and twisted sector}

Starting from the initial data described in the previous subsection, let us now discuss how to define the orbifold of our string model by a $0$-form finite subgroup $\Gamma$ of the full gauge group $G$. It might be possible to extend our procedure to higher form gauge symmetries, but we will not consider this generalization here.  Furthermore, for simplicity, we will only consider explicit examples where the symmetry group $\Gamma$ is cyclic, $\Gamma\cong \ZZ_N$.

If we want to compare our procedure to the standard `worldsheet orbifold', we also need to assume that the gauge group $\Gamma$ in spacetime corresponds, in some duality frame, to a group of global symmetries in the CFT describing the worldsheet of a fundamental string. The action of $\Gamma$ on the worldsheet of the fundamental string might not in principle be faithful; this is the case, for example, when $\Gamma$ is a discrete subgroups of a $U(1)$ gauge group associated with a R-R gauge fields in type II string models. 
In these cases, the final outcome of the `spacetime orbifold' procedure does not correspond to any  `worldsheet orbifold', at least in the given duality frame.  The examples we consider in the next sections only consider groups of symmetries that descend from global symmetry of the worldsheet CFT (in some duality frame); it would be very interesting to discuss the outcome of our construction in more general examples.

Vice versa, any group $\Gamma$ of global symmetries of the worldsheet theory, commuting with the BRST operator that defines the physical string states,\footnote{A worldsheet symmetry that does not commute with the BRST operator (for example not commuting with the worldsheet $\N=(1,1)$ super-Virasoro algebra in type II string theory) does not preserve the space of physical string states.} gives rise to a symmetry of perturbative string theory. Assuming that the symmetry is not broken at non-perturbative level, it must  correspond to a group of $0$-form gauge symmetries in spacetime. When $\Gamma$ is (a subgroup of) a global \emph{continuous} symmetry on the worldsheet, the existence of the corresponding gauge symmetry can also be inferred from the presence of a massless gauge boson in the physical string spectrum.

Since $\Gamma$ is a gauge symmetry in spacetime, it must be associated with a gauge field $A_1$. When $\Gamma$ is a subgroup of a continuous gauge group, for example $\Gamma\subset U(1)$, we let $A_1$ be the $U(1)$ gauge field and $F_2=dA_1$ the corresponding field strength. 

In general, a gauge theory with a gauge group $\Gamma$ should include Wilson lines or loops $W_R$ for each representation $R$ of $\Gamma$. Invariance under local gauge transformations requires an open Wilson line to be stretched all the way to infinity, i.e. with both ends at the boundary of spacetime, or to have one or both ends terminated on some suitably charged (point-like) operator, if it is present in the theory. In quantum gravity theories, the conjecture that the spectrum of charges is complete implies that all Wilson lines are `endable', i.e. that there are always point-like operators where a Wilson line can end in the interior of space time.

Let us now describe the theory obtained from an orbifold of this quantum gravity theory by the group $\Gamma$.

\begin{itemize}
	\item The orbifold theory should contain some of the degrees of freedom (fields or extended dynamical objects) of the original theory. In order to understand which degrees of freedom are retained in the orbifold, it is useful to reformulate the steps of the worldsheet orbifold from a spacetime point of view. 
	The first step in the worldsheet orbifold procedure is to project out the states and operators that are not invariant under the group $\Gamma$, seen as a group of global symmetries on the worldsheet. From the spacetime point of view, the string states that we eliminate correspond to degrees of freedom transforming in some non-trivial representation $R$ with respect to the gauge group $\Gamma$. More generally, we expect an analogous projection to occur for all the degrees of freedom localized on other extended objects such as D-branes, NS5-branes, etc.  From a spacetime perspective, therefore, we argue that the orbifold theory should contain only local degrees of freedom transforming in the trivial representation $R$ of the group $\Gamma$. \\
	Notice that, unless $\Gamma$ is a central subgroup of $G$, the $G$ gauge fields themselves transform  non-trivially under $\Gamma$. Thus, in this first step, the gauge group $G$ must be modified. In particular, only the gauge fields that commute with $\Gamma$ will be retained in the orbifold theory; see section \ref{s:reflection} for an example where the group $\Gamma$ is not in the centre of $G$.
	\item The local degrees of freedom that we eliminated in the first step are exactly the ones where Wilson lines in some non-trivial $\Gamma$-representation $R$ could end. Thus, if such Wilson lines were still present in the orbifold theory, then they would be `non-endable'. As discussed in section \ref{s:highform}, the presence of non-endable Wilson lines is related to the presence of a potential $1$-form `electric' global symmetry group, still isomorphic to $\Gamma$. The group acts on a Wilson line $W_R$ in the representation $R$ of $\Gamma$ by multiplication by $\Tr_R(g)$, which is just a phase for a cyclic group $\Gamma$. From a different perspective, the $1$-form global symmetry is generated by  $(D-2)$-dimensional topological operators $T_g$ (Gukov-Witten operators), labeled by elements $g\in \Gamma$.\footnote{For non-abelian groups $\Gamma$, the Gukov-Witten operators are labeled by conjugacy classes of elements in $\Gamma$.}\\
	It is widely believed that in a consistent theory of quantum gravity, any `potential' global symmetry should be either broken or gauged. In the parent string model, the $1$-form symmetry was broken due to the presence of charged local operators where the Wilson lines $W_R$ could end, which make the Gukov-Witten operators $T_g$ non-topological. As stressed above, in the orbifold theory these charged local operators are not present, so that the $1$-form symmetry cannot be broken. Therefore, it must be gauged. Gauging the global symmetry has two effects. First of all, the Wilson lines $W_R$ for non-trivial $\Gamma$-representation $R$ are not gauge invariant, and therefore are not present in the theory anymore.  The second effect of gauging the electric $1$-form symmetry is that we should introduce new configurations of the gauge fields where the transition functions close only up to gauge transformations in $\Gamma$. Effectively, this means that the gauge group in the orbifold theory is obtained by taking a quotient of the `parent' gauge group by the subgroup $\Gamma$.
	\item For each dynamical object (fundamental string D-brane, NS-brane), we should allow for the worldvolume fields to have non-trivial $\Gamma$-monodromy around any non-trivial cycle wrapped by the brane. The reason for this is the following. In the original theory, the partition function is obtained by summing over all possible (principal) $\Gamma$-bundles with connection in spacetime. For each such choice of $\Gamma$-bundle, the fields in the worldvolume of each dynamical object must be represented by sections of a suitable $\Gamma$-vector bundles, which is determined by the given principal $\Gamma$-bundle (more precisely, by the pullback of such principal bundle in spacetime to the worldvolume of the brane). In the orbifold theory, after gauging the $1$-form symmetry, all different principal $\Gamma$-bundles in spacetime are physically equivalent to each other and to the trivial $\Gamma$-bundle -- one has no way to observe the difference. This means that, in the worldvolume theory, one should include all possible $\Gamma$-vector bundle configurations, independently of any (unobservable) choice of $\Gamma$-principal bundle in spacetime. In particular, we should allow for fields with non-trivial $\Gamma$-monodromy even in the absence of non-trivial gauge background in the orbifold theory.      Furthermore, for each $g\in \Gamma$, we should include operators of codimension $2$ on the worldvolume of each string or brane, that create twist vortices such that the worldsheet fields have monodromy $g$ around the insertion point. On the fundamental string, these operators exactly correspond to $g$-twisted operators in the worldsheet orbifold construction.
	\item At the end of the day, we expect the content of the twisted sector in the final orbifold theory to be dictated by consistency conditions in quantum gravity. Some of these conditions might come from standard tadpole cancellation -- for example, in some cases, it is well known that the orbifold theory is well defined only after including a suitable number spacetime filling branes \cite{Sethi:1996es}. In our description of the twisted sector, the main consistency condition we are imposing is the absence of any global symmetries in the final orbifold theory. It has been recently argued that the structure of string theory is strongly constrained by such a requirement. In particular, based on the absence of all `Chern-Weil symmetries', the authors of \cite{Heidenreich:2020pkc} were able to deduce the coupling of D-branes and other extended objects to various $p$-form gauge fields, as well as the presence of dynamical worldvolume degrees of freedom. In a similar spirit, in the previous point, we used the absence of global symmetries to deduce the structure of the twisted sector in the orbifold theory. The analysis of \cite{Heidenreich:2020pkc} suggests that this strategy, together with tadpole cancellation, should be sufficient to reconstruct the orbifold theory completely -- or at least up to a limited number of choices.
\end{itemize}

In the following sections we will show how this procedure can be implemented in some simple examples. In particular, our goal will be to show that the final theory is the same that one would obtain via the worldsheet orbifold procedure.

\section{Orbifold of type II on $S^1$ by half-period shift}\label{s:halfperiod}

In this section, we consider the compactification of type IIA or type IIB string theory on a circle $S^1$. We do not fix the geometry (or even the topology) of the interior of the $D=9$ dimensional spacetime, but we assume that the asymptotic region at infinity is just flat $8+1$ dimensional Minkowski spacetime. As we review below, this theory has a $U(1)$ gauge symmetry corresponding (in a suitable duality frame) to translations along the internal circle $S^1$. Our goal is to apply (at least in a simplified setup) the procedure described in section \ref{s:stringorbifolds} to obtain the orbifold of this model by the $\ZZ_2$ subgroup of this $U(1)$ symmetry, corresponding to the translation by half a period along $S^1$.

For simplicity, we will focus on the bosonic NS-NS sector, and more specifically on the various gauge fields, and ignore the fermions and the RR sector. 
\subsection{The initial data}

Let us start from considering the low energy effective action. The NS-NS sector of type II $10$-dimensional superstring theory contains a dilaton scalar $\hat{\phi}$, the Kalb-Ramond (KR) 2-form $\hat{B}_{2}=\frac{1}{2}\hat{B}_{\hat{\mu}\hat{\nu}}dx^{\hat{\mu}}\wedge dx^{\hat{\nu}}$,
and a two-index traceless symmetric tensor $\hat{g}_{\hat{\mu}\hat{\nu}}$ describing
the graviton.

The low-energy effective action in the \textit{string frame} is given by
\begin{eqnarray}
S_{NS-NS}  &=&\frac{\hat{g}^2}{16\pi G_N^{(10)}} \intop d^{10}\hat{x} \sqrt{-\hat{g}} e^{-2\hat{\phi}}\left[\hat{R}+4\left(\partial\hat{\phi}\right)^{2}-\frac{1}{2\cdot 3!}\hat{H}_{3}^{2}\right]\,,\nonumber\\
	& = &\frac{\hat{g}^2}{16\pi G_N^{(10)}} \intop d^{10}\hat{x} e^{-2\hat{\phi}}\sqrt{-\hat{g}}\left(\hat{R}+4\left(\partial\hat{\phi}\right)^{2}\right)\nonumber\\
    &&-\frac{\hat{g}^2}{16\pi G_N^{(10)}} \intop e^{-2\hat{\phi}}\frac{1}{2}\hat{H}_{3}\wedge\hat{\ast}\hat{H}_{3}\,,\label{10dEA}
\end{eqnarray}
where $\hat{H}_{3}=d\hat{B}_{2}$ (locally) is the KR field strength. Here, $\hat{\ast}$ is the Hodge dual with respect to the string metric  $\hat{g}_{\hat{\mu}\hat{\nu}}$, $\hat{g}=e^{\hat{\phi}_0}$ is the \textit{string coupling constant } compensating for the asymptotic value of the dilaton, and $G_N^{(10)}$ is the 10-dimensional supergravity Newton constant. The normalization of the KR field in the effective action \eqref{10dEA} is such that the fundamental string couples to $\hat B_2$ via the standard Wess-Zumino term in the worldsheet action
\begin{equation}
-{T}\int_\Sigma\hat{B}_{2}\,,
\end{equation}
with $T$  the string tension and $\Sigma$ the worldsheet. 
The Kalb-Ramond field $B_2$ is the gauge boson associated with a $1$-form $U(1)$ gauge symmetry
\begin{equation}\hat{B}_{2}\rightarrow\hat{B}_{2}+d\,\hat{\Sigma}_1\,,\label{10KRgaugetransform}\end{equation}
where $\hat{\Sigma}_1= \hat{\Sigma}_{\hat{\mu}}d\hat{x}^{ \hat{\mu}}$ is a $1$-form gauge parameter. We have the quantization conditions
\begin{equation}
-{T}\int_{M^{(3)}} \frac{\hat{H}_{3}}{2\pi}\in \ZZ\,, \qquad -{T}\int_{M^{(2)}} \frac{d\hat{\Sigma}_1}{2\pi}\in \ZZ\,,
\label{10dBquantization}
\end{equation}
when evaluated in closed surfaces $M^{(p)}$. 

The bosonic sector also includes the Ramond-Ramond (RR) fields,which are differential forms of even (odd) rank in the IIB (IIA) theory. In the string frame they do not couple to the dilaton, but they couple to the KR 2-form due to the definitions of the field strengths and also to the presence of Chern-Simons (CS) topological terms. Although these couplings are responsible for the quite interesting phenomenology of extended objects in string theory, we shall not consider the whole RR sector in the following. 

Instead we shall consider the structure emerging from  a standard Kaluza-Klein (KK) ansatz for the NS-NS sector. Denoting the compact coordinate by $x^{\underline{9}}\equiv z\in[0,2\pi\ell_z]$ and assuming that all the fields are independent of it, we can rewrite the action in terms of the corresponding 9-dimensional vielbein, KR 2-form $B_2$ and dilaton field $\phi$, the KK vector $A_1$, the winding vector $B_1$ and the KK scalar field $k$. We omit the Einstein-Hilbert term, the dilaton and KK scalar kinetic terms
in order to focus on the relevant $p$-form gauge symmetry without cluttering our notation. Therefore the action is
\begin{align}
		S_{NS-NS} & =  -\frac{{g}^2}{16\pi G_N^{(9)}}\intop d^{9}x\sqrt{-g}e^{-2\phi}\left(\frac{1}{2\cdot 3!}\tilde {H}_{3}^{2}+\frac{1}{4}k^{2}\tilde F_{2}^{2}+\frac{1}{4}k^{-2}\tilde H_{2}^{2}+\ldots\right)\notag\\
		& =  -\frac{{g}^2}{16\pi G_N^{(9)}}\intop \frac{1}{2}e^{-2\phi}\left(\tilde{H}_{3}\wedge\ast \tilde{H}_{3}+k^{2}\tilde F_{2}\wedge\ast \tilde F_{2}+k^{-2}\tilde H_{2}\wedge\ast \tilde H_{2}+\ldots\right)\,,\label{NSNSaction}
\end{align}
where
\begin{eqnarray}
\tilde{H}_{3}&=&d\tilde B_{2}- \tilde A_1\wedge \tilde H_{2}\,,\\
\tilde F_2 & = & d\tilde A_1\,,\\
\tilde H_2 & = & d\tilde B_1\,,
\end{eqnarray}
and 
\begin{equation}
g=\hat{g}k_0^{-\frac12}\,,\quad G_N^{(9)}=G_N^{(10)}/\left(2\pi R_z\right)\,,
\end{equation}
with \(k_0\) the asymptotic value of the KK scalar field and 
\[
R_{z}=l_{z}k_{0}\,,
\]
the radius of the compact dimension $z$. ${\ast}$ is the Hodge dual with respect to the string metric  ${g}_{{\mu}{\nu}}$. 

The action \eqref{NSNSaction} is invariant under the  Nicolai-Townsend transformations \cite{Nicolai:1980td,Bergshoeff:1981um}
\begin{gather}
	\delta \tilde A_{1}=d\tilde\lambda,\qquad\delta \tilde B_{1}=d\tilde \sigma,\nonumber \\
	\delta \tilde B_{2}=d\tilde\Sigma_{1}+\tilde \lambda \tilde H_2\,.
	\label{eq:NicolaiGaugeTransformNorm}
\end{gather}
In particular one can see that $\delta \tilde{H}_{3}=0$.
The origin of the gauge transformations for the KK vector field $\tilde A_1$ is the general coordinate transformations of the compact coordinate $z$, $\delta z = -k_0^{-1}\tilde \lambda$, so that the period of the gauge parameter is $\tilde\lambda\sim \tilde\lambda+2\pi R_z$ and we have the quantization conditions
\be \frac{1}{R_z}\int_{M^{(2)}} \frac{d\tilde F_2}{2\pi}\in \ZZ\,,\qquad\qquad \frac{1}{R_z}\int_\gamma \frac{d\tilde \lambda}{2\pi}\in \ZZ\, .
\ee
Given the dimensional reduction definitions, 
\begin{equation}
\tilde B_{\mu}\equiv\hat{B}_{\mu\underline{z}}\,,\qquad \tilde\sigma=\hat{\Sigma}_{\underline{z}}\,,
\end{equation}
(which are taken to not depend on the compact dimension $z$), upon integrating over the closed manifolds $M^{(3)}=  S^1 \times N^{(2)} $ $M^{(2)}= S^1 \times \gamma $, with $S^1$ the circle in the compact dimension and $N^{(2)}$, $\gamma$ respectively a closed surface and line in the transverse directions,  \eqref{10dBquantization} implies
\begin{equation}
-{T}\left(2\pi R_z\right)\int_{N^{(2)}} \frac{\tilde {H}_{2}}{2\pi}\in \ZZ\,, \qquad -{T} \left(2\pi R_z\right)\int_{\gamma} \frac{d \tilde\sigma}{2\pi}\in \ZZ\,.
\end{equation} In particular, the gauge parameter $\tilde \sigma$ is defined modulo 
\begin{equation}
\tilde \sigma \sim \tilde \sigma + 2\pi R_z'= \tilde \sigma +\frac{2\pi \alpha'}{R_z} \,.
\end{equation} where
\begin{equation}
R_z'= \frac{\alpha'}{R_z}\,,
\end{equation}
is the radius of the T-dual circle.

Finally, by \eqref{10dBquantization}, the Kalb-Ramond $2$-form field $\tilde B_2$ and the gauge parameter $\tilde\Sigma_1$ are quantized as
\begin{equation}
 T\int_{N^{(3)}}\frac{d \tilde B_2}{2\pi}\in \ZZ\,, \qquad T\int_{N^{(2)}} \frac{{d}\tilde {\Sigma}_1}{2\pi}\in \ZZ\,,
\end{equation}

For our purpose, it is more convenient to normalize the  fields so that they have standard quantization conditions for $U(1)$ gauge fields. We use the redefinitions
\be A_1=\frac{\tilde A_1}{2\pi R_z}\,,\qquad B_1={R_z T}\tilde B_1\,, \qquad B_2=\frac{T}{2\pi}\tilde B_2\, .
\ee The gauge transformations have the same form as before
\begin{gather}
	\delta  A_{1}=d \lambda,\qquad\delta  B_{1}=d \sigma,\nonumber \\
	\delta  B_{2}=d \Sigma_{1}+  \lambda\,  H_2\,.
	\label{eq:NicolaiGaugeTransform}
\end{gather} but the respective gauge parameters
\be \lambda=\frac{\tilde\lambda}{2\pi R_z}\,,\qquad \sigma={R_z T}\tilde \sigma\,, \qquad \Sigma_1=\frac{T}{2\pi}\tilde \Sigma_1\, ,
\ee now have integral periods over closed manifolds
\be \int_\gamma d\lambda\in \ZZ\,,\qquad
 \int_\gamma d\sigma\in \ZZ\,,\qquad
  \int_{M^{(2)}} d\Sigma_1\in \ZZ\,.
\ee 
In terms of the field strengths $F_2=dA_1$, $H_2=dB_1$ and $H_3=dB_2-A_1\wedge H_2=\frac{T}{2\pi}\tilde H_3$, the action \eqref{NSNSaction} becomes
\be -\intop \left(\frac{1}{2g_{B_2}^2}{H}_{3}\wedge\ast {H}_{3}+\frac{1}{2g_{A_1}^2}F_{2}\wedge\ast F_{2}+\frac{1}{2g_{B_1}^2}H_{2}\wedge\ast H_{2}+\ldots\right)\,,\ee
where
\be  \frac{1}{2g_{B_2}^2}=\frac{\pi{g}^2}{4 G_N^{(9)}T^2}e^{-2\phi}\qquad \frac{1}{2g_{A_1}^2}=\frac{\pi{g}^2k^2R_z^2}{4 G_N^{(9)}}e^{-2\phi}\qquad \frac{1}{2g_{B_1}^2}=\frac{{g}^2k^{-2}}{16\pi G_N^{(9)}R_z^2T^2}e^{-2\phi}\,.
\ee

In the language of \cite{Cordova:2018cvg}, the gauge transformations \eqref{eq:NicolaiGaugeTransform} are an example of $2$-group structure of the form $G^{(0)}\times_\kappa G^{(1)}$, with $G^{(0)}=U(1)^{A_1}\times U(1)^{B_1}$ and $G^{(1)}=U(1)^{B_2}$, where the $0$-form gauge transformation of the fields $A_1$ and $B_1$ also affects the $1$-form gauge field $B_2$.

This string compactification is also invariant under the $\ZZ_2$ symmetry acting by reflection on the internal circle $S^1$. From the $9$-dimensional point of view, this symmetry acts by charge  conjugation on the $0$-form gauge fields
\[
A'_{\mu}=-A_{\mu},\qquad B'_{\mu}=-B_{\mu}.
\]
Strictly speaking, one should include this transformation as a discrete gauge symmetry, making the whole $0$-form symmetry group non-abelian. In this section, we will just focus on the normal abelian subgroup. We will consider this complication in the next section, in the more general context of toroidal compactifications.  

In the absence of external electric or magnetic charges, we get the Bianchi identities
\begin{align}
	d {H}_{3}+H_{2}\wedge F_{2}&=0,\\
	dF_2&=0\\
	dH_2&=0
\end{align}
and the equations of motion
\begin{align}
	d(g^{-2}_{B_2}\ast{H}_{3})  &= 0,\\
	d(g^{-2}_{A_1}\ast F_{2})-H_{2}\wedge g^{-2}_{B_2}\ast {H}_{3}  &= 0,\label{EOMF2pure}\\
	d (g^{-2}_{B_1}\ast H_{2})- F_{2}\wedge g^{-2}_{B_2}\ast {H}_{3} & =0,
\end{align}
where $\ast$ is taken with respect to the $D$-dimensional metric
$g_{\mu\nu}$.  

Let us now consider our supergravity theory in the presence of external electric or magnetic sources for these gauge fields. The Bianchi identities and equations of motion are modified by the presence of suitable $p$-form currents $j_p$ that are `localized', i.e supported on $\left(9-p\right)$-dimensional manifolds in spacetime. As discussed in  \cite{Heidenreich:2020pkc}, self-consistency of the equations require that the presence of new degrees of freedom localized on the world-volume of the extended objects coupled with the gauge fields.

Let us now discuss how the Bianchi identities and equations of motion are modified in the presence of such electric and magnetic currents. In string theory, these sources correspond to various kinds of extended dynamical objects. The presence of Chern-Simons terms and the non-standard form of some Bianchi identities put non-trivial consistency conditions on the couplings to sources. In particular, the extended objects are required to carry worldvolume degrees of freedom with suitable transformations under the action of the gauge group. More precisely, the equations in the presence of sources take the following form:
\begin{align}
	d {H}_{3}+H_{2}\wedge F_{2}&=j_4^{NS5}+\nabla_{B_1} \rho^{KK}\wedge j_3^{KK}+\nabla_{A_1} z^{NS5}\wedge j_3^{NS5} ,\label{BianchiH3}\\
	dF_2&=j_3^{KK}\label{BianchiF2}\\
	dH_2&=j_3^{NS5}\label{BianchiH2}\\
	d(g^{-2}_{B_2}\ast{H}_{3})  & =j_7^F,\label{EOMH3}\\
	d(g^{-2}_{A_1}\ast F_{2})-H_{2}\wedge g^{-2}_{B_2}\ast {H}_{3} & 
    =j_8^m - \nabla_{B_1} \tilde Z\wedge j_7^F - \nabla_{B_5} \tilde z_4^{NS5}\wedge j_3^{NS5},\label{EOMF2}\\
	d (g^{-2}_{B_1}\ast H_{2})- F_{2}\wedge g^{-2}_{B_2}\ast {H}_{3} & 
    =j_8^F - \nabla_{A_1} Z\wedge j_7^F - \nabla_{B_5}  \tilde\rho_4^{KK}\wedge j_3^{KK}.\label{EOMH}
\end{align}

In string theory, the currents admit the following interpretation:
\begin{itemize}
	\item $j^F_7$ and $j^F_8$ correspond to fundamental strings that are, respectively, localized at the point $2\pi \ell_z Z$ in the compact direction, or wrapping the compact direction and localized in the uncompactified ones. From the $9$-dimensional spacetime perspective, the former current is supported in the $2$-dimensional worldsheet spanned by the string; the latter current is supported on a line, which is the trajectory in the $9$-dimensional spacetime of a string wrapping the compact direction. Strings that are localized in the compact $S^1$ direction carry a worldsheet degree of freedom, the scalar field $Z$, that corresponds to the position of the string along $S^1$.  The scalar field $\tilde Z$ is defined as the worldsheet dual of $Z$,
	\begin{equation}\label{Hodgedual}
    g^{-1}_{A_1}\nabla_{A_1} Z = g^{-1}_{B_1}\star \nabla_{B_1}\tilde Z \, ,  
	\end{equation} where $\star$ denotes the worldsheet Hodge duality.
	\item $j^{NS5}_3$ and $j^{NS5}_4$ correspond to NS5 branes that are, respectively, localized at a point or wrapping the compact direction. Branes localized at the point carry a world-volume degree of freedom, the scalar field $z^{NS5}$, corresponding to the position of the brane in the compact direction. The $4$-form $\tilde z^{NS5}_4$ is the Hodge dual of $z^{NS5}$ in the worldvolume of the NS5 brane, in analogy with \eqref{Hodgedual}.
	\item $j_3^{KK}$ is a KK monopole. It is known \cite{Gregory:1997te} that KK monopoles carry a worldvolume degree of freedom, a scalar field $\rho^{KK}$, that gets shifted as $\rho^{KK}\to \rho^{KK}+\sigma_1$ under a $U(1)^{B_2}$ transformation with gauge parameter $\sigma_1$. The $4$-form $\tilde \rho^{KK}_4$ is the Hodge dual of $\rho^{KK}$ in the worldvolume of the KK monopole, in analogy with \eqref{Hodgedual}.
	\item $j_8^m$ is a current corresponding to units of momentum along the internal $S^1$ circle.
	\item The covariant derivatives are $\nabla_X Y=dY+X$, where $X$ is a $p$-form potential and $Y$ is a charged $p-1$ form field on the worldvolume;  $B_5$ is the magnetic dual of $ B_2$, i.e. the gauge field with field strength $dB_5=g_{B_2}^{-2}\ast H_3$.
\end{itemize}

The existence of worldvolume degrees of freedom is required for consistency of these equations. To see this, let us apply the differential $d$ to both sides of one these equations, for example \eqref{EOMF2}. We get
\begin{align}
	-j_3^{NS5}\wedge  g^{-2}_{B_2}\ast\tilde{H}_{3} - H_2\wedge j_7^F& =dj_8^m-(d\nabla_{B_1} \tilde Z)\wedge j_7^F- (d\nabla_{B_5} \tilde z_4^{NS5})\wedge j_3^{NS5},\label{dEOMF2}
\end{align} 
which is an identity using that \be d\nabla_{B_5} \tilde z_4^{NS5}=d(d\tilde z_4^{NS5}+B_5)=g^{-2}_{B_2}\ast\tilde{H}_{3} \ee and 
\be d\nabla_{B_1} \tilde Z=d(d\tilde Z+B_1)=H_2\ .
\ee Without the worldvolume fields $z_4^{NS5}$ and $Z$, one would not be able to write gauge invariant combinations such as $\nabla_{B_5} \tilde z_4^{NS5}$ and $\nabla_{B_1} \tilde Z$, and it would be impossible to cancel the terms on the left hand side of this equation. See for example \cite{Heidenreich:2020pkc} for a more detailed description of this mechanism.

In a low energy effective description of string theory, it is justified to consider these currents as external non-dynamical sources only when the corresponding objects are very heavy. On the other hand, our orbifold procedure does not depend on such dynamical details, but only on the structure of the gauge group and on their coupling to charged objects. Therefore, we can regard the equations above as a schematic way to encode such information, independently of the conditions under which these equations make sense in some suitable low energy limit.

\subsection{The projection}\label{s:projection}

We have described the `initial data' of the parent string theory. Now, we want to apply the procedure described in section \ref{s:stringorbifolds}, and consider the orbifold by the $\ZZ_2$ subgroup of the $U(1)^{A_1}$ gauge group; this group is generated by a half-period shift along the internal $S^1$ circle. 

The first step in the procedure is to project out all dynamical objects carrying non-trivial charge with respect to such $\ZZ_2$ gauge group. This projection will make the corresponding Wilson lines non-endable, and `restore' a $1$-form $\ZZ_2$ electric global symmetry acting on such Wilson lines. Naively, the operators generating such a $1$-form symmetry are of the form
$$ e^{\pi i \int_{M_7} g_{A_1}^{-2} *F_2}\ ,
$$ for any closed $7$-manifold $M_7$. Even without introducing the electric and magnetic sources, this $1$-form symmetry is broken due to the Chern-Simons term in the equation of motion. Indeed, if $M_7$ and $M_7'$ are the boundary components of a $8$-manifold $N_8$, i.e. $\partial N_8=M_7-M_7'$, then 
\be \int_{M_7} g_{A_1}^{-2} *F_2-\int_{M'_7} g_{A_1}^{-2} *F_2=\int_{N_8} d(g_{A_1}^{-2} *F_2)=\int_{N_8} - H_2\wedge g_{B_2}^{-2}\ast\tilde H^3\ .
\ee In the classification of \cite{Marolf:2000cb}, the current $d(g_{A_1}^{-2}*F_2)$ is called a Maxwell current, and it is gauge invariant and conserved, but it is not quantized nor `localized', i.e. it is not supported on a manifold of positive codimension in spacetime. 
Assuming that the gauge bundle for $U(1)^{B_1}$ is trivial, i.e. the class $[H_2]\in H^2(X,\ZZ)$ vanishes, then $B_1$ is a well defined $1$-form on $X$, and we can consider the operator
\be\label{Page}  e^{\pi i \int_{M_7} J_7^{Page}}\ ,
\ee where
\be\label{PageCurr} J_7^{Page}= g_{A_1}^{-2} *F_2+B_1\wedge g_{B_2}^{-2}* \tilde H_3\ .
\ee In the absence of external sources one has $$dJ_7^{Page}=0\ .$$
More generally, the quantity $dJ_7^{Page}$ (known as the Page current) is quantized, so that in principle one could restore a $1$-form $\ZZ_2$ symmetry by restricting to configurations where the number of quantized charge is even.
Unfortunately, the operator \eqref{Page} is not gauge invariant: under $B_1\to B_1+d\sigma$, $J_7^{Page}$ transforms as
\be  J_7^{Page}\to  J_7^{Page}+d\sigma \wedge g_{B_2}^{-2}* \tilde H_3\ .
\ee
The $1$-form $d\sigma$ is closed, but it is not exact as a real-valued $1$-form, because $\sigma$ is only defined in $\RR/\ZZ$ and in general does not lift to a real valued $0$-form. This means that $\int_{M_7}d\sigma \wedge g_{B_2}^{-2}* \tilde H_3$ is integral but not necessarily zero. As a consequence, the operator \eqref{Page} is not well defined, since a gauge transformation can change its sign. 

The presence of the Chern-Simons term leads to a second, closely related problem.
We have seen in the previous subsection that consistency of the Bianchi identities and equations of motions requires the presence of worldvolume degrees of freedom. For example, among the electric sources for $\ast F_2$ one must include terms such as $\nabla_{B_1}\tilde Z\wedge j_7^F$, with very precise normalization and quantization conditions. One cannot modify such contributions arbitrarily without modifying the Chern-Simons term $H_2\wedge g_{B_2}^{-2}* \tilde H_3$. In particular, the integral of $d\tilde Z\wedge j_7^F$ over a closed $8$-manifold can be any integer. Indeed, requiring $d\tilde Z$ to have even periods when integrated over loops would be incompatible with $U(1)^{B_1}$ gauge invariance, while imposing that $j_7^F$ is even would `restore' an additional $2$-form global $\ZZ_2$ symmetry with generator $e^{i\pi\int_{M^{(6)}} g_{B_2}^{-2}\ast H_3}$.
Even in the absence of external sources, if the term $H_2\wedge g_{B_2}^{-2}\ast H_3$ is not modified, one expects solitonic solutions for the massless fields carrying any possible electric charge for $U(1)^{A_1}$ compatible with the Dirac quantization conditions. In general, in string theory the distinction between solitons for the massless fields and fundamental extended objects is really an artefact of the low energy effective description. Thus, one can argue that there must be suitable solitonic solutions corresponding, for example, to the source term $d\tilde Z\wedge j_7^F$.

Our orbifold procedure instructs us to exclude all possible field configurations giving rise to an odd electric charge for the $U(1)^{A_1}$ gauge group. 
As  argued above, in order to implement such a projection, we are forced to modify the quantization conditions for the fields $B_1$ or $B_2$, in order to be consistent with the term $ H_2\wedge g_{B_2}^{-2}*\tilde H^3$ in the equations of motion for $\ast F_2$.
This modification can be obtained as follows. Consider the operator
\be\label{H2symm} e^{\pi i\int_{M_2} H_2}\ ,
\ee for any closed $2$-manifold $M_2$. In the absence of external sources, this defines a $6$-form $\ZZ_2$ global symmetry. In type II string theory, this symmetry is broken by any odd number of NS5 branes transverse to the internal $S^1$ circle. In the same spirit as in the procedure of section \ref{s:stringorbifolds}, the symmetry \eqref{H2symm} can be `restored' by requiring the number of NS5 branes to be even, for example by imposing
\be\label{NS5proj} j_3^{NS5}=2\tilde j_3^{NS5}\ ,
\ee where $\tilde j_3^{NS5}$ is an integral-valued cochain. Once such a restriction on $j_3^{NS5}$ is imposed, \eqref{H2symm} defines a global symmetry and can be gauged. As explained, for example, in \cite{Gaiotto:2014kfa}, one possible way to gauge a $(D-3)$-form $\ZZ_2$ symmetry is via a BF theory. We introduce some dynamical $(D-2)$-form $U(1)$ gauge field $\C_{D-2}$ and a $1$-form $U(1)$ gauge field $\B_1$, obeying the standard quantization conditions, and then add to the spacetime action the terms
\be 2\pi i\int_X \C_{D-2}\wedge (H_2-2 d\B_1)\ .
\ee The functional integral over $\C_{D-2}$ then forces
\be\label{eomZ2H2} B_1=2\B_1\ ,
\ee up to irrelevant gauge transformations. Note that we did do not introduce any additional propagating degrees of freedom, since the functional integral over $\B_1$ implies $d\C_{D-2}=0$.  

After this gauging, the integral $\int_{M_2} H_2$ over any closed manifold $M_2$ is even. Furthermore, the flux of $ H_2\wedge g_{B_2}^{-2}* \tilde H_3$ through any closed $8$-manifold $N$ is also even, so that one cannot construct a solitonic operator where the Wilson line for $\ZZ_2\subset U(1)^{A_1}$ can end just out of these fields.

In fact, gauging the `magnetic' $(D-3)$-form $\ZZ_2$ symmetry \eqref{H2symm} has the effect of modifying the gauge group $U(1)^{B_1}$ by replacing it by a double cover\footnote{In general, by $X.Y$ we denote a group containing $X$ as normal subgroup and such that the quotient by $X$ is isomorphic to $Y$.} $\ZZ_2.U(1)\cong U(1)$, that can be described as a group extension 
\be 1\longrightarrow \ZZ_2\longrightarrow \ZZ_2.U(1)\longrightarrow U(1)^{B_1}\longrightarrow 1\ .
\ee  While the double cover $\ZZ_2.U(1)$ is still isomorphic to $U(1)$ as an abstract group, it has different quantization conditions with respect to $U(1)^{B_1}$. In fact, only configurations carrying even magnetic charge for $U(1)^{B_1}$ lift to a well defined gauge bundle for the cover $\ZZ_2.U(1)$. On the other hand, the cover  $\ZZ_2.U(1)$ admits new representations corresponding to half-integral electric charges for $U(1)^{B_1}$, i.e. half-integral winding number along the circle $S^1$.   The double cover $\ZZ_2.U(1)$ is the group associated with the gauge field $\B_1$; therefore, we will denote such group by $U(1)^{\B_1}$. One can use \eqref{eomZ2H2} to replace $B_1$ by $2\B_1$ and $H_2$ by $2\Hh_2=2d\B_1$ everywhere; $\B_1$ has the standard quantization conditions for a $U(1)$ field.

As a bonus, after gauging the symmetry \eqref{H2symm}, the operator \eqref{Page} is now gauge invariant and conserved (in the absence of external sources). Indeed, 
using
\be J_7^{Page}= g^{-2}_{A_1} *F_2+B_1\wedge g^{-2}_{B_2}* \tilde H_3=g^{-2}_{A_1} *F_2+2\B_1\wedge g^{-2}_{B_2}* \tilde H_3\ ,
\ee we see immediately that a gauge transformation for $\B_1$ changes the integral \be\label{intF2} \int_{M_7} J_7^{Page} \ee by \emph{even} integers, so that \eqref{Page} is well defined. Notice that, when the gauge bundle for $U(1)^{\B_1}$ is non-trivial, the expression $2\B_1\wedge g^{-2}_{B_2}* \tilde H_3$ is not a well-defined real valued $7$-form, but should rather be considered as a $7$-cochain with values in $\RR/2\ZZ$. This is sufficient for the operator \eqref{Page} to be well-defined.

Let us now discuss how the current $J_7^{Page}$ (as an integral cochain modulo $2$) needs to be modified upon introducing electric and magnetic sources, as in eq.\eqref{EOMF2}.
We rewrite \eqref{EOMF2} as
\be\label{EOMF2b} d(g^{-2}_{A_1}\ast F_{2})-H_{2}\wedge g^{-2}_{B_2}\ast {H}_{3}-B_1\wedge j_7^F +B_5 \wedge j_3^{NS5} 
    =j_8^m+d\tilde Z\wedge j_7^F - d \tilde z^{NS5}\wedge j_3^{NS5}\ .
\ee
After imposing \eqref{NS5proj}, the term $d\tilde z^{NS5}\wedge j_3^{NS5}=2d\tilde z^{NS5}\wedge \tilde j_3^{NS5}$ on the right-hand side of \eqref{EOMF2} is already even. We need to impose that $j^8_m$ and $d\tilde Z \wedge j_7^F$ be even as well. The latter requirement amounts to impose that the integral $\oint_\gamma d\tilde Z$, where $\gamma$ is any closed circle on the worldsheet of a fundamental string, must be even (physically, this is just the center of mass momentum of the string in the compact direction). In fact, the change in the quantization conditions of $B_1$ forces $\tilde Z_1$ to takes values in $\RR/2\ZZ$ rather than $\RR/\ZZ$. We consider now an integral-valued $8$-cochain $\mathsf{H}_8\in C^8(X,\ZZ)$ in spacetime such that
\be \int_{M^{(8)}}(-H_{2}\wedge g^{-2}_{B_2}\ast {H}_{3}-B_1\wedge j_7^F +B_5 \wedge j_3^{NS5})=2\int_{M^{(8)}}\mathsf{H}_8\ , 
\ee for any closed $8$-manifold $M^{(8)}$. Such an integral cochain exists, because the right-hand side of \eqref{EOMF2b} is quantized and even. This implies that
\be -H_{2}\wedge g^{-2}_{B_2}\ast {H}_{3}-B_1\wedge j_7^F +B_5 \wedge j_3^{NS5}=2(\mathsf{H}_8+\delta b_7)\ ,
\ee for some real $7$-cochain $b_7$, where $\delta$ denotes the coboundary operator. The cochains $\mathsf{H_8}$ and $b_7$ satisfying these conditions are defined up to
\be \mathsf{H}_8-\delta \mathsf{n}_7\ ,\qquad b_7+\mathsf{n}_7+\beta_7\ ,
\ee where $\mathsf{n}_7$ is an integral $7$-cochain and $\beta_7$ is a closed $7$-cochain (a cocycle), i.e. $\delta\beta=0$. The shifts of $b_7$ by an integral cochain $\mathsf{n}_7$ or by a coboundary $\beta_7=d\alpha_6$ correspond to gauge transformations of $B_1$ and $B_5$.  On the other hand, the shift of $b_7$ by a cochain $\beta_7$ that is closed but not exact is physically meaningful: it corresponds to shifts of $B_1$ and $B_5$ by flat connections, i.e. to changes of holonomy.  
We can now define the current
\be\label{Page2} J_7^{Page}=g^{-2}_{A_1}\ast F_{2}+2b_7\ ,
\ee so that by \eqref{EOMF2b}
\be \delta J_7^{Page}=-2\mathsf{H}_8+j_8^m+d\tilde Z\wedge j_7^F - d \tilde z^{NS5}\wedge j_3^{NS5}\ .
\ee The right hand side of this equation is quantized and even. It follows that, with this definition of $J_7^{Page}$, the operator \eqref{Page} is gauge invariant and topological, so that it implements a $\ZZ_2$ global symmetry, and it can be gauged.

\subsection{Gauging and twisted sectors}

Let $\Gamma\cong \ZZ_2$ be the subgroup of the gauge group $U(1)^{A_1}$ corresponding to half-integral shifts; the operator \eqref{Page} is the generator of the $1$-form `electric' global symmetry corresponding to the gauge group $\Gamma$. Gauging \eqref{Page} has the effect of projecting out all closed Wilson lines measuring a non-trivial holonomy for $\Gamma$. Equivalently, such a gauging can be described as the insertions of a network of codimension $2$ defects $T_g$, $g\in \Gamma$, (Gukov-Witten operators for $\Gamma$), that shift the $A_1$ gauge field by a flat connection corresponding to monodromy $g$ around the defect. As usual in a gauging procedure, one has to sum over all possible ways of labeling the defects in the network by elements in $\Gamma$, subject the condition that junctions between defects is compatible with the group law. After this gauging, all configurations for the group $U(1)^{A_1}$ that differ  by a flat connection in $\Gamma\subset U(1)^{A_1}$ are physically equivalent and indistinguishable.

Gauging \eqref{Page} has also consequences for the configurations of the fields localized on the worldvolumes of dynamical objects, such as the fundamental strings and the branes. In the original theory, the insertions of a Gukov-Witten operator $T_g$, $g\in \Gamma$, on a codimension $2$ submanifold $N$ had physically observable consequence. For example, consider a fundamental string with worldsheet of the form $S^1\times \RR$, with the $\RR$ direction parallel to the defect $N$ and the circle $S^1$ encircling $N$. In this situation, the worldsheet fields are constrained to have monodromy $g$ around $S^1$. Furthermore, if the defect $T_g$ intersects the worldsheet of a fundamental string, it creates a vortex with monodromy $g$ for the worldsheet fields localized at the intersection point. Such vortices on the worldsheet are not allowed outside of the intersection points with some $T_g$. This shows that the position of $T_g$ in the original theory is physically observable. 

On the other hand, in the orbifold theory, gauge configurations with or without the insertion of $T_g$ must be physically equivalent. This means that, even when the background for the orbifold gauge group is trivial, one should include all possible monodromies $g\in \Gamma$ for the worldsheet fields along an $S^1$ circle. Furthermore, one is free to insert, at any point on the worldsheet, operators creating a vortex for worldsheet fields, independently of the configuration of the gauge fields in spacetime. 

These additional states and operators for the fundamental string correspond exactly to the twisted sector in the worldsheet orbifold. In our spacetime orbifold procedure,  we see immediately that analogous twisted sector operators should be introduced in the worldvolume of all dynamical objects that are present in the theory, and not just on the string worldsheet. 
Of course, the presence of such operators can be shown to be necessary in the worldsheet orbifold procedure as well, but their derivation is less direct. For example, for D-branes, it can be deduced by studying the boundary states in the orbifold worldsheet CFT.

\medskip 

Let us now discuss the effects of the spacetime orbifold procedure on the $0$-form gauge groups of the theory. We have already found that the factor $U(1)^{B_1}$ gets extended to a double cover $U(1)^{\B_1}$, with $B_1=2\B_1$. We have also excluded all objects carrying odd electric charge with respect to $U(1)^{A_1}$. After gauging \eqref{Page}, we should include new gauge bundles for the $A_1$ gauge field, where the transition functions on triple intersections only close up to transformations in $\Gamma\subset U(1)^{A_1}$. These gauge bundles carry `half-integral' magnetic flux, from the point of view of the original $A_1$ gauge field. We should also allow for the corresponding codimension $3$ 't Hooft operators with `half-integral' magnetic charge. 
In the original theory, these `badly quantized' 't Hooft operators could only arise as boundaries of some Gukov-Witten operator; after the orbifold, the presence of Gukov-Witten operators $T_g$, $g\in \Gamma$, is unobservable, so that the corresponding 't Hooft lines become genuine codimension $3$ operators. 
Finally, in order for the completeness conjecture to be satisfied by the orbifold theory, one should also include `half-integral' KK monopoles as dynamical objects, where the new 't Hooft operators can end. 
As discussed for example in \cite{Heidenreich:2021xpr,McNamara:2021cuo}, the completeness conjecture is required in order for all global symmetries to be broken. In our approach, the absence of global symmetries is a guiding principle for constructing the twisted sector of our theory.

At the end of the day, the modifications to the gauge group $U(1)^{A_1}$ correspond to quotienting $U(1)^{A_1}$ by its central subgroup $\Gamma\cong \ZZ_2$. As an abstract group,  $U(1)^{A_1}/\ZZ_2$ is still isomorphic to $U(1)$, but the quantization conditions on the charges are different. Indeed, only the even charge representations of $U(1)^{A_1}$ define representations of the quotient group $U(1)^{A_1}/\ZZ_2$. On the other hand,  $U(1)^{A_1}/\ZZ_2$ allows for more general magnetic charges, that are `half-integral' quantized from the point of view of $U(1)^{A_1}$. The half-integral KK monopoles we introduced are the dynamical objects carrying such magnetic charges. Their presence breaks a potential  $(D-3)$-form global symmetry (the spacetime quantum symmetry) which arises when gauging a $1$-form symmetry. Such a quantum symmetry would act on the newly introduced 't Hooft operators, if the latter were non-endable. 
We denote by $\A_1$ the $U(1)$ gauge field, with standard quantization conditions, of the quotient $U(1)^{A_1}/\ZZ_2\cong U(1)$. Such a field is related to $A_1$ by $2A_1=\A_1$. 

Even without any detailed analysis of the Bianchi identities and equations of motion and their coupling to the currents, the need to perform the gauging of both $e^{i\pi \int_{M_2} H_2}$ and $e^{i\pi\int_{M_7} J_7^{Page}}$ could be understood in terms of the $2$-group structure of the gauge group, which is encoded in the gauge transformations \eqref{eq:NicolaiGaugeTransform}. 
One cannot simply replace the gauge group $U(1)^{A_1}$ by its quotient $U(1)^{A_1}/\ZZ_2$ without modifying any other gauge field, because the new quantization of the gauge parameter $\lambda$ would not be compatible with the quantization of $B_2$ and $B_1$. In particular, the term $\lambda H_2$ in the gauge variation of $B_2$ in \eqref{eq:NicolaiGaugeTransform} would be inconsistent with the quantization conditions on $H_2$ and $dB_2$.
On the other hand, if we simultaneously replace $U(1)^{A_1}$ by its quotient $U(1)^{A_1}/\ZZ_2$ and $U(1)^{B_1}$ by the double cover $\ZZ_2.U(1)^{B_1}$, the quantization of the term $\lambda H_2$ in the gauge transformation of $B_2$ in \eqref{eq:NicolaiGaugeTransform} is left invariant, and there is no inconsistency. 
This argument is very robust, in the sense that it is not based on any effective description of the theory or on the existence of any particular source for the gauge fields, but is intrinsic to the structure of the gauge group.

\bigskip

In this section, we considered only a small part of the full gauge group -- the one that was more relevant form to understand the relation with the worldsheet orbifold procedure. Similar considerations should apply when all the other gauge fields are included, in particular  the ones arising in the RR sector. For example, the gauge fields $B_1$ and $B_2$ are related by U-duality to RR $U(1)$ gauge fields $C_1$ and $C_2$, whose corresponding electric charges are carried by suitable D-branes that are, respectively, point-like or string-like objects in the $9$-dimensional spacetime. 
All the arguments we described for the fields $A_1$, $B_1$, $B_2$ applies with few modifications for the fields triplet of fields $A_1$, $C_1$, $C_2$. 

In particular, the $2$-group structure of the gauge transformations, and the corresponding Chern-Simons terms in the action are completely analogous. This means that, as a first step in our orbifold procedure, we will have to `restore' and then gauge a $(D-3)$-form $\ZZ_2$ symmetry generated by $e^{i\pi \int_{M^{(2)}} G_2}$, $G_2=dC_1$, (the RR analogue of $e^{i\pi \int_{M^{(2)}}H_2}$), which is a magnetic symmetry for the gauge field $C_1$. 
As a consequence of this gauging, the $U(1)$ gauge group related to $C_1$ gets extended to a double cover $\ZZ_2.U(1)$. 
The current \eqref{Page2} receives corrections from the RR analogue of $b_7$. With these corrections included and both $e^{\pi i\int_{M^{(2)}} G_2}$ and $e^{\pi i\int_{M^{(2)}} H_2}$ gauged, the operator $e^{\pi i\int_{M_7} J_7^{Page}}$ is again topological, and we can proceed with gauging the corresponding $1$-form $\ZZ_2$ symmetry, as described above.

\section{Orbifold of toroidal compactification by coordinate inversion}\label{s:reflection}

As a second example, let us now consider a compactification of type II on a $d$-dimensional torus $T^d$, $d\le 6$ at some generic point in the moduli space, times an  uncompactified spacetime $X$ of dimension $D=10-d$, which we require to be asymptotically Minkoswski. We want to describe the orbifold by the order $2$ symmetry $\mathsf{C}$ that inverts all coordinates of the torus $T^d$.   As in the previous case, we focus on the bosonic NS-NS sector for simplicity.\footnote{In this oversimplified treatment, where we plainly ignore all R-R fields, as well as spacetime fermions, the dimension $d$ of the torus is not very important. In general, one should take into account that, for odd $d$, the coordinate inversion $k$ changes the orientation of the $10$-dimensional spacetime. Furthermore, when $d\equiv 2\mod 4$, $k$ lifts to an element of order $4$ in the spin group. Strictly speaking, the naive treatment in this section is mostly relevant for the case $d=4$, where no such subtleties arise.}

The $0$-form part of the gauge group contains a semidirect product
$$ G=\ZZ_2^C\ltimes \prod_{i=1}^d (U(1)^{A^i}\times U(1)^{B^i})\ ,
$$ where in the normal subgroup $U(1)^{A^i}\times U(1)^{B^i}$ the first $U(1)^{A_i}$ factor represents translations along the $i$-th circle $S^1$ of the torus $T^d$, while the second $U(1)^{B^i}$ factor represents translations in the T-dual circle. 
The associated conserved charges are internal momenta and fundamental string windings, respectively, and we denote the corresponding $1$-form gauge fields respectively by $A_1^i$ and $B_1^i$, $i=1,\ldots, d$, with field strengths $F_2^i=dA_1^i$ and $H_2^i=dB_1^i$. The group $\ZZ_2^C$ is the symmetry inverting all coordinates of $T^d$, and therefore acts by charge conjugation on all $U(1)$ factors; we denote by $\mathsf{C}$ the generator of $\ZZ_2^C$ and by $C$ the corresponding $\ZZ_2$ gauge field. Therefore, this $\ZZ_2^C$ is the subgroup of the gauge group corresponding to the symmetry we want to orbifold by. More precisely, we define $\mathsf{C}$ to be an element in $G$ with non-trivial image under the projection $G\to \ZZ_2^C$, and we identify $\ZZ_2^C$ with the subgroup of $G$ generated by $\mathsf{C}$. All such elements are conjugate to each other in $G$, and therefore the orbifold procedure does not depend on this choice.
For the moment, we ignore the fact that $G$ is part of a higher group structure -- this will turn out to be too naive, as we will see below.

The first step in our procedure is to eliminate all local operators where a Wilson line in a non-trivial representation of $\ZZ_2^C$ can end, so as to restore a global electric $1$-form symmetry acting on `non-endable' Wilson lines. In this case, this global $1$-form symmetry is broken even if we considered just a pure gauge theory for the group $G$. Indeed, because $G$ is non-abelian, the gauge fields themselves transform non-trivially under $\ZZ_2^C$. In particular, the field strengths of the $U(1)^d\times U(1)^d$ subgroup transform as
\be F_2^i\to -F_2^i\ ,\qquad\qquad H_2^i\to -H_2^i\ .
\ee Therefore, an operator $F_2^i(x)$ or $H_2^i(x)$ can be inserted at the ending point of a Wilson line corresponding to a non-trivial representation of $\ZZ_2^C\subset G$ to get a $G$-invariant operator. This fits with the general statement that, for a non-abelian gauge group $G$, the only electric global $1$-form symmetries that are possibly unbroken are the ones corresponding to the center of $G$ \cite{Gaiotto:2014kfa}.

Thus, in order to implement the first step in our procedure and restore an unbroken global $1$-form symmetry corresponding to $\ZZ_2^C$, we must restrict the gauge group $G$ to the maximal subgroup of symmetries $G_{res}\subset G$ commuting with $\ZZ_2^C$, i.e. to the centralizer
$$ G_{res}:=C_G(\ZZ_2^C)=\{g\in G\mid g\mathsf{C}=\mathsf{C}g\}\ .
$$ It is easy to check that the only symmetries commuting with $\mathsf{C}$, apart from $\mathsf{C}$ itself, are generated by the half period shifts along any internal direction, as well as the half-period shifts in the T-dual torus, so that
\be\label{restrgroup}  G_{res}=\ZZ_2^C\times (\ZZ_2\times \ZZ_2)^d\ .
\ee
The fact that the gauge group must be restricted to this centralizer can be also seen at the level of the gauge fields. We want to restrict our path integral only to gauge bundles and connections that commute with $\ZZ_2^C$. At the level of the field strengths, this means that we have to impose $F_2^i=-F_2^i$ and $H_2^i=-H_2^i$ for all $i=1,\ldots, d$, which implies
\be F_2^i=0\ ,\qquad H_2^i=0\ ,\qquad i=1,\ldots, d\ ,
\ee i.e. all $U(1)^d\times U(1)^d$ connections must be flat.\footnote{At the level of the Chern classes $c_1\in H^2(X,\ZZ)$ of the various $U(1)$ gauge bundles, the condition that the gauge field configuration commutes with $\ZZ_2^C$ implies $c_1=-c_1$, i.e. $2c_1=0$.  Thus, if $H^2(X,\ZZ)$ contains a non-zero $2$-torsion class, a topologically non-trivial gauge bundle might still be allowed. For simplicity, we will ignore this possibility here.} For what concerns the gauge $1$-forms $A_1^i$ and $B_1^i$, one has to impose
\be A_1^i\sim -A_1^i\ ,\qquad B_1^i\sim -B_1^i\ ,\qquad i=1,\ldots, d\ ,
\ee where $\sim$ denotes equality up to $U(1)^d\times U(1)^d$ gauge transformations. Besides the flatness conditions $dA_1^i=0=dB^i_1$ that we already discussed, this implies that $2A_1^i$ and $2B_1^i$ must be pure gauge for all $i=1,\ldots, d$. For flat $U(1)$ connections, this is equivalent to requiring that
\be \oint_\gamma 2A_1^i\in \ZZ\ ,\qquad \oint_\gamma 2B_1^i\in \ZZ\ ,\qquad i=1,\ldots, d\ ,
\ee for all $1$-cycles (closed $1$-chains) $\gamma$. Equivalently, we require that the cohomology classes with representatives $2A_1^i$ and $2B_1^i$ are all integral
\be
[2A_1^i]\in H^1(X,\ZZ)\ ,\qquad [2B_1^i]\in H^1(X,\ZZ)\ ,\qquad i=1,\ldots, d\ ,
\ee where $X$ is the spacetime. These conditions do not imply that the gauge fields $A_1^i$ and $B_1^i$ are themselves trivial, since their integral along any closed curve in spacetime is allowed to be half-integral. This means that the holonomies $e^{2\pi i\oint_\gamma A_1^i}$ and $e^{2\pi i\oint_\gamma B_1^i}$ must be signs, i.e. they must take values in the $\ZZ_2^d\times \ZZ_2^d$ subgroup of $U(1)^d\times U(1)^d$. But this means that the gauge group is restricted to \eqref{restrgroup}, as expected.

Naively, this argument seems to show that, at least for what concerns the gauge fields, the first part of our procedure is completed: because $\ZZ_2^C$ is, by construction, central in the restricted gauge group $G_{res}=C_G(\ZZ_2^C)$, the Wilson lines corresponding to non-trivial $\ZZ_2^C$ representations are non-endable in the pure $G_{res}$ gauge theory. However, as mentioned above, this reasoning neglects the fact that the $0$-form group $G$ is part of a higher group structure. In particular, the $U(1)^d\times U(1)^d$ $0$-form gauge transformations affect also the B-field $B_2$, which from the $10-d$ spacetime point of view is the $2$-form gauge field of a $U(1)^{(1)}$ $1$-form symmetry. 

To illustrate this point, let us focus for a moment on the case $d=1$, so that we have only a $U(1)\times U(1)$ continuous $0$-form gauge group, with gauge fields $A_1$ and $B_1$. The gauge transformations of $A_1$, $B_1$ and $B_2$ are the same as in \eqref{eq:NicolaiGaugeTransform}
\be\label{twogroupgauge} A_1\to A_1+d\lambda\ ,\qquad B_1\to B_1+d\sigma\ ,\qquad B_2\to B_2+B_1\wedge d\lambda+d\Sigma_1\ ,
\ee where $\lambda$ and $\sigma$ are $0$-forms with values in $\RR/\ZZ$, and $\Sigma_1$ is (locally) a $1$-form defined modulo
$$ \Sigma_1\sim \Sigma_1+d\rho\ ,
$$ for any $\RR/\ZZ$ valued $0$-form $\rho$.\footnote{To give a more precise definition of $\lambda$, $\sigma$, and $\Sigma_1$, let us consider an open cover $\{U_i\}_{i\in I}$ of the spacetime $X=\bigcup_i U_i$ such that each $U_i$ is topologically trivial. On each patch $U_i$, $\lambda$ can be represented by a real-valued function $\lambda_i$. On  double intersections $U_i\cap U_j$, we require the transitions functions to be constant integral functions, i.e. $\lambda_i-\lambda_j\in \ZZ$. Similarly, on each local patch $U_i$, $\Sigma_1$ can be represented by a real $1$-form. On intersections $U_i\cap U_j$ we allow the corresponding local $1$-forms to differ by $d\rho_{ij}$, for some $\RR/\ZZ$-valued $0$-forms $\rho_{ij}$. In turn, if $U_i\cap U_j$ are topologically trivial, we can always lift all transition functions $\rho_{ij}$ to real-valued $0$-forms satisfying $\rho_{ij}+\rho_{jk}+\rho_{ki}\in \ZZ$ on any triple intersection $U_i\cap U_j\cap U_k$.}

The transformation of the $B_2$ field can be recast in a different form by some field redefinition $B_2\to B_2-\alpha A_1\wedge B_1$, $\alpha\in \RR$, which must be accompanied by the addition of a term $\alpha A_1\wedge B_1$ to the worldsheet action. No such field redefinition can cancel completely the dependence of the $B_2$ gauge transformation on the $0$-form gauge parameters $\sigma$ and $\lambda$; this is the signature of a higher (in this case $2$-) group structure.

The fact that $\lambda$ and $\sigma$ are valued in $\RR/\ZZ$ rather than $\RR$ means that the integrals of $d\lambda$ and $d\sigma$ along a closed curve $\gamma$ are not necessarily $0$, but rather that
$$ \oint_\gamma d\lambda\in \ZZ\ ,\qquad \oint_\gamma d\sigma\in \ZZ\ .
$$ As a cross-check, notice that a transformation \eqref{twogroupgauge} of $A_1$ and $B_1$, with parameters $\lambda$ and $\sigma$ quantized as above, does not affect the fields strengths $F_2$ and $H_2$, nor the $U(1)$-valued holonomies $e^{2\pi i\oint_\gamma A_1}$ and $e^{2\pi i\oint_\gamma B_1}$, as expected for a gauge transformation. Similarly, a gauge transformation with parameter $\Sigma_1$ does not affect $dB_2$, and changes the integral of $B_2$ over any closed $2$-manifold only by integers.

Now, the generator of the $\ZZ_2^C$ symmetry acts on a triple of fields $(A_1,B_1,B_2)$ as
\be (A_1,B_1,B_2)\to (-A_1,-B_1,B_2)\ .
\ee Notice that $B_2$ is not affected by the inversion of the coordinates of the internal $T^d$ torus, since it corresponds to the components of the $10$-dimensional Kalb-Ramond field $B_{\mu\nu}$ where none of the indices $\mu,\nu$ is along $T^d$.  As argued above, in order to implement the first step in the orbifold procedure, we must restrict the gauge fields to configurations such that
\be (A_1,B_1,B_2)\sim (-A_1,-B_1,B_2)\ ,
\ee where in this case $\sim$ denotes equivalence up to the gauge transformations \eqref{twogroupgauge}. Because
\be (-A_1,-B_1,B_2)=(A_1-2A_1,B_1-2B_1,B_2)\ ,
\ee this restriction implies, as we argued above, that $2A_1$ and $2B_1$ must be pure gauge, i.e. that
\be\label{Z2gauge} 2A_1=d\lambda\ ,\qquad 2B_1=d\sigma\ ,
\ee for some $\RR/\ZZ$-valued $0$-forms $\lambda$ and $\sigma$. However, this condition is necessary but not sufficient to guarantee that the field configuration $(A_1,B_1,B_2)$ is invariant under $\ZZ_2^C$. Indeed, by applying \eqref{twogroupgauge} with $\lambda$ and $\sigma$ as in \eqref{Z2gauge} and $\Sigma_1=0$, we obtain
\be (-A_1,-B_1,B_2)=(A_1-2A_1,B_1-2B_1,B_2)\sim (A_1,B_1,B_2+2A_1\wedge B_1)\ ,
\ee which is in general different from $(A_1,B_1,B_2)$. One can wonder whether the additional term
\be 2A_1\wedge B_1=\frac{1}{2}d\lambda\wedge d\sigma\ ,
\ee can be canceled by a $B_2$ gauge transformation with a suitable parameter $\Sigma_1$, but this is not the case. 
Indeed, the integral of $\frac{1}{2}d\lambda\wedge d\sigma$ over a closed $2$-manifold is valued in $\frac{1}{2}\ZZ$, but not necessarily in $\ZZ$. This means that there is no way of choosing the gauge parameter $\Sigma_1$, satisfying the correct quantization conditions, and such that $d\Sigma_1=\frac{1}{2}d\lambda\wedge d\sigma$. 
For example, if we choose $\Sigma_1$ on each local patch $U_i$ to be equal to $\frac{1}{2}\lambda_id\sigma$, then in general the transitions functions $\rho_{ij}=\frac{1}{2}(\lambda_i-\lambda_j)\sigma_j$ only satisfy $\rho_{ij}+\rho_{jk}+\rho_{ki}\in \frac{1}{2}\ZZ$ on triple intersections, so that $\Sigma_1$ is not a valid gauge parameter. This conclusion is not changed by any field redefinition $B_2\to B_2-\alpha A_1\wedge B_1$, because $A_1\wedge B_1$ is invariant under gauge transformations with parameters \eqref{Z2gauge}.

We conclude that the condition that $2A_1$ and $2B_1$ are pure gauge is not sufficient to ensure that the field configuration is invariant under the $\ZZ_2^C$ action. One needs also to require that the period of $2A_1\wedge B_1$ over any closed $2$-manifold $M^{(2)}$ is an integer, rather than a half-integer, i.e. that
\be\label{restrict} \oint_{M^{(2)}}2A_1\wedge B_1\in \ZZ\ .
\ee
Notice that, because of \eqref{Z2gauge}
both $A_1$ and $B_1$ are closed $dA_1=0=dB_1$ as $\RR/\ZZ$-valued $1$-forms, so that
\be d(2A_1\wedge B_1)=0\ .
\ee 
This means that $2A_1\wedge B_1$ is a conserved current corresponding to a global $(D-3)$-form $\ZZ_2$ symmetry, whose corresponding defect is $e^{2\pi i\int_{M^{(2)}} 2A_1\wedge B_1}$, where $D=10-d$ is the dimension of the uncompactified spacetime.\footnote{We stress that integrals such as $\int_\gamma A_1$, $\int_\gamma B_1$ and $\int_{M^{(2)}} 2A_1\wedge B_1$ are only well-defined modulo integers. This reflects the fact that $A_1$ and $B_1$ are $\RR/\ZZ$-valued, and there might be an obstruction to lift them to real valued $1$-forms.} Therefore, after restricting to the gauge group $\ZZ_2^C\times \ZZ_2\times \ZZ_2$, one way to implement the further restriction \eqref{restrict} is to gauge this global symmetry. As in section \ref{s:projection}, we introduce some $(D-2)$-form $U(1)$ gauge field $\B_{D-2}$ and a $1$-form $U(1)$ gauge field $\A_1$, obeying the standard quantization conditions, and then add to the spacetime action the terms
\be 2\pi i\int_X \B_{D-2}\wedge (4A_1\wedge B_1-2 d\A_1)\ .
\ee The functional integral over $\B_{D-2}$ and $\A_1$ then forces
\be\label{eomZ2} 2A_1\wedge B_1=d\A_1\ ,\qquad d\B_{D-2}=0\ .
\ee  The fact that $\A_1$ is a $1$-form $U(1)$ gauge field implies that $d\A_1$ is a globally defined closed (but not necessarily exact) real $2$-form with integral periods. Therefore, the first equation implies \eqref{restrict}. 

The logic here is completely analogous to the example in section \ref{s:projection}: in order to implement the first step of the orbifold procedure, we need to gauge a  $(D-3)$-form $\ZZ_2$ global symmetry. In section \ref{s:projection}, the corresponding conserved current was $H_2$ (mod $2$), while in this case it is $2A_1\wedge B_1$ (mod $1$). We stress that this higher form global symmetry only exists when we ignore the various dynamical objects in string theory (strings, branes, etc.); in a complete theory, we expect these symmetries to be broken. For example, in the previous example, the $\ZZ_2$ symmetry with current $H_2$ is broken once we introduce an odd number of NS5-branes transverse to the compactified direction. In this example, we also expect the conservation of $2A_1\wedge B_1$ to be broken by some dynamical objects\footnote{We are not aware of any explicit description of such objects in string theory. It would be interesting to provide more details about them.} that create a non-trivial background for $2A_1\wedge B_1$.  Therefore, in order to obtain our orbifold theory, one also needs first of all to eliminate such objects, so that $2A_1\wedge B_1$ is conserved, and then proceed with the gauging.

As in the previous example, the effect of gauging the $(D-3)$-form symmetry is to extend the $0$-form gauge group by a $\ZZ_2$ factor, corresponding to the newly introduced $1$-form gauge field $\A_1$. We denote by $\ZZ_2^{\A}$ this group, and by $Q$ its generator.

This construction generalizes trivially to the case of a higher dimensional internal torus $T^d$, $d\ge 1$. In this case, after restricting the gauge group to \eqref{restrgroup}, the field $B_2$ transforms under charge conjugation by 
\be\label{twogrou} B_2\to B_2+\sum_{i=1}^d 2A_1^i\wedge B_1^i\ ,
\ee if we keep all $A_1^i$ and $B_1^i$ fixed. Therefore, in order to eliminate all field configurations in non-trivial representations of $\ZZ_2^C$, we have to gauge the $(D-3)$-form $\ZZ_2$ symmetry associated with the conserved current $\sum_{i=1}^d 2A_1^i\wedge B_1^i$.

Once all point-like operators that transform non-trivially under charge conjugation have been projected out, the Wilson lines for non-trivial representations of $\ZZ_2^C$ are non-endable, and the theory has a $1$-form $\ZZ_2$ symmetry acting by a minus sign on such Wilson lines. This symmetry must be gauged in order to obtain a consistent orbifold theory. After this gauging, the  non-endable Wilson lines are not physical operators anymore, as they are not gauge invariant, and the $\ZZ_2^C$ subgroup of the $0$-form gauge group is effectively quotiented out.

What is the final $0$-form gauge group in the orbifold theory? We quotiented out the factor $\ZZ_2^C$ from the original $G_{res}=\ZZ_2^C\times \prod_i (\ZZ_2^{A_1^i}\times \ZZ_2^{B_1^i})$, and we introduced a new order $2$ group $\ZZ_2^{\A_1}$. Therefore, the final gauge group is a $\ZZ_2^{\A_1}$ extension of $G_{res}/\ZZ_2^C=\prod_i (\ZZ_2^{A_1^i}\times \ZZ_2^{B_1^i})$, i.e. a group $\tilde G$ fitting in an exact sequence
\be 1\longrightarrow \ZZ_2^{\A_1}\longrightarrow \tilde G\longrightarrow \prod_i (\ZZ_2^{A_1^i}\times \ZZ_2^{B_1^i})\longrightarrow 1\ .
\ee

It turns out that, due to the non-trivial $2$-group structure of the gauge group in the original theory, the extension $\tilde G$ is not simply a product $\ZZ_2^{\A_1}\times \ZZ_2^d\times \ZZ_2^d$. To understand this point, let us go back to the group $G_{res}=\ZZ_2^C\times \prod_i (\ZZ_2^{A_1^i}\times \ZZ_2^{B_1^i})$. We want to describe a flat gauge bundle for $G_{res}$ in terms of defects, similarly to what we did in section \ref{s:2DCFT} for $2$-dimensional CFTs. More precisely, a $G_{res}$ flat connection is represented by a network of $(D-1)$-dimensional defects $\Li_g$ in the spacetime $X$, with each defect labeled by an element of $g\in G_{res}$. Roughly speaking, the network is supported on the Poincar\'e dual to the $G_{res}$-valued $1$-cocycle representing the class in $H^1(X,G_{res})$ associated with the flat $G_{res}$-connection. The network partitions the spacetime into different regions, and the defects represent the transition functions between adjacent regions. The junctions between defects must be compatible with the group product in $G_{res}$. Continuous deformations of the network of defects can be identified with appropriate gauge transformations.

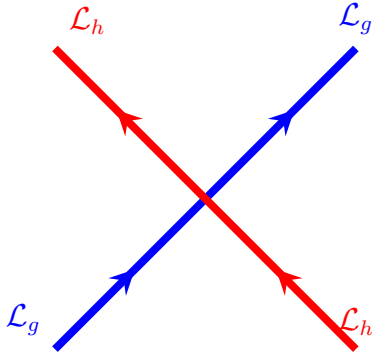
\begin{figure}
\begin{center}
\begin{tikzpicture}[line width=3 pt,
>=latex,
decoration={
markings,
mark=at position 1.5cm with {\arrow[]{stealth}},
mark=at position 4.5cm with {\arrow[]{stealth}}}
]
\draw [blue,postaction={decorate}] (-2,-2) node[above left] {$\Li_{g}$} -- (2,2) node[above] {$\Li_{g}$} ;
\draw [red,postaction={decorate}](2,-2) node[above] {$\Li_{h}$} -- (-2,2) node[above right] {$\Li_{h}$};
\end{tikzpicture}
\end{center}
\caption{\small The group $G_{res}$ is abelian. Thus, the fusion of any two defects $\Li_g$ and $\Li_h$, $g,h\in G_{res}$ is commutative $\Li_{g}\Li_{h}=\Li_{h}\Li_{g}$, which means that $\Li_{g}$ and $\Li_{h}$   can  simply `cross each other'. In this figure, the lines represent codimension $1$ defects.}\label{f:simplecross}
\end{figure}

Because the group $G_{res}$ is abelian, any two defects $\Li_{g}$  and  $\Li_{h}$ , $g,h\in G_{res}$ can simply cross each other, as in figure \ref{f:simplecross}.

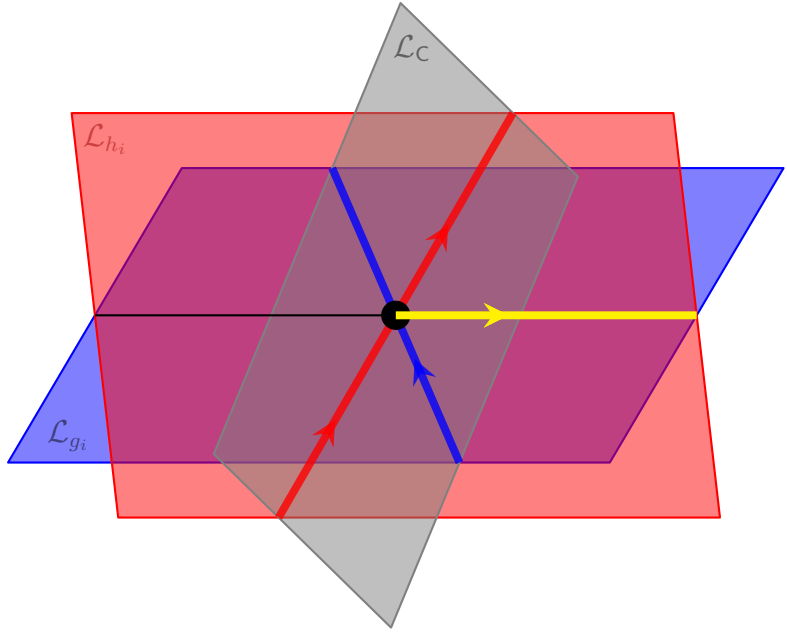
\begin{figure}
    \centering
    \begin{tikzpicture}[line width=.8 pt,
decoration={
markings,
mark=at position 1.5cm with {\arrow[]{stealth}},
mark=at position 4.5cm with {\arrow[]{stealth}}},
label/.style={postaction={decorate,decoration={markings,
mark=at position .25 with \node #1 ;}}}
]
\draw[name path=blueplane,fill=blue,draw=blue,fill opacity=.5, line join=round]
($(-4,0,3) + tan(15)*(0,-3,0)$)  -- 
($(4,0,3) + tan(15)*(0,-3,0)$) node [pos=.1, above] {$\Li_{g_i}$} --
($(4,0,-3) + tan(15)*(0,3,0)$)  --
($(-4,0,-3) + tan(15)*(0,3,0)$)  --cycle;
\draw[name path=redplane,fill=red,draw=red,text=black!40!red,fill opacity=.5, line join=round]
($(-4,-3,0) + tan(15)*(0,0,-3)$)  -- 
($(4,-3,0) + tan(15)*(0,0,-3)$) --
($(4,3,0) + tan(15)*(0,0,3)$)--
($(-4,3,0) + tan(15)*(0,0,3)$) node [below right] {$\Li_{h_i}$} --cycle;
\draw[name path=grayplane,fill=gray,draw=gray,fill opacity=.5, line join=round]
($(0,-3,3) + tan(20)*(3,0,0)$)  -- 
($(0,-3,-3) + tan(50)*(-3,0,0)$) --
($(0,3,-3) + tan(20)*(-3,0,0)$) node [pos=.9, right] {$\Li_{\mathsf{C}}$} --
($(0,3,3) + tan(50)*(3,0,0)$)  --cycle;
\draw [name intersections={of=grayplane and blueplane},blue,opacity=.8,line width=3 pt,
postaction={decorate}] 
(intersection-4) -- 
(intersection-2)
;
\draw [name intersections={of=grayplane and redplane},red,opacity=.8,line width=3 pt,
,postaction={decorate}] 
(intersection-1) -- 
(intersection-3)
;
\draw (-4,0,0) -- (0,0,0) node[circle,fill=black]{};
\draw [yellow,line width=3 pt,label={[below left]{$\Li_{Q}$}}, postaction={decorate}] (0,0,0) -- (4,0,0);

\end{tikzpicture}
    \caption{\small A triple intersection between three codimension $1$ defects, namely $\Li_{g_i}$ (blue), $\Li_{h_i}$ (red) and $\Li_\mathsf{C}$ (grey). Due to the non-trivial $2$-group structure, the B-field on the two sides of the $\Li_\mathsf{C}$ defect differs by a $2$-cochain that is Poincar\'e dual to the $(D-2)$-dimensional intersection of $\Li_{g_i}$ and $\Li_{h_i}$ (the yellow line). } 
    \label{f:defcross}
\end{figure}

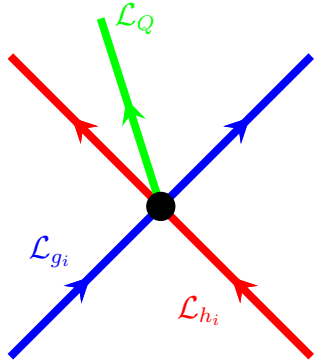
\begin{figure}
\begin{center}

\begin{tikzpicture}[line width=3 pt,
decoration={
markings,
mark=at position 1.5cm with {\arrow[]{stealth}},
mark=at position 4.5cm with {\arrow[]{stealth}}},
label/.style={postaction={decorate,decoration={markings,
mark=at position .25 with \node #1 ;}}}
]
\draw [blue,label={[above left]{$\Li_{g_i}$}},postaction={decorate}] (-2,-2) node[above] {} -- (2,2) node[above] {} ;
\draw [red,label={[below left]{$\Li_{h_i}$}}, postaction={decorate}] (2,-2) node[above] {} --  (-2,2) node[above] {};
\draw [green, postaction={decorate}](0,0) node[circle,fill=black]{} -- (-.8,2.5) node[right] {$\Li_Q$};
\end{tikzpicture}

\end{center}    
\caption{\small After gauging $2A_1\wedge B_1$, the $(D-2)$-dimensional intersection of a $\Li_{g_i}$ (blue) and a $\Li_{h_i}$ (red) defects must be the boundary of a $\Li_Q$ defect (green). The presence of this $\Li_Q$ defect can be deduced by the fact that a small Wilson line $e^{2\pi i\oint_{\gamma_1} \A_1}$ encircling the $\Li_{g_i}$ and $\Li_{h_i}$ intersection is non trivial. This means that, after gauging, the fusion between the $\Li_{g_i}$, $\Li_{h_i}$, $\Li_Q$ defects is not commutative, i.e. the corresponding group is non-abelian. In this picture, the lines represent codimension $1$ defects.}\label{f:nonabelian}
\end{figure}

Let us denote by $g_i$ and $h_i$ the generators of $\ZZ_2^{A_1^i}$ and $\ZZ_2^{B_1^i}$, respectively. Consider three defects $\Li_{g_i}$, $\Li_{h_i}$ and $\Li_\mathsf{C}$ in generic positions, such that each pair of defects cross in a $(D-2)$-dimensional intersection, and with a non-trivial $(D-3)$-dimensional triple intersection (see figure \ref{f:defcross}).
The fact that $B_2$ shifts by $2A_1^i\wedge B_1^i$ (see eq.\eqref{twogrou}) under the action of $\mathsf{C}$ means the $B_2$-field on the two sides of the $\Li_\mathsf{C}$-defect `jumps' by an half-integer flux localized at the intersection between the $\Li_{g_i}$ and $\Li_{h_i}$ defects.   
In other words, if $S^2$ is a small $2$-sphere encircling the $(D-3)$-dimensional triple intersection between a $\Li_{g_i}$, a $\Li_{h_i}$ and a $\Li_\mathsf{C}$ defect, then the $B_2$ flux across $S^2$ is $e^{2\pi i\int_{S^2} B_2}=-1$ (at least in the limit where the radius of $S^2$ is infinitesimal).
	
When we gauge the symmetry with current $2A_1\wedge B_1$, we get a new $\ZZ_2$ gauge field $\A_1$; we denoted by $Q$ the corresponding gauge group generator. An integral $\oint_{\gamma_1}\A_1$ along a closed curve $\gamma_1$ just counts the number (mod $2$) of intersections of the curve $\gamma_1$ with $\Li_Q$ defects. The equation of motion \eqref{eomZ2} implies that if we take a very small closed curve $\gamma_1$ that encircles the $(D-2)$-dimensional intersection of a $g_i$ and $h_i$ defect, then $e^{2\pi i\oint_{\gamma_1}\A_1}=-1$.   This means that, after gauging $2A_1\wedge B_1$, there must be a $\Li_Q$ defect starting from any intersection of a $\Li_{g_i}$ and a $\Li_{h_i}$ defect (see figure \ref{f:nonabelian}).  In other words, if a $\Li_{g_i}$ defect crosses a $\Li_{h_i}$ defect, then on the other side it must emerge as a $\Li_{g_iQ}$ defect. This implies that the $0$-form group has become non-abelian, with relations
	\be g_ih_i=h_ig_iQ\ .
	\ee
Because there are no other couplings between $\A_1$ and the $A_1^i$ and $B_1^i$ gauge fields, we conclude that the other group relations are simply
\be Qg_i=g_iQ, \qquad Qh_i=h_iQ,\qquad  g_i^2=h_i^2=Q^2=1\ .
\ee When $d=1$, the group $\tilde G$ with generators $g_1,h_1,Q$ satisfying these relations is known as the dihedral group $D_8$ of order $8$. For general $d$, the resulting group is known as the $2^{1+2d}$ extra-special group (see for example chapter 5 of \cite{Atlas}). 

This result fits nicely with the worldsheet orbifold construction and the geometric intuition. It is well known that the $\ZZ_2$ orbifold of a free boson on a circle $S^1$ has symmetry group $D_8$ (see for example \cite{Chang:2020imq,Thorngren:2021yso} for a derivation in terms of topological defects). The same result generalizes straightforwardly to the case of CFT with target $T^d/\ZZ_2$, whose symmetry group is an extra-special group $2^{1+2d}$. In this worldsheet orbifold, the $\prod_i(\ZZ_2^{A_1^i}\times \ZZ_2^{B_1^i})$ inherited from the parent CFT, gets extended by the $\ZZ_2$ group generated by the worldsheet quantum symmetry. Comparison with our spacetime construction leads us to identify the worldsheet quantum symmetry with the gauge symmetry $Q$ associated with the gauge field $\A_1$. Our spacetime derivation of the group structure tells us immediately that there must be $2^d$ worldsheet twisted ground states, since this is the dimension of the smallest irreducible representation of the extra-special group $2^{1+2d}$ where the central element $Q$ acts non-trivially. This matches with the number of fixed points of a geometric $T^d/\ZZ_2$ orbifold.

Again, this structure is very similar to what we found in the example of section \ref{s:projection}. In that case, the $\ZZ_2$-gauge field $\B_1$ was coupled to the current $H_2$ (modulo $2$). As a consequence of introducing the $\B_1$ field and coupling to $H_2$, the $U(1)^{B_1}$ gauge group, whose electric charge is the winding number along the internal $S^1$, gets extended by $\ZZ_2$ in a non-trivial way -- it is isomorphic to $U(1)$, and not to the direct product $\ZZ_2\times U(1)$. This extended gauge group admits half-integral winding charges (in the normalization of the original theory) to which the gauge field $\B_1$ was coupled. But only the twisted fundamental strings carry non-integral winding number, so that the $\ZZ_2$ gauge symmetry extending $U(1)^{B_1}$ was, once again, the worldsheet quantum symmetry. 

From the point of view of the worldsheet orbifold, the fact that the extension of $\prod_i(\ZZ_2^{A_1^i}\times \ZZ_2^{B_1^i})$ by the quantum symmetry is non-trivial is closely related to the presence of a mixed 't Hooft anomaly, in the original worldsheet theory, between $\prod_i(\ZZ_2^{A_1^i}\times \ZZ_2^{B_1^i})$ and the $\ZZ_2^C$ symmetry we want to gauge  \cite{Bhardwaj:2017xup,Tachikawa:2017gyf}. In a sense, the mechanism described above is the spacetime version of this phenomenon.

\bigskip

In this section, we focused on the fundamental string and on the gauge fields that are coupled to it, such as $A_1^i$, $B_1^i$, $B_2$, in order to show the relationship between the spacetime and the worldsheet orbifold procedures. However,  the discussion would be completely analogous if we focused on D1-branes instead.  One just needs to replace the fields $B_1^i$ and $B_2$ by the fields $C_1^i$ and $C_2$ that one obtains from the RR $2$-form of $10$-dimensional type IIB theory by dimensional reduction on $T^d$. For each $i=1,\ldots,d$, the D-string worldsheet currents that couple to the $A_1^i$ and $C_1^i$ gauge fields are related to each other by two dimensional Hodge duality. This means that the two global symmetries on the D-string worldsheet related to $A_1^i$ and $C_1^i$ potentially have a mixed 't Hooft anomaly. As explained in more detail at point $1$ in next subsection, this potential anomaly is canceled by imposing the correct gauge transformations for the field $C_2$. This argument inevitably leads to the same gauge transformations as in \eqref{twogroupgauge}, after replacing $B_1$ by $C_1$ and $B_2$ by $C_2$. From this point on, all our derivation, which was valid for the fundamental string, can be repeated without essential differences for the D-string and the gauge fields coupled to it.

In order to get the complete information about the orbifold theory, one should extend our arguments so as to include all gauge fields in the original theory, as well as their couplings to the various dynamical objects, and take into account the complicated higher group structure of their gauge transformations.

\section{Discussion and conclusions}\label{s:conclusions}

In this last section, we comment on various aspects of the orbifold procedure, that emerged in the examples of sections \ref{s:halfperiod} and \ref{s:reflection}. We also point out several directions of future investigation.
\begin{enumerate}
	\item  As pointed out in section \ref{s:2DCFT}, for a 2D CFT there might be an obstruction in gauging a certain subgroup $\Gamma$ of a finite group of symmetries $G$, which is measured by the restriction to $\Gamma$ of a class\footnote{Strictly speaking, this is true only for bosonic CFTs. Anomalies in superconformal theories possibly have further `layers' \cite{GuWen14,Johnson-Freyd:2017ble}. Here, we are assuming that such layers vanish, which is true in many interesting examples, included the ones considered in this article.} $[\alpha]\in H^3(G,U(1))$. 
	From a string theoretical viewpoint, one generally expects the global symmetries on the worldsheet, giving rise to exact symmetries of the string amplitudes, to correspond to gauge symmetries in spacetime, and therefore to be coupled to suitable dynamical gauge fields. 
	As a consequence, it must be possible to generalize the worldsheet theory so as to describe a string moving in some non-trivial background for these gauge fields. 
	This seems at odds with the possible presence of a 't Hooft anomaly $[\alpha]$. Indeed, a 't Hooft anomaly can be interpreted as an obstruction to coupling the worldsheet theory to background gauge fields in a gauge invariant way. In quantum field theory a 't Hooft anomaly is not an inconsistency of the theory. In string theory, the situation is very different: from the spacetime point of view, these gauge fields are dynamical, and the coupling to the string worldsheet must be gauge invariant in order to be consistent.\\  
	What saves the day is that the anomaly can be canceled by a non-trivial transformation of the $B_2$ field, through a version of the Green-Schwarz mechanism. In particular, let us view the $G$-bundle on spacetime $X$ as the pull-back $\phi^*$ of the tautological bundle on the universal classifying space $BG$ of $G$ through some map $\phi:X\to BG$. The class $[\alpha]$ can be described as a class in the cohomology $H^3(BG,U(1))$ of the universal classifying space $BG$ of $G$, so that the pullback $\phi^*\alpha$ defines a $U(1)$-valued $3$-cocycle in spacetime. Then, in order to cancel the anomaly, it is sufficient to assume that the gauge invariant field strength for $B_2$ is 
	\be\label{modifiedH3} H_3=dB_2+\phi^*\alpha\ .\ee Indeed, let us assume that the worldsheet is a closed manifold $\Sigma$, and let $M$ be a $3$-manifold such that $\partial M=\Sigma$. The worldsheet action contains the term
	\be \int_{\Sigma} B_2=\int_M dB_2=\int _M H_3-\int_M \phi^*\alpha\ .
	\ee In the last expression, $\int _M H_3$ is gauge invariant, while the gauge variation of $\int_M \phi^*\alpha$ exactly cancels the 't Hooft anomaly, similarly to the anomaly inflow mechanism. The expression \eqref{modifiedH3} implies that the spacetime action contains non-trivial Chern-Simons terms, arising from the kinetic term from the $B_2$ field, and ultimately that the spacetime gauge group is not plainly the product of a $0$-form symmetry $G$ with gauge field $A$ and a $U(1)$ $1$-form symmetry with gauge field $B_2$, but that these symmetries are mixed in a non-trivial $2$-group structure. Eq.\eqref{modifiedH3} is the origin of the non-standard Bianchi identity $dH+F_2\wedge H_2=0$. Formally, the term $F_2\wedge H_2$ is analogous to the term $\Tr(F\wedge F)$ appearing in the Bianchi identity of the field strength $H_3$ in heterotic strings; in that case as well, the modification can be related to the cancellation of a sigma model anomaly \cite{Rohm:1985jv,Moore:1984dc,Lerche:1987sg,Lerche:1987qk}.\\
 This argument shows that the Postnikov class defining the $2$-group structure, as described in \cite{Benini:2018reh}, is exactly $[\alpha]\in H^3(BG,U(1))$. Thus, the obstruction to gauging a certain worldsheet symmetry group $\Gamma\subseteq G$ translates, in spacetime, with the fact that $\Gamma$ is not really present as a group of `pure' $0$-form gauge symmetries, but only within a non-trivial higher group structure.\\ For example, in a string theory compactified on a circle, considered in section \ref{s:halfperiod}, there is a well-known mixed 't Hooft anomaly in the worldsheet theory between the translations along $S^1$, associated with the gauge field $A_1$, and the translations along the T-dual circle, associated with $B_1$.\footnote{This example does not exactly fit into the general case described above, because the group $G=U(1)^{A_1}\times U(1)^{B_1}$ is continuous rather than finite. In this case, the anomaly is a class in $H^4(G,\ZZ)$,  and $\alpha=A_1\wedge H_2$ is a locally defined $3$-form such that the closed $4$-form $d\alpha=F_2\wedge H_2$  represents the anomaly. When the group $G$ is finite, there is an isomorphism $H^3(G,U(1))\cong H^4(G,\ZZ)$, so that the anomaly can be described directly as the $3$-cohomology class $[\alpha]$.  } This means that, in the absence of the coupling with $B_2$, the worldsheet theory cannot be coupled in a gauge invariant way to both $A_1$ and $B_1$. This anomaly is canceled by the coupling to $B_2$, once the Nicolai-Townsend  transformations \eqref{eq:NicolaiGaugeTransform} are imposed.\\ Similarly, in the example of section \ref{s:reflection}, the worldsheet CFT has a mixed 't Hooft anomaly  for the group $G=\ZZ_2^C\times \ZZ_2^d\times\ZZ_2^d$ containing a term $\sum_{i=1}^d C_1\wedge A_1^i\wedge B_1^i$. A derivation of this fact in the case of a single free boson on $S^1$ is given in appendix A.2 of \cite{Thorngren:2021yso}, and the $d>1$ case is a straightforward generalization. The worldsheet fermions do not contribute to this anomaly because the $\ZZ_2^d\times \ZZ_2^d$ group acts trivially on them. Such a class can be represented by a cocycle $\alpha:G\times G\times G\to U(1)$ such that $\alpha(\mathsf{C},g^i,h^i)=-1$ for all $i=1,\ldots,d$. This leads to the non-trivial transformation \eqref{twogrou} of the $B_2$ field under $\mathsf{C}$ in the presence of a non-trivial background for $A_1^i$ and $B_1^i$.
	
	\item 
	Let $\A_1$ be the $\ZZ_n$ gauge field corresponding to a group of $0$-form symmetries $\Gamma$, which arises from global symmetries on the worldsheet. If $\ZZ_n$ is the subgroup of some $U(1)$ gauge symmetry, then we take $\A_1$ to be the $U(1)$ gauge field. As discussed in the previous point, there is potentially a $2$-group structure, dictated by the 't Hooft anomalies in the string worldsheet, that leads to a number of Chern-Simons terms in the low energy effective action in spacetime, arising from the kinetic term $H_3\wedge *H_3$ for the $B_2$-field. In the two examples we studied, these Chern-Simons terms have the form
	\be \int_X \A_1\wedge R_2\wedge *H_3\ ,\ee
	where $R_2$ is a suitable $2$-form field. In particular, in the examples of section \ref{s:halfperiod} and \ref{s:reflection}, $R_2$ was equal to $H_2=dB_1$ and  to $2A_1\wedge B_1$, respectively. When such terms are present, then a non-trivial flux $\int_{M^{D-1}}R_2\wedge *H_3$ over $(D-1)$-dimensional manifold can provide a non-zero charge for the gauge field $\A_1$. 

	Now, the first step in our orbifold procedure amounts to eliminating all charged operators for the gauge field $\A_1$. This means that we have to modify our theory in such a way that the fluxes of $R_2\wedge *H_3$ are $0$ mod $n$, i.e. that they carry trivial charge under the gauge group $\Gamma\cong \ZZ_n$. In the first example of section \ref{s:halfperiod}, where $R_2=H_2=dB_1$ and $n=2$, this modification was obtained by first restricting the NS5 charges, which represent the magnetic charges for $B_1$, to even ones, so that $H_2$ is a conserved current mod $2$. Next, one can gauge the $(D-3)$-form $\ZZ_2$ global symmetry \eqref{H2symm}, corresponding to this current.
	The effect of gauging \eqref{H2symm} is to change the quantization of the magnetic fluxes carried by $H_2$, so that they are always even; equivalently, the $0$-form $U(1)$ gauge group associated with $B_1$ is extended by a $\ZZ_2$ group.
	A similar mechanism applies to the example of section \ref{s:reflection}. More generally, we expect a similar procedure to be necessary every time a Chern-Simons term involving the $\Gamma$-gauge field is present.
	\item The discussion in the point $1$ was focused on the worldsheet of the fundamental string. However, a similar argument should hold for the worldvolume of each $d$-dimensional dynamical object. Any such object is coupled to a number of $p$-form gauge fields. In particular, it usually carries an electric charge with respect to a $d$-form $U(1)$ gauge field $B_d$, analogous to the $B$-field $B_2$ for the fundamental string. The worldvolume fields, in general, transform in some (possibly) non-trivial representation of the $0$-form gauge group $G$, and therefore the worldvolume theory must be coupled to the corresponding spacetime $G$ gauge fields. In principle there might be a 't Hooft anomaly for the $G$ group action in the $d$-dimensional worldvolume theory, and that would make the coupling with the dynamical $G$-gauge fields inconsistent.\\ In the example of the worldsheet theory for the fundamental string, the anomaly cancels when one considers the coupling with the $B_2$ field and its non-trivial transformation. By analogy, we expect a similar cancellation to occur when one considers the coupling of the $d$-dimensional object to all $p$-form gauge fields in the theory, provided that such $p$-form fields transform non-trivially under the $0$-form $G$ gauge parameters \cite{Hsieh:2020jpj}.\\ This argument suggests that an intricate higher group structure, mixing the $0$-form gauge group $G$ with higher form gauge symmetries, is necessary in order for the string theory to be consistent. The precise mixing should be dictated by the requirement of cancellation of all potential anomalies in the worldvolume field theories of all dynamical objects. For example, in the two examples of section \ref{s:reflection} and \ref{s:halfperiod}, by S-duality, we expect the worldsheet theory of the  D-string to have anomalies that are completely analogous to the ones observed for the fundamental string, and that are canceled by a non-trivial transformation of the RR $2$-form field $C_2$ that is coupled to the D-string. This means that the $B_2$ and $C_2$ fields appear in the same Chern-Simons terms in the effective action, as expected by S-duality.\\ The discussion of point $2$ should be also generalized accordingly. Suppose the action contains Chern-Simons terms of the form
	\be \int_X \A_1\wedge R\ ,
	\ee where $\A_1$ is the gauge field of the group $\Gamma\cong \ZZ_n$  we want to orbifold by, and $R$ is a $(D-1)$-form built in terms of the various $p$-form fields. Then in order to implement the orbifold by $\Gamma\cong \ZZ_n$, one needs to modify the theory in such a way that the fluxes of $R$ through any $(D-1)$-dimensional manifold are quantized in multiples of $n$, so that they do not carry any charge under the $\ZZ_n$ symmetry. This modification will in general modify the gauge symmetry of the model, by either quotienting/extending various factors in the original group by $\ZZ_n$.
	\item In general, the gauging of a $p$-form (abelian) global symmetry $\Gamma$ in a $D$-dimensional quantum field theory leads to a $(D-p-2)$-form global symmetry in the orbifold theory for the Pontryagin dual group $\hat \Gamma={\rm Hom}(\Gamma,U(1))$ \cite{Gaiotto:2014kfa}. The worldsheet orbifold procedure is a gauging of a $0$-form symmetry in a two dimensional CFT, so that the quantum symmetry is again a global $0$-form symmetry on the worldsheet. On the other hand, our procedure involves, as an intermediate step, the gauging of a $1$-form symmetry in a $D$-dimensional spacetime, with respect to which the quantum symmetry is a $(D-3)$-form global symmetry in spacetime. Therefore, the worldsheet and spacetime quantum symmetries are not related to each other. In fact, their fate in the final, consistent, orbifold string theory must be different. The global \emph{worldsheet} quantum symmetry becomes a new $0$-form gauge symmetry in spacetime. On the other hand, the $(D-3)$-form global \emph{spacetime} quantum symmetry is broken once we introduce the dynamical objects where the newly introduced codimension $3$ 't Hooft operators can end.
	\item Let us focus on the worldsheet quantum symmetry $Q$. This is a global symmetry on the worldsheet, so that, under the assumption that it extends to a symmetry of the whole string theory, it must be coupled to a gauge field in spacetime. In the examples we have studied, the additional gauge field arises automatically when one considers the procedure described at point $2$. Let us consider the example of section \ref{s:halfperiod}. As described in section \ref{s:projection}, modifying the quantization of the $H_2$ fluxes is equivalent to considering an extension of the $U(1)$ gauge group associated with $B_1$ by a $\ZZ_2$ group. The objects charged under this new $\ZZ_2$ gauge field are the ones carrying half-integral winding along the internal $S^1$ circle, i.e. exactly the twisted sector of the fundamental string. This is sufficient to identify the new $\ZZ_2$ symmetry with the worldsheet quantum symmetry. The same mechanism occurs for the example of section \ref{s:reflection}.\\ It would be interesting to understand how general this phenomenon is. Since the Chern-Simons terms are related to the potential anomalies of the worldsheet global symmetries, the procedure described at point $2$ might not be necessary if the 't Hooft anomaly for the full group of global worldsheet symmetries vanished.  
	Similar considerations should apply for the quantum symmetries that arise in the worldvolume theories describing the other dynamical objects in string theory.
	\item As discussed in section \ref{s:2DCFT}, the group of global symmetries of a 2D CFT can be generalized to include the fusion category of (possibly non-invertible) topological defects. This generalization is very natural when one considers orbifolds, for several reasons. First of all, the orbifold construction can be defined with respect to objects in the fusion category that do not come from the usual symmetries (generalized orbifolds) \cite{Frohlich:2006ch,Frohlich:2009gb,Carqueville:2012dk,Brunner:2013ota,Brunner:2013xna}. Furthermore, even if one started from a `parent' theory where all defects are invertible, and therefore form a group under fusion, after orbifolding by some subgroup of symmetries the defects of the original theory might induce some non-invertible defects on the orbifold \cite{Frohlich:2006ch,Frohlich:2009gb,Carqueville:2012dk,Brunner:2013ota,Brunner:2013xna,Bhardwaj:2017xup,Tachikawa:2017gyf}.\\ An example related to the model in section \ref{s:reflection} is the following. Consider the 2D CFT given by a sigma model with target a torus $T^d$, that has a $\ZZ^C_2\ltimes (U(1)^d\times U(1)^d)$ group of symmetries. As mentioned in section \ref{s:reflection}, after orbifolding by the $\ZZ^C_2$ symmetry that reverses all directions of the torus, only a $\ZZ_2^d\times \ZZ_2^d$ subgroup of the `parent' symmetry group survives as a proper global symmetry group in the `child' theory. However, the topological defects implementing the $U(1)^d\times U(1)^d$ symmetries do not disappear, but combine to form a continuum of non-invertible topological defects. Therefore, the orbifold CFT has a complicated fusion category of defects, where the `group-like' global symmetry is only a small subcategory \cite{Fuchs:2007tx,Chang:2020imq,Thorngren:2021yso}. Much like the $\ZZ_2^d\times \ZZ_2^d$ group of global symmetries, these non-invertible defects are `transparent' to the BRST charge whose cohomology describes the physical string states, and their existence puts non-trivial selection rules on the (perturbative) string amplitudes. So, it would seem reasonable to consider them on the same footing as the $\ZZ_2^d\times \ZZ_2^d$ group even at the level of the string theory.  \\
	There is one further reason for considering the category of topological defects, rather than groups of symmetries. The worldsheet orbifold procedure that produces a CFT $\C'$ starting from a theory $\C$, can always be reversed, i.e. one can re-obtain the `parent' theory $\C$ starting from the `child' $\C'$. But this reverse procedure is, in general, an orbifold by a non-invertible subcategory of defects, rather than by a group of symmetries \cite{Frohlich:2006ch,Brunner:2013ota,Bhardwaj:2017xup}.\\
	In the context of string theory, it is natural to ask how non-invertible topological defects on the worldsheet manifest themselves in spacetime. Because groups of worldsheet symmetries generalize so naturally to the category of defects, it is tempting to speculate that the group of $0$-form gauge symmetries in spacetime should get somehow extended to some `gauge fusion category' in spacetime. To the best of our knowledge, such a concept has not been developed so far. Providing a reasonable definition of such a structure would be extremely interesting, and might be a necessary step in order to describe an inverse of the spacetime orbifold procedure.
	\item One can consider the orbifold procedure in a spacetime that is asymptotically AdS, rather than Minkowski.  In this situation, holography provides an `intrinsic' description of the quantum gravity theory in terms of the dual QFT on the boundary. In general, one expects the gauge symmetries in the bulk to be dual to global symmetries on the boundary.\\ Therefore, there is a very natural candidate for the orbifold of the string model by some group of symmetries: one simply considers the orbifold of the boundary QFT by the corresponding global symmetry, and defines the string orbifold in the bulk as the dual of such a boundary orbifold QFT. Apparently, this approach completely solves our problem, since it provides a precise definition of the string orbifold model that makes sense even in regimes where there is no perturbative or semiclassical description of the bulk gravity theory. On the other hand, whenever an effective description of the bulk gravity theory is available, we expect to be able to provide an equivalent description of the boundary orbifold purely from the bulk point of view. Unfortunately, the task of translating the boundary orbifold procedure in terms of the bulk gravity theory is not necessarily obvious -- in a sense, we have just moved from one difficult problem to another.
    Nevertheless, the holographic viewpoint supports our proposal that a description independent spacetime orbifold procedure should exist.
	\item It is natural to ask how our proposal might generalize to orbifolds by a non-abelian group $\Gamma$. One can argue that such a generalization might be quite non-trivial, and require a slightly different approach to the spacetime orbifold procedure.\\
	Indeed, the proposal in this article involves `restoring' and then gauging a $1$-form global symmetry, namely the electric $1$-form symmetry associated to the $0$-form gauge symmetry $\Gamma$. On the other hand, $1$-form symmetries can only be abelian. In particular, a potential electric $1$-form global symmetry can only exist for the centre of the gauge group; this is the reason why, in section \ref{s:reflection}, our first step in the orbifold was to restrict the gauge group to the centralizer of $\ZZ_2^C$. When $\Gamma$ is non-abelian, restricting to the centralizing subgroup does not make sense, since $\Gamma$ does not centralizes itself. The correct approach, in this case, seems to be a restriction to the normalizer, rather than centralizer, i.e. the largest subgroup of the gauge group containing $\Gamma$ as a normal subgroup. After this restriction, one can project out all dynamical objects transforming in representations of the gauge group where the normal subgroup $\Gamma$ acts non-trivially, thus making the corresponding Wilson lines non-endable. In the non-abelian case, the presence of non-endable Wilson does not automatically lead to a group of $1$-form global symmetries. However, it does imply that the theory contains some (in general, non-invertible) topological operators, which are a generalization of the standard group-like global symmetries \cite{Rudelius:2020orz,Heidenreich:2021xpr,McNamara:2021cuo}. The analogue of our procedure, in this case, would be to `gauge' such topological defects, so as to obtain a theory with no (invertible or non-invertible) global symmetries. Thus, as discussed at point $6$ above, in order to include the non-abelian case in our procedure, a better understanding of the gauging of non-invertible symmetries in general spacetime dimensions seems to be necessary.
	\item All examples we discussed in this article involved the orbifold by some finite group of $0$-form symmetries with trivial action on the higher form symmetries. More generally, as explained in section \ref{HGS}, one could consider $0$-form symmetries acting by non-trivial automorphisms $\rho$ on the higher form gauge groups. For example, a $0$-form symmetry might act by $B_2\to -B_2$ on the Kalb-Ramond B-field, or mix it non-trivially with a R-R $2$-form gauge field. 
	This generalization would be very important because, as far as we know, all examples where `duality does not commute with orbifolds' belong to this class. We hope to address this more general case in future work.
	\item In our examples, we haven't discussed the possible obstructions to the orbifold procedure, i.e. the conditions under which the procedure we describe can possibly yield a consistent string model. From the worldsheet orbifold perspective, the main consistency condition for an orbifold by $\Gamma$ to exist, is that the cohomology class representing the 't Hooft anomaly vanishes when restricted to $\Gamma$. As discussed in point $1$ above, the potential 't Hooft anomalies for the worldsheet global symmetry are encoded in the particular higher group transformation of the B-field $B_2$. From the perspective of the spacetime orbifold procedure, the higher group structure of the full group of spacetime gauge symmetries is part of the initial data. Therefore, trying to take a worldsheet orbifold by an anomalous symmetry translates, from the spacetime perspective, to orbifolding by a transformation that is not a symmetry, because it ignores the higher group structure.\\ On the other hand, there are potentially further obstructions to orbifolding that might be difficult to interpret from the point of view of the worldsheet QFT. In particular, it is known that, in some cases, the worldsheet orbifold procedure leads to a string model with a tadpole, that needs to be canceled by inserting an appropriate number of spacetime filling branes \cite{Sethi:1996es,Sen:1996na}. \\
	In QFT,  a systematic classification of all possible obstruction to orbifolds is available. For strings orbifolds, it might be that cancellation of all potential tadpoles is the only condition required for the orbifold string model to be consistent. On the other hand, while the consistency conditions have been studied in a large number of examples, a classification of orbifold obstructions as systematic and rigorous as in QFT does not exist, as far as we know. We hope that our approach might help in shedding light on this issue in the future.
	\item An important open problem, closely related to the previous point, is a systematic construction of all consistent `twisted sectors'  for any given parent string model $A$ and orbifold group $\Gamma$. Let us discuss this point by analogy with the case of orbifolds in 2D CFT, which is quite well understood. Given a consistent CFT $A$, with a finite group of global symmetries $\Gamma$, one can construct the orbifold CFT (or orbifold CFTs) by first projecting on a $\Gamma$-invariant subtheory $A^\Gamma$, which is common to both the parent and child theory (or children theories). The subtheory $A^\Gamma$ is well defined on the sphere, in the sense that the OPE of any two operators in $A^\Gamma$ closes within $A^\Gamma$, but not on the torus or higher genus Riemann surfaces -- for example, the partition function is not modular invariant. The orbifold theory is a way to complete $A^\Gamma$ to a fully consistent CFT. This process can be made very rigorous and precise. For example, if $A$ is a holomorphic CFT, then it can be viewed as a vertex operator algebra  (VOA), and $A^\Gamma$ as a subVOA of $A$. In general, the category of modules of a rational VOA forms a modular tensor category (MTC) \cite{Fuchs:2002cm,Huang2005}. The latter can also be described in terms of a 3D topological quantum field theory \cite{Reshetikhin:1991,Turaev:1992}. In order for a VOA $A$ to define a fully consistent CFT by itself, its MTC must contain only one simple object (the algebra itself). In this case, the MTC of the invariant subVOA $A^\Gamma$ can be obtained as the double of the fusion category of topological defects corresponding to the group $\Gamma$ \cite{Bhardwaj:2017xup} -- each simple object is labeled by a pair consisting of a conjugacy class in $\Gamma$ and an irreducible $\Gamma$-representation. The modular tensor category is the only information we need to know to classify and construct all possible completions of $A^\Gamma$ to a fully consistent CFT, as well as the possible obstructions ('t Hooft anomalies) -- no other information about $A^\Gamma$ is needed. One of the main properties of a MTC is that it has a finite number of simple objects. For example, if $\Gamma\cong \ZZ_N$ the MTC of $A^\Gamma$ has $N^2$ simple objects, and any completion of $A^\Gamma$ to a consistent CFT involves a suitable collection of $N$ of them.   \\
	Our formulation of the spacetime orbifold of a $D$-dimensional quantum gravity theory follows a similar pattern. Namely, we first projected our `parent model' $A$ to a subtheory $A^\Gamma$ and then tried to complete it to a consistent orbifold theory by adding a suitable twisted sector. One natural open question is about the nature of the `subtheory' $A^\Gamma$. It cannot be a consistent theory of quantum gravity, but should be sufficiently well defined to be able to speak about its global symmetries. \\
	In a sense, we expect the common subtheory $A^\Gamma$ to contain  most of the information about both the parent theory and its orbifold -- roughly speaking, it can be obtained by restricting the parent theory to trivial representations of a finite group, and such a group admits only finitely many irreducible representations. The point of view that we tried to pursue in this article is that the `parent' and the `child' string models are different consistent ways to gauge or break all global symmetries of the common subtheory $A^\Gamma$. By analogy with 2D CFTs, one might expect that only a finite amount of additional information (the analogue of a MTC) is needed to  `complete' $A^\Gamma$ to a fully consistent theory.\\
	What is the analog of MTC for the subtheory $A^\Gamma$? Given our current understanding of the problem, an answer to this question is out of reach. The best we can do at this point is to propose some vague speculation, based on analogy with quantum field theory.\\
	In a $D$-dimensional QFT, both the category of global symmetries and the 't Hooft anomalies are expected to be described by suitable $(D+1)$-dimensional topological quantum field theory \cite{Freed:2022qnc}. In $D=2$ CFT, the MTC can be described as a topological field theory in  $D+1=3$ dimensions. One can speculate that for a $D$-dimensional quantum gravity theory,  the same role might be played again by a topological theory, either in $(D+1)$ or in $D$ dimensions -- the latter hypothesis might fit better with holography, in the cases where the $D$-dimensional quantum gravity theory is dual to a $(D-1)$-dimensional QFT. Topological field theory descriptions of the symmetries arising from string and M-theory compactifications have been recently proposed, see for example \cite{DelZotto:2015isa,Apruzzi:2021nmk,Apruzzi:2022dlm,DelZotto:2022joo,Cvetic:2022imb,Apruzzi:2022rei,GarciaEtxebarria:2022vzq,Heckman:2022muc,Morrison:2020ool,Bhardwaj:2020phs,Albertini:2020mdx}.
	\\
 One first step in trying to make these ideas more precise would be to provide a reasonable definition of the subtheory $A^\Gamma$. Next, one should try to describe the (topological?) theory playing the role of the MTC for holomorphic 2D CFTs. Any further progress in this direction would be a major advance in our understanding of symmetries and orbifolds in string theory.
\end{enumerate}

\medskip

\noindent {\bf Acknowledgements.} We thank C. Blair, J. Heckman, H. Parra de Freitas, and E. Sharpe for useful comments. S.G. and R.V. are supported by a grant from BIRD-2021 (PRD-2021). R.V. would like to thank the Isaac Newton Institute for Mathematical Sciences, Cambridge, for support and hospitality during the programme ``New connections in Number Theory and Physics'' where work on this paper was undertaken, and to thank the participants to the programme for stimulating discussions.


\begin{thebibliography}{99}

%
\bibitem{Kapustin:2014gua}
A.~Kapustin and N.~Seiberg,
``Coupling a QFT to a TQFT and Duality,''
JHEP \textbf{04}, 001 (2014)
doi:10.1007/JHEP04(2014)001
[arXiv:1401.0740 [hep-th]].

\bibitem{Gaiotto:2014kfa}
	D.~Gaiotto, A.~Kapustin, N.~Seiberg and B.~Willett,
	``Generalized Global Symmetries,''
	JHEP \textbf{02}, 172 (2015)
	doi:10.1007/JHEP02(2015)172
	[arXiv:1412.5148 [hep-th]].
	


\bibitem{Cordova:2018cvg}
C.~C\'ordova, T.~T.~Dumitrescu and K.~Intriligator,
``Exploring 2-Group Global Symmetries,''
JHEP \textbf{02} (2019), 184
doi:10.1007/JHEP02(2019)184
[arXiv:1802.04790 [hep-th]].


\bibitem{Cordova:2022ruw}
C.~Cordova, T.~T.~Dumitrescu, K.~Intriligator and S.~H.~Shao,
``Snowmass White Paper: Generalized Symmetries in Quantum Field Theory and Beyond,''
[arXiv:2205.09545 [hep-th]].

\bibitem{Gaiotto:2020iye}
D.~Gaiotto and J.~Kulp,
``Orbifold groupoids,''
JHEP \textbf{02}, 132 (2021)
doi:10.1007/JHEP02(2021)132
[arXiv:2008.05960 [hep-th]].

\bibitem{Freed:2022qnc}
D.~S.~Freed, G.~W.~Moore and C.~Teleman,
``Topological symmetry in quantum field theory,''
[arXiv:2209.07471 [hep-th]].



\bibitem{Polchinski:2003bq}
J.~Polchinski,
``Monopoles, duality, and string theory,''
Int. J. Mod. Phys. A \textbf{19S1} (2004), 145-156
doi:10.1142/S0217751X0401866X
[arXiv:hep-th/0304042 [hep-th]].


\bibitem{Banks:2010zn}
T.~Banks and N.~Seiberg,
``Symmetries and Strings in Field Theory and Gravity,''
Phys. Rev. D \textbf{83}, 084019 (2011)
doi:10.1103/PhysRevD.83.084019
[arXiv:1011.5120 [hep-th]].

\bibitem{Heidenreich:2020pkc}
B.~Heidenreich, J.~McNamara, M.~Montero, M.~Reece, T.~Rudelius and I.~Valenzuela,
``Chern-Weil global symmetries and how quantum gravity avoids them,''
JHEP \textbf{11}, 053 (2021)
doi:10.1007/JHEP11(2021)053
[arXiv:2012.00009 [hep-th]].


\bibitem{Heidenreich:2021xpr}
B.~Heidenreich, J.~McNamara, M.~Montero, M.~Reece, T.~Rudelius and I.~Valenzuela,
``Non-invertible global symmetries and completeness of the spectrum,''
JHEP \textbf{09}, 203 (2021)
doi:10.1007/JHEP09(2021)203
[arXiv:2104.07036 [hep-th]].

\bibitem{McNamara:2021cuo}
J.~McNamara,
``Gravitational Solitons and Completeness,''
[arXiv:2108.02228 [hep-th]].

\bibitem{Rudelius:2020orz}
T.~Rudelius and S.~H.~Shao,
``Topological Operators and Completeness of Spectrum in Discrete Gauge Theories,''
JHEP \textbf{12} (2020), 172
doi:10.1007/JHEP12(2020)172
[arXiv:2006.10052 [hep-th]].



\bibitem{McNamara:2019rup}
J.~McNamara and C.~Vafa,
``Cobordism Classes and the Swampland,''
[arXiv:1909.10355 [hep-th]].




\bibitem{Frohlich:2009gb}
J.~Frohlich, J.~Fuchs, I.~Runkel and C.~Schweigert,
``Defect lines, dualities, and generalised orbifolds,''
in {\it 16th {I}nternational {C}ongress on {M}athematical {P}hysics}, (2010)
doi:10.1142/9789814304634\_0056
[arXiv:0909.5013 [math-ph]].


\bibitem{Brunner:2013ota}
I.~Brunner, N.~Carqueville and D.~Plencner,
``Orbifolds and topological defects,''
Commun. Math. Phys. \textbf{332} (2014), 669-712
doi:10.1007/s00220-014-2056-3
[arXiv:1307.3141 [hep-th]].

\bibitem{Brunner:2013xna}
I.~Brunner, N.~Carqueville and D.~Plencner,
``A quick guide to defect orbifolds,''
Proc. Symp. Pure Math. \textbf{88} (2014), 231-242
doi:10.1090/pspum/088/01456
[arXiv:1310.0062 [hep-th]].


\bibitem{Carqueville:2012dk}
N.~Carqueville and I.~Runkel,
``Orbifold completion of defect bicategories,''
Quantum Topol. \textbf{7} (2016), 203
doi:10.4171/QT/76
[arXiv:1210.6363 [math.QA]].






\bibitem{Bhardwaj:2017xup}
L.~Bhardwaj and Y.~Tachikawa,
``On finite symmetries and their gauging in two dimensions,''
JHEP \textbf{03} (2018), 189
doi:10.1007/JHEP03(2018)189
[arXiv:1704.02330 [hep-th]].


\bibitem{Tachikawa:2017gyf}
Y.~Tachikawa,
``On gauging finite subgroups,''
SciPost Phys. \textbf{8} (2020) no.1, 015
doi:10.21468/SciPostPhys.8.1.015
[arXiv:1712.09542 [hep-th]].


\bibitem{Dixon:1985jw}
L.~J.~Dixon, J.~A.~Harvey, C.~Vafa and E.~Witten,
``Strings on Orbifolds,''
Nucl. Phys. B \textbf{261} (1985), 678-686
doi:10.1016/0550-3213(85)90593-0


\bibitem{Dixon:1986jc}
L.~J.~Dixon, J.~A.~Harvey, C.~Vafa and E.~Witten,
``Strings on Orbifolds. 2.,''
Nucl. Phys. B \textbf{274} (1986), 285-314
doi:10.1016/0550-3213(86)90287-7


\bibitem{Narain:1986qm}
K.~S.~Narain, M.~H.~Sarmadi and C.~Vafa,
``Asymmetric Orbifolds,''
Nucl. Phys. B \textbf{288} (1987), 551
doi:10.1016/0550-3213(87)90228-8


\bibitem{Vafa:1986wx}
C.~Vafa,
``Modular Invariance and Discrete Torsion on Orbifolds,''
Nucl. Phys. B \textbf{273} (1986), 592-606
doi:10.1016/0550-3213(86)90379-2


\bibitem{Hamidi:1986vh}
S.~Hamidi and C.~Vafa,
``Interactions on Orbifolds,''
Nucl. Phys. B \textbf{279} (1987), 465-513
doi:10.1016/0550-3213(87)90006-X


\bibitem{Hohm:2013vpa}
O.~Hohm and H.~Samtleben,
``Exceptional Field Theory I: $E_{6(6)}$ covariant Form of M-Theory and Type IIB,''
Phys. Rev. D \textbf{89} (2014) no.6, 066016
doi:10.1103/PhysRevD.89.066016
[arXiv:1312.0614 [hep-th]].

\bibitem{Hohm:2013uia}
O.~Hohm and H.~Samtleben,
``Exceptional field theory. II. E$_{7(7)}$,''
Phys. Rev. D \textbf{89} (2014), 066017
doi:10.1103/PhysRevD.89.066017
[arXiv:1312.4542 [hep-th]].

\bibitem{Hohm:2014fxa}
O.~Hohm and H.~Samtleben,
``Exceptional field theory. III. E$_{8(8)}$,''
Phys. Rev. D \textbf{90} (2014), 066002
doi:10.1103/PhysRevD.90.066002
[arXiv:1406.3348 [hep-th]].

\bibitem{Berman:2020tqn}
D.~S.~Berman and C.~D.~A.~Blair,
``The Geometry, Branes and Applications of Exceptional Field Theory,''
Int. J. Mod. Phys. A \textbf{35} (2020) no.30, 2030014
doi:10.1142/S0217751X20300148
[arXiv:2006.09777 [hep-th]].

\bibitem{Blair:2018lbh}
C.~D.~A.~Blair, E.~Malek and D.~C.~Thompson,
``O-folds: Orientifolds and Orbifolds in Exceptional Field Theory,''
JHEP \textbf{09} (2018), 157
doi:10.1007/JHEP09(2018)157
[arXiv:1805.04524 [hep-th]].

\bibitem{Pantev:2005rh}
T.~Pantev and E.~Sharpe,
``Notes on gauging noneffective group actions,''
[arXiv:hep-th/0502027 [hep-th]].

\bibitem{Pantev:2005wj}
T.~Pantev and E.~Sharpe,
``String compactifications on Calabi-Yau stacks,''
Nucl. Phys. B \textbf{733} (2006), 233-296
doi:10.1016/j.nuclphysb.2005.10.035
[arXiv:hep-th/0502044 [hep-th]].

\bibitem{Pantev:2005zs}
T.~Pantev and E.~Sharpe,
``GLSM's for Gerbes (and other toric stacks),''
Adv. Theor. Math. Phys. \textbf{10} (2006) no.1, 77-121
doi:10.4310/ATMP.2006.v10.n1.a4
[arXiv:hep-th/0502053 [hep-th]].


\bibitem{Sethi:1996es}
S.~Sethi, C.~Vafa and E.~Witten,
``Constraints on low dimensional string compactifications,''
Nucl. Phys. B \textbf{480} (1996), 213-224
doi:10.1016/S0550-3213(96)00483-X
[arXiv:hep-th/9606122 [hep-th]].

\bibitem{Vafa:1995gm}
C.~Vafa and E.~Witten,
``Dual string pairs with N=1 and N=2 supersymmetry in four-dimensions,''
Nucl. Phys. B Proc. Suppl. \textbf{46} (1996), 225-247
doi:10.1016/0920-5632(96)00025-4
[arXiv:hep-th/9507050 [hep-th]].

\bibitem{Sen:1995ff}
A.~Sen and C.~Vafa,
``Dual pairs of type II string compactification,''
Nucl. Phys. B \textbf{455} (1995), 165-187
doi:10.1016/0550-3213(95)00498-H
[arXiv:hep-th/9508064 [hep-th]].

\bibitem{Sen:1996na}
A.~Sen,
``Duality and orbifolds,''
Nucl. Phys. B \textbf{474} (1996), 361-378
doi:10.1016/0550-3213(96)00291-X
[arXiv:hep-th/9604070 [hep-th]].

\bibitem{Hull:1998he}
C.~M.~Hull,
``The Nonperturbative SO(32) heterotic string,''
Phys. Lett. B \textbf{462} (1999), 271-276
doi:10.1016/S0370-2693(99)00802-3
[arXiv:hep-th/9812210 [hep-th]].

\bibitem{Bergshoeff:1998re}
E.~Bergshoeff, E.~Eyras, R.~Halbersma, J.~P.~van der Schaar, C.~M.~Hull and Y.~Lozano,
``Space-time filling branes and strings with sixteen supercharges,''
Nucl. Phys. B \textbf{564} (2000), 29-59
doi:10.1016/S0550-3213(99)00483-6
[arXiv:hep-th/9812224 [hep-th]].

\bibitem{Dijkgraaf:1989pz}
R.~Dijkgraaf and E.~Witten,
``Topological Gauge Theories and Group Cohomology,''
Commun. Math. Phys. \textbf{129} (1990), 393
doi:10.1007/BF02096988

\bibitem{Roche:1990hs}
P.~Roche, V.~Pasquier and R.~Dijkgraaf,
``QuasiHopf algebras, group cohomology and orbifold models,''
Nucl. Phys. B Proc. Suppl. \textbf{18} (1990), 60-72

\bibitem{Bantay:1990yr}
P.~Bantay,
``Orbifolds and Hopf algebras,''
Phys. Lett. B \textbf{245} (1990), 477-479
doi:10.1016/0370-2693(90)90676-W

\bibitem{DLM1997}
C.~Dong, H.~Li, G.~Mason, ``Regularity of rational vertex operator algebras,''
Adv. Math. {\bf 132} (1997), no. 1, 148–166. 
doi:10.1006/aima.1997.1681
[arXiv:q-alg/9508018].

\bibitem{DLM1998}
C.~Dong, H.~Li, G.~Mason, ``Twisted representations of vertex operator algebras,'' Math. Annalen \textbf{310} (1998), 571–600 
doi:10.1007/s002080050161
[arXiv:q-alg/9509005].

\bibitem{DLM1998b}
C.~Dong, H.~Li, G.~Mason, ``Twisted representations of vertex operator algebras and associative algebras,'' Internat.~Math.~Res.~Notices 8 (1998),
389–397
doi:10.1155/S1073792898000269
[arXiv:q-alg/9702027].

\bibitem{Carnahan:2016guf}
S.~Carnahan and M.~Miyamoto,
``Regularity of fixed-point vertex operator subalgebras,''
[arXiv:1603.05645 [math.RT]].

\bibitem{Frohlich:2006ch}
J.~Frohlich, J.~Fuchs, I.~Runkel and C.~Schweigert,
``Duality and defects in rational conformal field theory,''
Nucl. Phys. B \textbf{763} (2007), 354-430
doi:10.1016/j.nuclphysb.2006.11.017
[arXiv:hep-th/0607247 [hep-th]].

\bibitem{Burbano:2021loy}
I.~M.~Burbano, J.~Kulp and J.~Neuser,
``Duality defects in E$_{8}$,''
JHEP \textbf{10}, 186 (2022)
doi:10.1007/JHEP10(2022)187
[arXiv:2112.14323 [hep-th]].



\bibitem{Choi:2021kmx}
Y.~Choi, C.~Cordova, P.~S.~Hsin, H.~T.~Lam and S.~H.~Shao,
``Noninvertible duality defects in 3+1 dimensions,''
Phys. Rev. D \textbf{105} (2022) no.12, 125016
doi:10.1103/PhysRevD.105.125016
[arXiv:2111.01139 [hep-th]].

\bibitem{Kaidi:2021xfk}
J.~Kaidi, K.~Ohmori and Y.~Zheng,
``Kramers-Wannier-like Duality Defects in (3+1)D Gauge Theories,''
Phys. Rev. Lett. \textbf{128} (2022) no.11, 111601
doi:10.1103/PhysRevLett.128.111601
[arXiv:2111.01141 [hep-th]].

\bibitem{Choi:2022zal}
Y.~Choi, C.~Cordova, P.~S.~Hsin, H.~T.~Lam and S.~H.~Shao,
``Non-invertible Condensation, Duality, and Triality Defects in 3+1 Dimensions,''
[arXiv:2204.09025 [hep-th]].








\bibitem{Kapustin:2005py}
A.~Kapustin,
``Wilson-'t Hooft operators in four-dimensional gauge theories and S-duality,''
Phys. Rev. D \textbf{74}, 025005 (2006)
doi:10.1103/PhysRevD.74.025005
[arXiv:hep-th/0501015 [hep-th]].

\bibitem{Goddard:1976qe}
P.~Goddard, J.~Nuyts and D.~I.~Olive,
``Gauge Theories and Magnetic Charge,''
Nucl. Phys. B \textbf{125}, 1-28 (1977)
doi:10.1016/0550-3213(77)90221-8

\bibitem{Gukov:2006jk}
S.~Gukov and E.~Witten,
``Gauge Theory, Ramification, And The Geometric Langlands Program,''
[arXiv:hep-th/0612073 [hep-th]].

\bibitem{Gukov:2008sn}
S.~Gukov and E.~Witten,
``Rigid Surface Operators,''
Adv. Theor. Math. Phys. \textbf{14}, no.1, 87-178 (2010)
doi:10.4310/ATMP.2010.v14.n1.a3
[arXiv:0804.1561 [hep-th]].

	
\bibitem{Aharony:2013hda}
O.~Aharony, N.~Seiberg and Y.~Tachikawa,
``Reading between the lines of four-dimensional gauge theories,''
JHEP \textbf{08}, 115 (2013)
doi:10.1007/JHEP08(2013)115
[arXiv:1305.0318 [hep-th]].

\bibitem{Kapustin:2013qsa}
A.~Kapustin and R.~Thorngren,
``Topological Field Theory on a Lattice, Discrete Theta-Angles and Confinement,''
Adv. Theor. Math. Phys. \textbf{18}, no.5, 1233-1247 (2014)
doi:10.4310/ATMP.2014.v18.n5.a4
[arXiv:1308.2926 [hep-th]].

\bibitem{Gukov:2013zka}
S.~Gukov and A.~Kapustin,
``Topological Quantum Field Theory, Nonlocal Operators, and Gapped Phases of Gauge Theories,''
[arXiv:1307.4793 [hep-th]].

\bibitem{Brennan:2022tyl}
T.~D.~Brennan, C.~Cordova and T.~T.~Dumitrescu,
``Line Defect Quantum Numbers \& Anomalies,''
[arXiv:2206.15401 [hep-th]].


\bibitem{Delmastro:2022pfo}
D.~Delmastro, J.~Gomis, P.~S.~Hsin and Z.~Komargodski,
``Anomalies and Symmetry Fractionalization,''
[arXiv:2206.15118 [hep-th]].

\bibitem{Benini:2018reh}
F.~Benini, C.~C\'ordova and P.~S.~Hsin,
``On 2-Group Global Symmetries and their Anomalies,''
JHEP \textbf{03} (2019), 118
doi:10.1007/JHEP03(2019)118
[arXiv:1803.09336 [hep-th]].

\bibitem{Bhardwaj:2022scy}
L.~Bhardwaj and D.~S.~W.~Gould,
``Disconnected 0-Form and 2-Group Symmetries,''
[arXiv:2206.01287 [hep-th]].

\bibitem{Sharpe:2015mja}
E.~Sharpe,
``Notes on generalized global symmetries in QFT,''
Fortsch. Phys. \textbf{63}, 659-682 (2015)
doi:10.1002/prop.201500048
[arXiv:1508.04770 [hep-th]].

\bibitem{Bhardwaj:2022dyt}
L.~Bhardwaj, M.~Bullimore, A.~E.~V.~Ferrari and S.~Schafer-Nameki,
``Anomalies of Generalized Symmetries from Solitonic Defects,''
[arXiv:2205.15330 [hep-th]].















\bibitem{Baez:2004in}
J.~Baez and U.~Schreiber,
``Higher gauge theory: 2-connections on 2-bundles,''
[arXiv:hep-th/0412325 [hep-th]].

\bibitem{Baez:2005qu}
J.~C.~Baez and U.~Schreiber,
``Higher gauge theory,''
[arXiv:math/0511710 [math.DG]].



\bibitem{Brennan:2020ehu}
T.~D.~Brennan and C.~Cordova,
``Axions, higher-groups, and emergent symmetry,''
JHEP \textbf{02}, 145 (2022)
doi:10.1007/JHEP02(2022)145
[arXiv:2011.09600 [hep-th]].

\bibitem{Hidaka:2020iaz}
Y.~Hidaka, M.~Nitta and R.~Yokokura,
``Higher-form symmetries and 3-group in axion electrodynamics,''
Phys. Lett. B \textbf{808}, 135672 (2020)
doi:10.1016/j.physletb.2020.135672
[arXiv:2006.12532 [hep-th]].

\bibitem{Hidaka:2020izy}
Y.~Hidaka, M.~Nitta and R.~Yokokura,
``Global 3-group symmetry and 't Hooft anomalies in axion electrodynamics,''
JHEP \textbf{01}, 173 (2021)
doi:10.1007/JHEP01(2021)173
[arXiv:2009.14368 [hep-th]].


\bibitem{Fraiman:2022aik}
B.~Fraiman and H.~Parra De Freitas,
``Unifying the 6D $\mathcal{N}=(1,1)$ String Landscape,''
[arXiv:2209.06214 [hep-th]].

\bibitem{Harlow:2018jwu}
D.~Harlow and H.~Ooguri,
``Constraints on Symmetries from Holography,''
Phys. Rev. Lett. \textbf{122} (2019) no.19, 191601
doi:10.1103/PhysRevLett.122.191601
[arXiv:1810.05337 [hep-th]].







\bibitem{Harlow:2018tng}
D.~Harlow and H.~Ooguri,
``Symmetries in quantum field theory and quantum gravity,''
Commun. Math. Phys. \textbf{383} (2021) no.3, 1669-1804
doi:10.1007/s00220-021-04040-y
[arXiv:1810.05338 [hep-th]].

\bibitem{Nicolai:1980td}
H.~Nicolai and P.~K.~Townsend,
``N=3 Supersymmetry Multiplets with Vanishing Trace Anomaly: Building Blocks of the N\ensuremath{>}3 Supergravities,''
Phys. Lett. B \textbf{98}, 257-260 (1981)
doi:10.1016/0370-2693(81)90009-5

\bibitem{Bergshoeff:1981um}
E.~Bergshoeff, M.~de Roo, B.~de Wit and P.~van Nieuwenhuizen,
``Ten-Dimensional Maxwell-Einstein Supergravity, Its Currents, and the Issue of Its Auxiliary Fields,''
Nucl. Phys. B \textbf{195}, 97-136 (1982)
doi:10.1016/0550-3213(82)90050-5
\bibitem{Gregory:1997te}
R.~Gregory, J.~A.~Harvey and G.~W.~Moore,
``Unwinding strings and t duality of Kaluza-Klein and h monopoles,''
Adv. Theor. Math. Phys. \textbf{1}, 283-297 (1997)
doi:10.4310/ATMP.1997.v1.n2.a6
[arXiv:hep-th/9708086 [hep-th]].

\bibitem{Marolf:2000cb}
D.~Marolf,
``Chern-Simons terms and the three notions of charge,''
[arXiv:hep-th/0006117 [hep-th]].

\bibitem{Atlas}
J.H.~Conway, R.T.~Curtis, S.P.~Norton, R.A.~Parker and R.A.~Wilson,
``Atlas of finite groups,''
Oxford University Press (1985).

\bibitem{Chang:2020imq}
C.~M.~Chang and Y.~H.~Lin,
``Lorentzian dynamics and factorization beyond rationality,''
JHEP \textbf{10} (2021), 125
doi:10.1007/JHEP10(2021)125
[arXiv:2012.01429 [hep-th]].

\bibitem{Thorngren:2021yso}
R.~Thorngren and Y.~Wang,
``Fusion Category Symmetry II: Categoriosities at $c$ = 1 and Beyond,''
[arXiv:2106.12577 [hep-th]].

\bibitem{GuWen14}
Z.-C.~Gu and X.-G.~Wen, ``Symmetry-protected topological orders for interacting fermions: Fermionic topological nonlinear $\sigma$ models and a special group supercohomology theory,'' Phys. Rev. B, 90(115141)  (2014), doi:10.1103/PhysRevB.90.115141. [arXiv:1201.2648].

\bibitem{Johnson-Freyd:2017ble}
T.~Johnson-Freyd,
``The Moonshine Anomaly,''
Commun. Math. Phys. \textbf{365} (2019) no.3, 943-970
doi:10.1007/s00220-019-03300-2
[arXiv:1707.08388 [math.QA]].


\bibitem{Moore:1984dc}
G.~W.~Moore and P.~C.~Nelson,
``Anomalies in Nonlinear $\sigma$ Models,''
Phys. Rev. Lett. \textbf{53} (1984), 1519
doi:10.1103/PhysRevLett.53.151


\bibitem{Rohm:1985jv}
R.~Rohm and E.~Witten,
``The Antisymmetric Tensor Field in Superstring Theory,''
Annals Phys. \textbf{170} (1986), 454
doi:10.1016/0003-4916(86)90099-0


\bibitem{Lerche:1987sg}
W.~Lerche, B.~E.~W.~Nilsson and A.~N.~Schellekens,
``Heterotic String Loop Calculation of the Anomaly Cancelling Term,''
Nucl. Phys. B \textbf{289} (1987), 609
doi:10.1016/0550-3213(87)90397-X

\bibitem{Lerche:1987qk}
W.~Lerche, B.~E.~W.~Nilsson, A.~N.~Schellekens and N.~P.~Warner,
``Anomaly Cancelling Terms From the Elliptic Genus,''
Nucl. Phys. B \textbf{299} (1988), 91-116
doi:10.1016/0550-3213(88)90468-3

\bibitem{Hsieh:2020jpj}
C.~T.~Hsieh, Y.~Tachikawa and K.~Yonekura,
``Anomaly inflow and $p$-form gauge theories,''
[arXiv:2003.11550 [hep-th]].





\bibitem{Fuchs:2007tx}
J.~Fuchs, M.~R.~Gaberdiel, I.~Runkel and C.~Schweigert,
``Topological defects for the free boson CFT,''
J. Phys. A \textbf{40} (2007), 11403
doi:10.1088/1751-8113/40/37/016
[arXiv:0705.3129 [hep-th]].



\bibitem{Fuchs:2002cm}
J.~Fuchs, I.~Runkel and C.~Schweigert,
``TFT construction of RCFT correlators 1. Partition functions,''
Nucl. Phys. B \textbf{646} (2002), 353-497
doi:10.1016/S0550-3213(02)00744-7
[arXiv:hep-th/0204148 [hep-th]].

\bibitem{Huang2005}
Y.-Z.~Huang,
``Vertex operator algebras, the Verlinde conjecture, and modular tensor categories,''
Proceedings of the National Academy of Sciences
{\bf 102} (2005),
15
doi:10.1073/pnas.0409901102,
[arXiv:math/0412261 [math.QA]].

\bibitem{Reshetikhin:1991}
N.~Reshetikhin, V.~G.~Turaev, ``Invariants of 3-manifolds via link polynomials and quantum groups,'' Invent Math {\bf 103}, 547–597 (1991) 
https://doi.org/10.1007/BF01239527.

\bibitem{Turaev:1992} 
V.~G.~Turaev,
 ``Modular categories and 3-manifold invariants,''
 International Journal of Modern Physics B {\bf 06}, 11n12, pp. 1807-1824 (1992)
 doi:10.1142/S0217979292000876.

\bibitem{DelZotto:2015isa}
M.~Del Zotto, J.~J.~Heckman, D.~S.~Park and T.~Rudelius,
``On the Defect Group of a 6D SCFT,''
Lett. Math. Phys. \textbf{106} (2016) no.6, 765-786
doi:10.1007/s11005-016-0839-5
[arXiv:1503.04806 [hep-th]].


\bibitem{Apruzzi:2021nmk}
F.~Apruzzi, F.~Bonetti, I.~G.~Etxebarria, S.~S.~Hosseini and S.~Schafer-Nameki,
``Symmetry TFTs from String Theory,''
[arXiv:2112.02092 [hep-th]].



\bibitem{Apruzzi:2022dlm}
F.~Apruzzi,
``Higher Form Symmetries TFT in 6d,''
[arXiv:2203.10063 [hep-th]].

\bibitem{DelZotto:2022joo}
M.~Del Zotto, I.~G.~Etxebarria and S.~Schafer-Nameki,
``2-Group Symmetries and M-Theory,''
[arXiv:2203.10097 [hep-th]].

\bibitem{Cvetic:2022imb}
M.~Cveti\v{c}, J.~J.~Heckman, M.~H\"ubner and E.~Torres,
``0-Form, 1-Form and 2-Group Symmetries via Cutting and Gluing of Orbifolds,''
[arXiv:2203.10102 [hep-th]]

\bibitem{Apruzzi:2022rei}
F.~Apruzzi, I.~Bah, F.~Bonetti and S.~Schafer-Nameki,
``Non-Invertible Symmetries from Holography and Branes,''
[arXiv:2208.07373 [hep-th]].

\bibitem{GarciaEtxebarria:2022vzq}
I.~Garc\'\i{}a Etxebarria,
``Branes and Non-Invertible Symmetries,''
[arXiv:2208.07508 [hep-th]].

	


\bibitem{Heckman:2022muc}
J.~J.~Heckman, M.~H\"ubner, E.~Torres and H.~Y.~Zhang,
``The Branes Behind Generalized Symmetry Operators,''
[arXiv:2209.03343 [hep-th]].

\bibitem{Morrison:2020ool}
D.~R.~Morrison, S.~Schafer-Nameki and B.~Willett,
``Higher-Form Symmetries in 5d,''
JHEP \textbf{09}, 024 (2020)
doi:10.1007/JHEP09(2020)024
[arXiv:2005.12296 [hep-th]].

\bibitem{Bhardwaj:2020phs}
L.~Bhardwaj and S.~Sch\"afer-Nameki,
``Higher-form symmetries of 6d and 5d theories,''
JHEP \textbf{02}, 159 (2021)
doi:10.1007/JHEP02(2021)159
[arXiv:2008.09600 [hep-th]].

\bibitem{Albertini:2020mdx}
F.~Albertini, M.~Del Zotto, I.~Garc\'\i{}a Etxebarria and S.~S.~Hosseini,
``Higher Form Symmetries and M-theory,''
JHEP \textbf{12}, 203 (2020)
doi:10.1007/JHEP12(2020)203
[arXiv:2005.12831 [hep-th]].











































\end{thebibliography}
\end{document}